\documentclass[conference]{IEEEtran}
 \IEEEoverridecommandlockouts

\usepackage{amsmath}
\usepackage{bm}
\usepackage{mdframed}
\usepackage{amsthm}

\usepackage[utf8]{inputenc}
\usepackage{graphicx}
\usepackage[colorinlistoftodos]{todonotes}
\usepackage{amssymb}
\usepackage{longtable}
\usepackage{nohyperref}
\hypersetup{bookmarks=false}
\usepackage[noadjust]{cite}
\allowdisplaybreaks
\usepackage{lineno}
\usepackage{lipsum}
\usepackage{multicol}
\usepackage{subcaption}
\usepackage{enumitem}

\usepackage{xcolor}
\usepackage{nicefrac}
\usepackage{comment}

\usepackage[resetcount, ruled]{algorithm2e}

\newcommand{\dtp}[2]{\langle {#1}, {#2} \rangle}
\def\c{\mathcal}
\def\OL{\texttt{OpL}}
\def\p{\tilde}
\def\g{\nabla}

\usepackage{booktabs}
\newcommand{\sumT}{\sum_{t=1}^T}

\begin{document}
\title{Optimistic Learning for Communication Networks}

\author{\IEEEauthorblockN{George Iosifidis\IEEEauthorrefmark{1}, Naram Mhaisen\IEEEauthorrefmark{1}, Douglas J. Leith\IEEEauthorrefmark{2}}\\
	\vspace{-4.5mm}
	\IEEEauthorblockA{
		\IEEEauthorrefmark{1}Delft University of Technology, The Netherlands\\
		\IEEEauthorrefmark{2}Trinity College Dublin, Republic of Ireland}
}

\maketitle
\thispagestyle{plain}
\pagestyle{plain}

\begin{abstract}

AI/ML-based tools are at the forefront of resource management solutions for communication networks. Deep learning, in particular, is highly effective in facilitating fast and high-performing decision-making whenever representative training data is available to build offline accurate models. Conversely, online learning solutions do not require training and enable adaptive decisions based on runtime observations, alas are often overly conservative. This extensive tutorial proposes the use of optimistic learning (\OL) as a decision engine for resource management frameworks in modern communication systems. When properly designed, such solutions can achieve fast and high-performing decisions—comparable to offline-trained models—while preserving the robustness and performance guarantees of the respective online learning approaches. We introduce the fundamental concepts, algorithms and results of \OL, discuss the roots of this theory and present different approaches to defining and achieving optimism. We proceed to showcase how \OL~can enhance resource management in communication networks for several key problems such as caching, edge computing, network slicing, and workload assignment in decentralized O-RAN platforms. Finally, we discuss the open challenges that must be addressed to unlock the full potential of this new resource management approach. 

\end{abstract}
\IEEEpeerreviewmaketitle


\section{Introduction} \label{sec:introduction}

Optimizing communication networks\footnote{We use the term here in its general form, referring to both single and multi-hop systems (networks), with both wired and wireless links.} through rigorous models and principled  algorithms has been a longstanding focus in academia and industry. Yet, despite significant advancements, it remains an open challenge due to the constant evolution of these systems driven by new technologies, growing user base and new types of services. At the same time, the urgency for resource efficiency in communications has intensified, motivated by both economic and environmental considerations. A new promising tool that can contribute in pursuing this goal is the emerging field of online learning and, specifically, the theory  of online convex optimization and its recent extension known as optimistic learning (\OL).

\subsection{Facets of Network Control: from Optimization to Learning}

The framework of Network Utility Maximization (NUM) was the outcome of a systematic effort to create a general toolbox for optimizing communication systems \cite{kelly1997, kelly1998}. In NUM, the network controller (NC) has prior access to user demands and system parameters (e.g., link delays), and formulates an optimization problem that defines the desired system operation, including the optimization criteria (e.g., throughput) and its operational and resource constraints. The problem is solved \emph{offline} and the system is then operated based on the obtained solution. NUM has been developed using convex optimization models and algorithms \cite{convex-algos-bertsekas}, and its optimality guarantees (due to convexity) and decomposability \cite{chiang-tutorial-num} have render it a powerful tool for designing system architectures and protocols, including with a cross-layer approach \cite{chiang-layering}.

The \emph{stochastic} NUM framework\footnote{Some queuing control policies for networks, e.g., \cite{maxweight}, have in fact preceded the development of NUM tools.} \cite{chiang-SNUM} extended this  methodology to dynamic systems where the user demands and system parameters vary with time, based on some stationary random process. In this case, instead of solving offline a problem and applying its solution one-off, the NC makes decisions $\bm x_t$ in a time-slotted fashion after observing the system state at the start of each slot $t$. The seminal Max-weight and Back-pressure policies (i.e., rules) \cite{maxweight}, and their Drift-plus-Penalty extensions \cite{neely-driftplus}, guarantee optimal performance and stability of the involved queues (e.g., backlog of requests), while being agnostic to the statistics of the perturbations. In effect, these policies ensure performance commensurate to that of an ideal benchmark policy $\bm x^\star$ that the controller would have designed if it had access to all future system and user parameters. SNUM has been instrumental in optimizing wireless networks  \cite{tassiulas-book} and various other systems (e.g., smart grid), that can be modeled as networks of queues \cite{neely-book}.

Despite its success, there is growing consensus that (S)NUM cannot serve as the primary optimization toolbox for future communication networks, whose design and mission become increasingly complex. Namely, these systems need to support a broader range of new services, including semantic communications, joint control-compute-and-communication schemes, multimedia and multi-modal communications, and ultra-reliable low-latency communications, among others \cite{6g-iot}. Second, future networks are expected to serve a larger and more diverse user base that includes cyber-physical systems (e.g., robots and vehicles), IoT nodes and embedded devices, with diverse needs and resource constraints \cite{6g-ericsson}. Third,  modern networks are already more heterogeneous than ever, as they include different types of equipment, multi-tier architectures with overlapping serving points (or cells), etc.  Finally, the growing softwarization of networks introduces unprecedented control flexibility but also exacerbates performance and operational expenditure (OpEx) volatility, which is compounded by the inherent uncertainty in resource sharing within the virtualized computing platforms \cite{bib:andres-magazine}.

These developments increase significantly the size and complexity of the problems that must be addressed in network optimization. In particular, assuming the user and system-related parameters are a priori given, or that they evolve according to some stationary process, does not remain a practical assumption. For example, in small cell wireless networks, user churn is non-stationary and often unpredictable \cite{paschos-caching-magazine}; and virtualized base stations in mobile networks exhibit platform and data dependent throughput and energy consumption, which cannot be modeled accurately \cite{jose-icc21}. In other words, there is lack of information for the values of the  user and system parameters, while the functions $f_t$ that are connecting these parameters with the performance metrics of interest in each slot $t$, more often than not, are unknown to the NC.

From a network management perspective, this change has significant implications. Essentially, it transforms the various network optimization problems that the NC must address into learning problems, where decisions must be made under information asymmetry -- specifically, determining $\bm x_t$ without knowing $f_t(\bm x)$ --  while user dynamics and other system perturbations remain highly volatile. This necessitates the development of novel learning-based NUM tools.

\subsection{Offline, Online \& Optimistic Learning}  

In this new era, Machine Learning (ML), and in particular \emph{Deep Learning} (DL), has emerged as a promising approach for network control and resource management \cite{patras-dl-survey}. By leveraging the abundance of raw measurements in these systems, DL can automate the prediction of future parameter values (e.g., channel gains) \cite{cormac-deeplearning} and enable the controller to recover the objective functions $f_t(\bm x)$ that need to be optimized in each slot $t$. Additionally, DL can generate control decisions $\bm x_t$ by solving large-scale problems in near-real-time \cite{dl-for-optimization-in-6G}. Indeed, DL, and ML in general, have been proposed as a replacement for traditional optimization techniques in network management \cite{dl-for-optimization-in-6Gb} and for addressing specific problems such as traffic engineering \cite{DL-TrafficEngineering}, design of intelligent services \cite{dl-end-to-end6G}, and optimization of PHY-layer communications \cite{dl-hoydis}. However, the effectiveness of these solutions depends on the availability of representative training data, which is not always guaranteed; data collection may be costly, or the problem itself may be dynamic and non-stationary. These challenges make the typical ML training cycle prohibitively slow and resource-intensive.

At the other end of the spectrum of AI-based tools for NUM lies the paradigm of \emph{online learning} which does not require pre-processing or offline operations. Instead, it adapts at runtime to the system and environment conditions using real-time observations. In particular, learning algorithms that rely on online convex optimization (OCO) \cite{hazan-book} are principled and provide guarantees for the performance of the dynamic decisions. In this case, the system operation is modeled as an online learning process over $T$ slots, where the NC commits its decision $\bm x_t$ at the \emph{start} of each slot $t$; observes the outcome (function $f_t$ and perturbations) at the end of the slot; and updates its learning rule accordingly. OCO algorithms ensure that the performance achieved by the sequence of decisions $\bm x_t$, $t=1,\ldots, T$, approaches gradually that of the ideal (but unknown) benchmark $\bm x^\star$. This is formally captured using the metric of \emph{regret} $\c R_T$, and one is interested in upper bounds of this gap that are fast-diminishing with the learning horizon, $\lim_{T\rightarrow \infty} \c R_T/T=0$, and at the same time scale gracefully with the problem's dimensionality.

OCO is appealing for network management, see \cite{giannakis-oco-iot,mertikopoulos-tutorial}, for several reasons. First, it builds on online versions of seminal algorithms like gradient descent, which have underpinned previous NUM frameworks, thereby it inherits key properties such as optimality, decomposability, and scalability. Second, it is transparent and interpretable; namely, the regret bounds explicitly reveal how various system parameters influence $\c R_T$. Finally, OCO ensures performance guarantees across diverse perturbation models, including adversarial settings, rendering it a versatile modeling tool. On the other hand, this robustness comes at a cost: it relies on inherently cautious learning, treating the function landscape as entirely unknown and optimizing for worst-case scenarios. While this conservatism safeguards performance in adversarial conditions, it can lead to overly slow adaptation in more predictable scenarios. In fact, the NC often has at least short-term foresight into system and user demands, making such extreme caution unnecessary. Put differently, while OCO’s regret bounds hold universally, its learning performance may lag behind optimal adaptation in scenarios where the problem structure allows it.

Given the complementary strengths and limitations of offline and online learning, a natural question from a network management perspective is whether we can develop a framework that combines their benefits without inheriting their drawbacks. This paper argues that \emph{optimistic learning} (\OL) provides a compelling answer, achieving the best of both worlds by integrating offline-trained predictors (either for functions or actions) into online learning algorithms. The key advantage of \OL~is its ability to significantly accelerate convergence to benchmark performance when predictions are accurate, while maintaining the robustness of traditional OCO methods in cases where predictions are unreliable. This adaptability makes \OL~particularly well-suited for network management, where prior training data is often available but may not always generalize perfectly to real-time conditions. Crucially, \OL~does not depend blindly on such prior models; instead, it evaluates their reliability using real-time observations and seamlessly transitions to pure online learning when necessary. This versatility allows NCs to deploy solutions that perform efficiently across diverse conditions, eliminating the need for rigid trade-offs between robustness and performance.

\section{Motivating Example \& Paper Organization}

We begin the main part of this tutorial with a motivating example on transmission power control in wireless networks with highly volatile channels. Despite its simplicity, this scenario effectively highlights the advantages of \OL~compared to legacy OCO-based learning algorithms and to static and stochastic NUM frameworks. Accordingly, we outline the organization of the material in this paper, summarizing the main contents and key messages of each section, and explaining how they relate to one another. 

\begin{figure}
	\centering
	\begin{subfigure}[t]{0.7\columnwidth}
		\centering
		\includegraphics[width=0.75\textwidth]{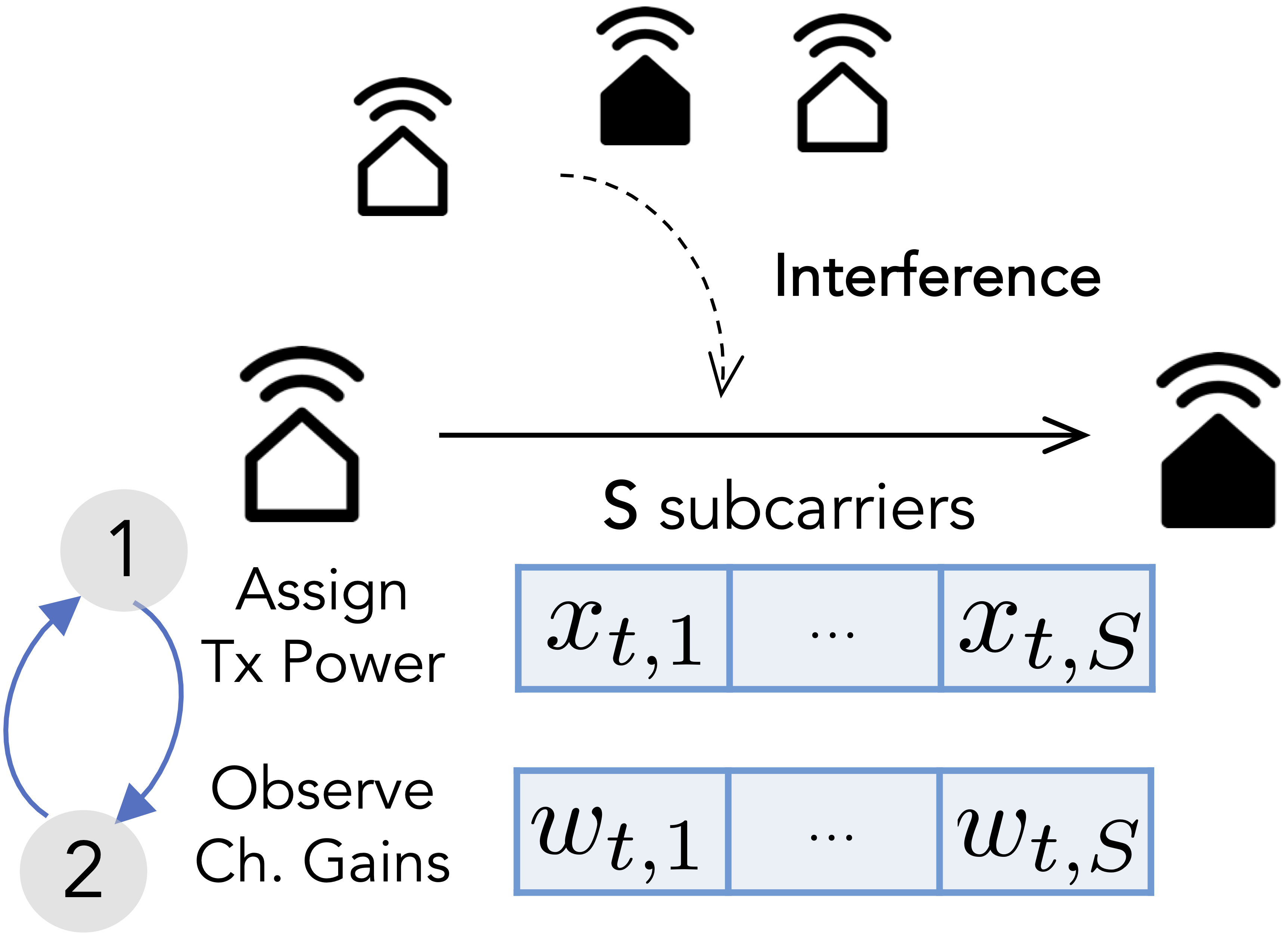}
		\caption{}
	\end{subfigure}%
	\hfill 
	\begin{subfigure}[t]{0.75\columnwidth}
		\centering
		\includegraphics[width=0.8\textwidth, page=1]{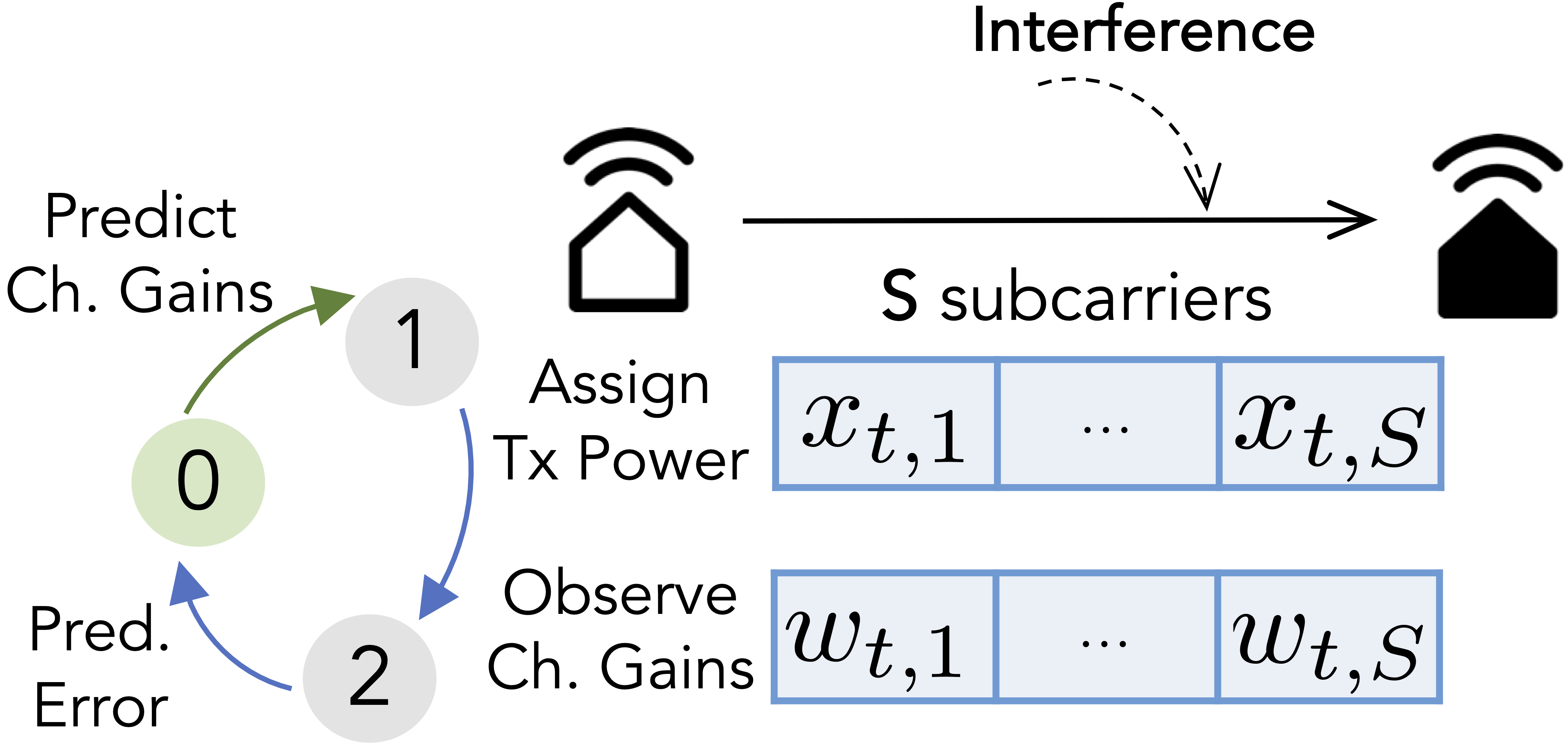}
		\caption{}
	\end{subfigure}
	\caption{\small{\textbf{(a):} OCO-based transmission control in a wireless network with fast-changing channel gains $\bm w_t$, \cite{mertikopoulos-iot}. \textbf{(b)}: Optimistic Learning-based transmission control in a wireless networks, using channel gain predictions $\bm{ \p w}_t$.} }
	\label{fig:motivating-example}
\end{figure}

\subsection{Warm-up Example: Learning how to Transmit}

We use a model inspired by  \cite{mertikopoulos-iot} and refer the reader to the pertinent studies in \cite{srikant-power-control, mertikopoulos-jsac-powercontrol, mertikopoulos-allerton-powercontrol, neely-iot} for similar examples. We consider a system comprising $M$ transmitters and $N$ receivers that communicate over a set $\c S$ of orthogonal channels (subcarriers). Each device can use any subset of channels to transmit to its intended receiver. We study the communication of one such pair of devices and our goal is to decide the transmission power in each of the $S$ channels, so as to maximize the effective throughput while minimizing the transmission power. We consider the dynamic version of the problem (time-slotted model) where the transmission decisions are updated at the beginning of each slot in order to adapt to the vector of time-varying channel gains $\bm w_t \!=\!(w_{t,s}, s\!\in\! \c S)$, where $w_{t,s}$ is the gain of channel $s$ during slot $t$. These parameters are shaped by the distance of the communicating nodes, the interference from other concurrent transmissions, and by various volatile exogenous noise sources. As such, in the general case, they may not only vary across slots, but also change drastically within a slot (e.g., due to fast fading). The objective function of the transmitter in this scenario can be written as:
\begin{align}
f_t(\bm x_t)=\sum_{s=1}^S x_{t,s}-\sum_{s=1}^S\log\big(1+ w_{t,s}x_{t,s}	\big)
\end{align}
where $x_{t,s}$ is the transmission power in subcarrier $s$ during $t$; and the transmission vector $\bm x_t$ belongs to set: 
\begin{align}
\c X=\bigg\{x_s\geq 0, \  s\in \c S; \  \sum_{s=1}^S x_s\leq P_{max} \bigg\}. \notag
\end{align}

This is a key problem in wireless networks and has been extensively studied through the lens of static and stochastic NUM \cite{tassiulas-book}, \cite{srikant-tutorial}, \cite{chiang-layering}. Alas, these approaches impose quite strict assumptions on the perturbation model governing the evolution of the channel qualities. In specific, MaxWeight policies require the channel gains $\bm w_t $ to be observable at the start of each slot and follow an i.i.d. stationary process, so as to ensure the gap between the performance of the dynamic decisions $\bm x_t, t=1,
\ldots, T$ and that of the benchmark policy $\bm x^\star$ converges  to zero. 

On the other hand, OCO-based algorithms can guarantee the convergence of this gap (the \emph{regret}, formally defined in the next section) even when the channels change arbitrarily \emph{and} without requiring to know their values before deciding the transmission $\bm x_t$ in each slot $t$. Therefore, using only past observed gradients, $\bm g_t=\g f_t(\bm x_t)$, an OCO algorithm can ensure an upper bound on the regret growth: 
\begin{align}
\c R_T &= \sumT\! f_t(\bm x_t) - \min_{\bm x\in \c X}\sumT \!f_t(\bm x)  \leq K\sqrt{T},
\label{eq:regret-bound-example}
\end{align}
where $K$ is a constant parameter (independent of $T$) that depends on the diameter $D_{\c X}$ of the decision space $\c X$,  and on the upper bound $L$ of gradients. The second term is the benchmark which yields the optimal performance, had we known all future channels (\emph{best-in-hindsight}); clearly, it is a hypothetical and unknown value. This result states that the average performance loss (or regret) becomes asymptotically zero, i.e., $\lim_{T\rightarrow\infty} \c R_T/T=0$, and this condition is achieved independently of the way the channel qualities change -- notice the steps in Fig. \ref{fig:motivating-example}(a).  

Now, suppose there exists a mechanism, e.g., a deep learning model, that can provide predictions for $\bm w_t$ at the \emph{beginning} of each slot $t$. In this case, we can construct a prediction for the objective function, denoted $\p f_t(\bm x)$, \emph{before} the power assignment, and use it to optimize the $t$-slot decision; see Fig. \ref{fig:motivating-example}(b). This optimistic approach transforms the regret bound to: 
\[
\c R_T \leq K^{\prime}\sqrt{\sumT\|\bm g_t - \bm{\p g}_t\| },
\]
where $K^{\prime}$ is a constant (possibly different than $K$) that depends on $D_{\c X}$ and $L$, and $\|\cdot\|$ a norm reflecting the properties of $\c X$. This new bound states that the regret of the algorithm is of the same order $\c O(\sqrt T$) as the standard OCO bound \eqref{eq:regret-bound-example} when all predictions are inaccurate (worst-case), but shrinks to the impressive bound $\c O(1)$ when the predictions are accurate. Additionally, the \OL~algorithm does not require prior knowledge of the predictor's accuracy but simply assesses its performance (prediction errors) dynamically while using it.

In what follows, we provide the theoretical underpinnings of this remarkable learning mechanism and present applications to representative network management problems.

\begin{figure*}
	\centering
	\includegraphics[scale=0.27]{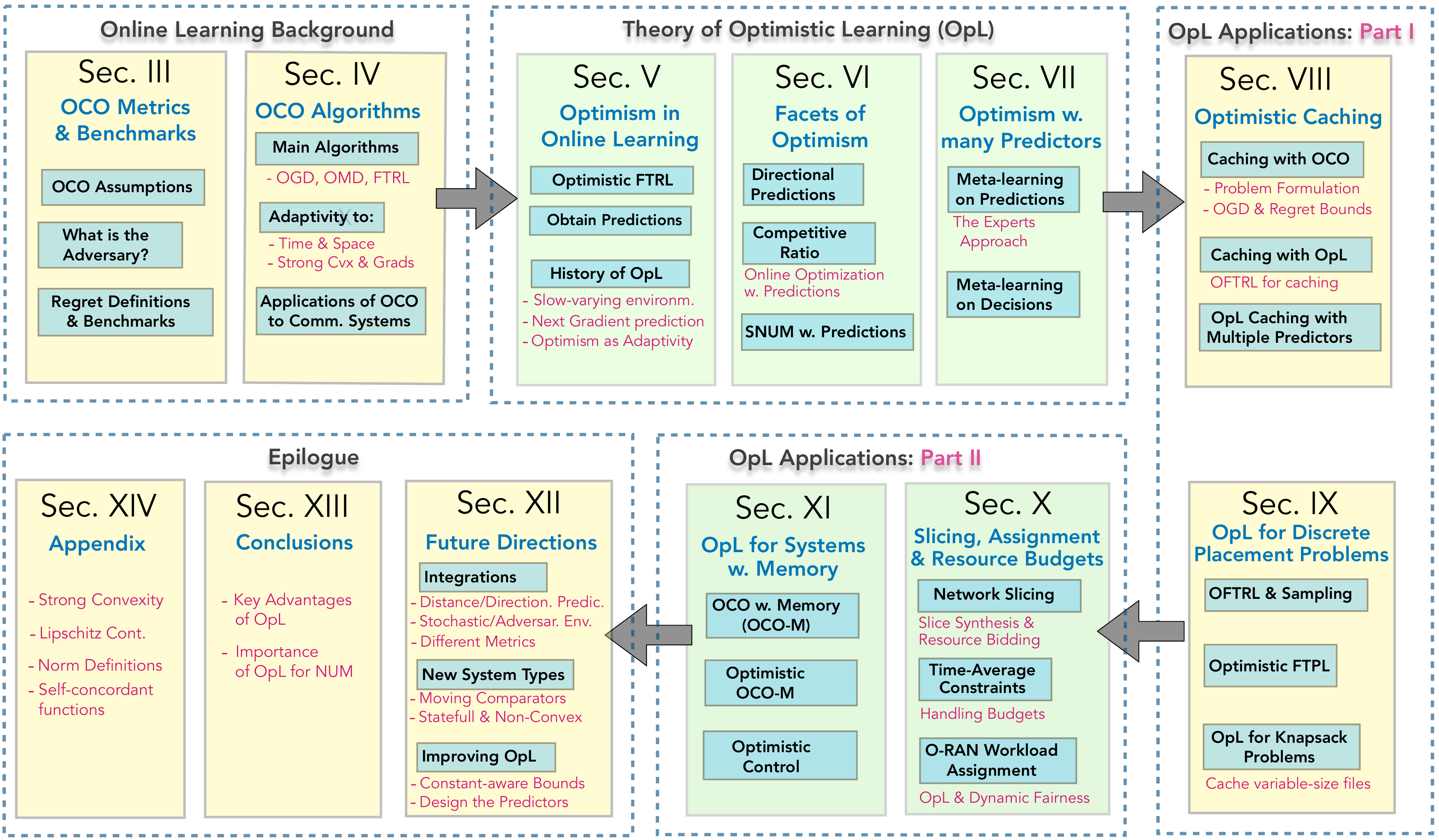} 
	\caption{Paper Organization: Sections and Main Results.}
	\label{fig:paper-organization}
\end{figure*}

\subsection{Organization}
This tutorial is structured as follows.  Section \ref{sec:oco-metrics} introduces the fundamental assumptions, concepts, and metrics such as the Regret and its benchmarks, which are central to the theory of online convex optimization. Section \ref{sec:oco-algorithms} introduces key algorithmic frameworks in OCO, examines their regret bounds, and categorizes different forms of adaptivity in OCO decision-making. This latter discussion allows the reader to study in a unified way an important group of algorithms. Building on these concepts, Section \ref{sec:optimistic-basics} introduces the core idea of this tutorial: optimistic learning. It explains the fundamental mechanics of this approach, analyzes its regret bounds, and demonstrates how it enhances standard OCO regret guarantees. This section also provides a historical perspective on how optimistic learning emerged, while discussing practical issues—largely overlooked until now—regarding how predictions can be obtained. Section \ref{sec:different-optim} explores parallel threads of optimistic learning where the predictions are defined differently (i.e., not as approximate gradients); and competitive ratio metrics replace regret-based evaluations. Section \ref{sec:experts} discusses how to leverage multiple predictors and introduces the meta-learning framework necessary for this problem. It discusses two distinct approaches to addressing this challenge.

After laying the foundations, Section \ref{sec:opt-caching} introduces dynamic caching as an OCO problem and presents the first optimistic learning caching algorithm. This includes scenarios involving a single cache, a network of caches and the case of multiple predictors. Section \ref{sec:opt-discrete-caching} addresses \OL~for discrete (i.e., non-splittable) placement problems, such as whole-file caching or service caching, and presents two solutions leveraging the key algorithmic frameworks Follow-the-Regularized-Leader (FTRL) and Follow-the-Perturbed-Leader (FTPL). Section \ref{sec:opt-applications} discusses three additional categories of problems that can be addressed using \OL. The first involves joint composition and reservation of sliced resources; the second is related to workload assignment in O-RAN systems, and the last focuses on a general network control solution that optimizes a performance metric under  budget constraints -- i.e., enhancing the constrained-OCO framework with optimism. This latter concept lays the foundation for addressing queuing control problems through optimistic learning, leveraging the connection between convex optimization and SNUM. The final section with applications, Sec. \ref{sec:opt-memory}, examines systems that exhibit memory, where decisions influence future performance functions beyond the immediate next time slot. This section explains how optimism can improve performance despite this inertia that affects the observability (and thus the learning capability) of such systems. To make this section more informative, we provide a brief discussion on the pertinent frameworks of OCO with memory (OCO-M) and of Non Stochastic Control via OCO. 

Finally, Sec. \ref{sec:future-directions} presents the most pressing open challenges that need to be tackled so as to make \OL~even more versatile and directly usable in network management problems. These future directions related to creating algorithms that integrate multiple ideas, hence becoming applicable independently of the problem specifics (e.g., both for stochastic and adversarial settings); extending \OL~to new types of systems, and improving the \OL~theory per se. We summarize the conclusions in Sec. \ref{sec:conclusions} and include lastly an Appendix which defines few key mathematical concepts that are used throughout the paper.

\section{OCO Metrics \& Benchmarks }\label{sec:oco-metrics}

This section provides a condensed introduction to online convex optimization (OCO) that has served as the modeling and optimization engine of online and optimistic learning. For a more detailed treatment of the topic we refer the interested reader to the pioneering tutorials \cite{hazan-book}, \cite{shai-book} and the more recent monograph \cite{orabona-book}.

\subsection{Notation \& Assumptions}

We denote with small bold typeface the vectors and with large calligraphic letters the sets. We write $\{\bm x_t\}$ and $\{\bm x_t\}_t$ for a sequence of vectors and use subscripts to index them; $\|\cdot\|$ denotes a general norm and $\|\cdot\|_{\star}$ its dual norm. The inner product of two vectors $\bm x$ and $\bm y$ is defined as $\dtp{\bm x}{\bm y}$. We denote with $\|\cdot\|_2^2$, $\|\cdot\|_1$, $\|\cdot\|_{\infty}$ the Euclidean ($\ell_2$), Manhattan ($\ell_1$) and infinity ($\ell_{\infty}$) norms. More generally, we use $\|\cdot\|_{(t)}=\sqrt{\sigma_t}\|\cdot\|$ to denote a norm that is $\sigma_t$-strongly convex, for some positive parameter $\sigma_t$, and we denote with $\|\cdot\|_{(t),\star}$ the respective dual norm . Vector $\bm g_t$ denotes the gradient $\nabla f_t(\bm x_t)$, and we denote with $g_{t,i}$ the $i$th element of that vector. We use the shorthand notation $\bm c_{1:t}$ for $\sum_{\tau=1}^t \bm c_\tau$, and $\bm{\p x}$ denotes a prediction for vector $\bm x$. When clear from the context, few symbols are redefined and used with different meaning across the sections.

Throughout the tutorial we use asymptotic order of growth for the regret, to describe the performance of the learning algorithms. For detailed definitions and examples of asymptotic growth metrics, we refer the reader to \cite[Chapter 3]{cormen2009introduction}, and we provide below the main definitions for completeness.
\begin{itemize}
	\item \emph{Asymptotic Upper Bound}. We write $\c R_T=\c O(T^c)$, for some $c\geq 0$, if there exists a constant $M$ such that $\c R_T\leq M T^c$, for large enough $T$, i.e., $\forall T\geq T_0$, for some $T_0$. 
	\item \emph{Asymptotic Lower Bound}. We write $\c R_T= \Omega(T^c)$, $c\geq 0$, if there is a constant $M$ such that $\c R_T\geq M T^c$, $\forall T\geq T_0$. 	
	\item \emph{Asymptotic Tight Bound}. We write $\c R_T= \Theta(T^c)$, $c\geq 0$, if there are constants $M_1$, $M_2$, such that $M_1 T^c\leq \c R_T\leq M_2 T^c$, $\forall T\geq T_0$. 		
\end{itemize}
Since our metric of interest is the convergence of time-average regret to zero, i.e., $\lim_{T\rightarrow \infty} \c R_T/T=0$, we are interested in algorithms for which their upper bounds satisfy $c<1$, where, ideally, parameter $c$ should be as small as possible. The lower bounds, on the other hand, are extremely informative in the sense that, when available for some problem, we are able to determine the performance gap of our algorithm from the best possible result. For instance, we will see that for a typical (i.e., without additional assumptions) OCO problem, there is no algorithm that achieves regret which convergence rate smaller than $c\!=\!1/2$. Hence, an algorithm that achieves $\c R_T=\c O(T^{1/2})$ is called \emph{order-optimal}. We note, however, that algorithms with the same \emph{convergence rate} might differ in their constant factors, i.e., the regret bound parameters that do not depend on $T$. For many practical problems, this difference has huge performance implications.

The analysis of \OL\ algorithms requires the same minimal assumptions that typically apply to OCO algorithms:
\begin{enumerate}
\item[\textbf{A1:}]	The decision set $\c X$ is compact and convex with diameter: $ \|\bm x - \bm y\|\leq D, \ \forall \bm x, \bm y \in \c X, $.
\item[\textbf{A2:}]	Functions $\{f_t\}_t$ are convex and Lipschitz continuous with constants $L_{f_t}\leq L, \forall t$. It follows that $\|\nabla f_t(\bm x)\|\leq L$, $\forall t, \bm x\in \c X$, and $\|f_t(\bm x)\|\leq F, \forall \bm x \in \c X$ since the decision set is compact.
\end{enumerate}

Additional assumptions are imposed in some cases, such as the functions or the decision sets being strongly convex. On the other hand, as we will see, we can also drop some of these assumptions, e.g., the boundedness of its diameter or even the convexity of $\c X$, and still design effective learning algorithms.

\subsection{The Concept of Adversary}

\begin{figure}[t]
		\centering
		\includegraphics[width=\columnwidth]{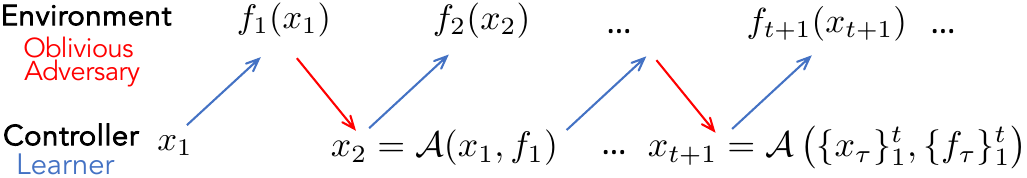}
	\caption{\small{The typical OCO template for the interaction between a learner and an oblivious adversary. The learning algorithm $\c A$ observes the past decisions of the learner and past decisions (i.e., functions) of the adversary and yields the next action.} } 	\label{fig:oco-sequence}
\end{figure}

A common way to introduce OCO is as a non-cooperative repeated game between a decision-maker (\emph{learner}) and an \emph{adversary}. At the beginning of each slot $t$, the learner selects its action $\bm x_t$, and subsequently the adversary determines the cost function\footnote{We follow the traditional OCO terminology, where the problems of interests are minimization problems; thus, the functions to be optimized are assumed to represent some type of cost or penalty.} $f_t(\cdot)$, thereby defining the cost $f_t(\bm x_t)$ that the learner will incur in that slot. The key aspect is that the learner selects $\bm x_t$ without prior knowledge of the cost function for that slot. It only has information about its own past decisions and the cost functions selected by the adversary up to slot $t-1$, which it can feed into some algorithm $\c A$ to determine its next action $\bm x_t$, see Figure \ref{fig:oco-sequence}. The concept of the adversary is central to OCO, and actually the modeling power of this theory stems from it. If a learner manages to perform well when its costs are determined by an adversary that aims to disrupt its learning, then it can also perform well in benign scenarios, e.g., when the cost functions are fixed or when they depend on some stationary process that generates the cost parameters. 

The adversary can be used  to model everything outside the learner: the evolving demands of users for some service, the wireless channel fluctuations, or even the decisions of exogenous optimization/learning processes that run in the background and affect $\{f_t\}_t$. There are, however, different levels of adversity, which naturally determine how much one can learn. In the simplest case, the adversary devises its decisions (which shape $f_t$) without considering the past responses of the learner. This is known as the \emph{oblivious adversary}, presented in the figure above. One may consider stronger adversaries, often termed \emph{adaptive}, which generate cost functions by considering both the current but also (some of) the previous responses of the learner. Technically, this means that the cost function $f_t$ depends not only on $\bm x_t$, but also on the past 
$m$ decisions, i.e., $f_t\left(\bm x_t, \bm x_{t-1}, \ldots, \bm x_{t-m} \right)$. The stronger the adversary, the larger the value of $m$, and the more challenging it becomes for the learner to find a sequence $\{\bm x_t\}_{t}$ such that the regret remains sublinear. We refer the reader to \cite{bianchi-switching} for a detailed discussion on adversarial models.

\begin{table}[t]
	\centering
	\begin{tabular}{c || c } 
		\textbf{Symbol} & \textbf{Meaning} \\ [0.5ex] 
		\toprule
		$\c R_T$ & Accumulated Regret over $T$ slots    \\ [0.5ex] 
		$T$ & Time horizon and number of slots, $t=1,\ldots,T$   \\ [0.5ex] 
		$f_t$ & Objective function during slot $t$ \\ [0.5ex] 
		$\p f_t$ & Prediction for the objective function of slot $t$ \\ [0.5ex] 		
		$\bm g_t$ & Gradient of the objective function at $\bm x_t$, $\g f_t(\bm x_t)$ \\ [0.5ex] 		
		$\bm{\p g}_t$ & Prediction for gradient of slot $t$ \\ [0.5ex] 				
		$\bm x_t$ & Network decision at the beginning of slot $t$  \\ [0.5ex] 
		$\c X$ & Decision space from which vectors $\bm x$ are selected \\ [0.5ex] 		
		$D_{\c X}$ & Diameter of $\c X$, measured with some norm. \\ [0.5ex] 				
		$L$ & Upper bound on Lipschitz constant of functions $\{f_t\}_t$. \\ [0.5ex] 												
		$\Pi_{\c X}(\bm x)$ & Projection operation of vector $\bm x$ onto set $\c X$. \\ [0.5ex] 				
		$ {Co}({\c X})$ & Convex hull of set $\c X$. \\ [0.5ex] 		
		$\bm x^\star$, $\bm x_t^\star$ & Optimal decisions for entire horizon; for slot $t$  \\ [0.5ex] 
		$\eta_t$ & Learning rate during slot $t$  \\ [0.5ex] 		
		$r_t, \sigma_t$ & Regularization function and parameter for slot $t$  \\ [0.5ex] 				
		$\epsilon_t$ & Prediction error for gradients of slot $t$ \\ [0.5ex] 								
		\bottomrule
	\end{tabular}
	\caption{Main Notation}
\end{table}

\subsection{Definitions of Regret}\label{subsec:definitions-of-regret}

In this context, the performance of learning algorithms is captured through the metric of regret. In its most commonly used form, it is defined as:
\begin{align}
\c R_T={\sup_{\{f_t\}_{t=1}^T}\bigg[}\sum_{t=1}^T \Big(f_t(\bm x_t) -  f_t(\bm x^\star) \Big){\bigg]} \label{def:static-regret}
\end{align}
where  $\bm x^\star$ denotes the benchmark, i.e., the performance-maximizing decision we could have devised had we known the entire future (all functions):
\begin{align}
\bm x^\star = \arg\min_{\bm{x} \in \c X} \sum_{t=1}^T f_t(\bm x).
\end{align}
Clearly, this is a hypothetical policy that one cannot hope to know in practice without possessing clairvoyant powers. 

An algorithm that generates the decision sequence $\{\bm x_t\}_{t}$ is said to achieve no-regret if $\lim_{T\rightarrow \infty}\mathcal{R}_T/T=0$. Naturally, we are also interested in achieving this no-regret condition as fast as possible. Nevertheless, there exists a lower bound of $\c R_T=\Omega\big(\sqrt T\big)$, meaning that no algorithm, in the general case, can learn the benchmark's performance at a faster rate. Finally, we note the role of $\sup$, which ensures that this regret bound applies to any possible realization of the problem, i.e., any possible sequence of cost functions.

The benchmark is key to understanding the efficacy of the learning algorithm; however is neither used in the algorithm's design nor required for its implementation. By analogy, in the static NUM framework, the optimization algorithms aim to find the optimal operating point \cite{chiang-layering}, and in stochastic NUM the optimality of the dynamic decision policies is defined relative to a (possibly randomized) policy that one could devise with access to the joint probability distributions of all system perturbations \cite{tassiulas-book}. In other words, the static and stochastic NUM frameworks adopt essentially a static benchmark for static or stationary problems, respectively.

The OCO framework, on the other hand, is much more flexible and allows for greater choice in selecting the benchmark type, which in turn shapes the design of the learning algorithms. Specifically, beyond the static regret in \eqref{def:static-regret}, one can select the more competitive \emph{dynamic regret} metric:
\begin{align}
	\c R_T^d={\sup_{\{f_t\}_{t=1}^T}\bigg[}\sum_{t=1}^T \Big(f_t(\bm x_t) -  f_t(\bm x_t^\star) \Big){\bigg]}, \label{def:dynamic-regret}
\end{align}
where the benchmark is allowed to change at each slot and is selected with one-slot-ahead knowledge:
\begin{align}
	\bm x_t^\star = \arg\min_{\bm{x} \in \c X} f_t(\bm x).
\end{align}
This is a more challenging benchmark to match. Indeed, one can readily prove that $\c R_T \leq \c R_T^d$. This implies that whenever an algorithm ensures sublinear (and thus asymptotically zero average) dynamic regret, it also ensures sublinear static regret. Finally, one can also use the \emph{adaptive regret} benchmark, which minimizes the local regret over every possible time window during $T$, to better capture the dynamics of changing environments \cite{hazan-adaptive-regret}; or the most competitive benchmark $\{\bm x_t^\star\}_{t=1}^T$ that is chosen with full knowledge of the entire horizon and is allowed to change across slots. As expected, achieving no-regret against such more demanding benchmarks requires stricter assumptions about the problem properties. In some cases, the algorithm designer can select the benchmark freely, whereas in other problems, the type of benchmark is dictated by the problem properties. For instance, when learning user association policies in mobile networks, it is only meaningful to focus on dynamic benchmarks since the ideal associations naturally change as users move \cite{mike-handover}.

\section{OCO Algorithms} \label{sec:oco-algorithms}

After introducing OCO and its key concepts, we now turn our attention to solution algorithms. Understanding the mechanics and differences among the different OCO algorithms is non-trivial and, in fact, has been the subject of extensive research per se. One of the excellent references here is \cite{mcmahan-survey} which focuses on FTRL (introduced below) and explains how its variants can be mapped to other algorithms. Similarly, \cite{beck-md} introduces the Mirror Descent (MD) algorithm in an elegant fashion that explains its relation to Gradient Descent (GD); while \cite{tim-blog} provides an informal but rich comparison among the online version of MD (OMD) and FTRL. 

\begin{algorithm}[t]
	\begin{small}
		\nl \textbf{Input}: $x_1\!\in\!\mathcal X$.\\
		\nl  \textbf{Output}: $\{\bm{x}_t\}_t$.\\%
		\nl \For{ $t=1,2,\ldots, T$  }{
			\nl Apply $\bm{x}_t$ and incur cost $f_t(\bm x_t)$; \\
			\nl Observe the new cost function $f_t(\cdot)$ and its gradient $\bm g_t$; \\
			\nl Update the learning rate $\eta_t$; \\
			\nl Calculate the next action using \eqref{eq:ogd-update}
		}
		\caption{\small{Online Gradient Descent (OGD)}}
		\setlength{\intextsep}{0pt} 
		\label{alg:ogd}
	\end{small}	
\end{algorithm}

\subsection{Main Decision Rules}\label{sec:oco-algorithms-rules}

\begin{figure}[t]
	\centering
	\includegraphics[width=0.81\columnwidth]{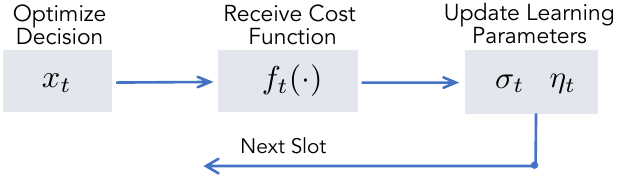}
	\caption{\small{Sequence of events in OCO: \emph{(i)} The learner makes its decision $\bm x_t$; \emph{(ii)} the adversary decides the cost function $f_t(\cdot)$ and the learner observes it; \emph{(iii)} the learner updates its parameters ($\sigma_t$ or $\eta_t$).} } \label{fig:oco-template}
\end{figure}

We start by presenting the main iteration formulas, i.e., the decision rules of the three most common OCO algorithms.

\subsubsection{Online Gradient Descent} OGD was introduced in  \cite{zinkevich-icml03} and its basic iteration is:
\begin{align}
\bm x_{t+1} &= \Pi_{\c X} \Big(\bm{x}_t - \eta_t\nabla f_t(\bm{x}_t)\Big) \quad \text{or, equivalently:}\notag \\
\bm x_{t+1} &= \arg\min_{\bm{x} \in \c{X}} \Big\{ \dtp{\nabla f_t(\bm{x}_t)}{\bm x} +\frac{1}{2\eta_t}\|\bm{x}-\bm{x}_t\|_2^2\Big\},  \label{eq:ogd-update}
\end{align}
where $\c X$ is the set of possible decisions, and $\Pi_{\c X}$ the projection operation. A meaningful interpretation of \eqref{eq:ogd-update} is the following: at the end of each slot $t$, we decide the next action $\bm x_{t+1}$ by moving along the direction of the gradient (i.e., the maximum-improvement direction) while trying to stay close to the previous point $\bm x_t$. The distance to the latter is measured using the Euclidean $\ell_2$ norm, and its effect on the overall decision is weighted by the learning rate $\eta_t\geq 0$. The decision steps when applying OGD can be seen in Fig. \ref{fig:oco-template} and a high-level blueprint of its execution is shown in Algorithm \ref{alg:ogd}.

\subsubsection{Online Mirror Descent} The definition of OMD follows from the above interpretation of OGD, by noticing that one can employ different methods to measure the distance between $\bm x_{t+1}$ and $\bm x_t$ so as to reflect the geometry of the decision space $\c X$ \cite{beck-md}. The OMD update introduced in \cite{warmuth-omd}, is given by:
\begin{align}
	\bm x_{t+1} = \arg\min_{\bm{x} \in \c{X}} \Big\{ \dtp{\nabla f_t(\bm{x}_t)}{\bm x} +\frac{1}{\eta_t} B_{\psi}(\bm x, \bm x_t) \Big\},  \label{eq:omd-update}
\end{align}
where $B_{\psi}$ is merely a distance-measuring tool. This function, known as the Bregman Divergence with respect to the strongly convex and differentiable function $\psi$, is defined as:
\begin{align}
	B_{\psi}(\bm x, \bm y)=\psi(\bm x) - \psi(\bm y) - \dtp{\nabla \psi(\bm y)}{\bm x-\bm y}.
\end{align}
Different choices of $\psi$ lead to different OMD algorithms. For instance if $\psi(\bm x)=\frac{1}{2}\|\bm x\|_2^2$, we recover the OGD expression. Another common choice is the negative entropy, i.e., $\psi(\bm x)=\sum_{i=1}^Nx_i\ln x_i$ which, as discussed later, leads to useful closed-form expression for the OMD decision rule. The main advantage of OMD lies in its ability to mitigate the impact of problem dimensionality on the regret bound through appropriate selection of this distance-measuring function. This property makes OMD particularly appealing for certain large-scale optimization problems.

\begin{figure}[t!]
	\centering
	\begin{subfigure}[t]{0.48\textwidth}
		\centering
		\includegraphics[width=0.68\textwidth]{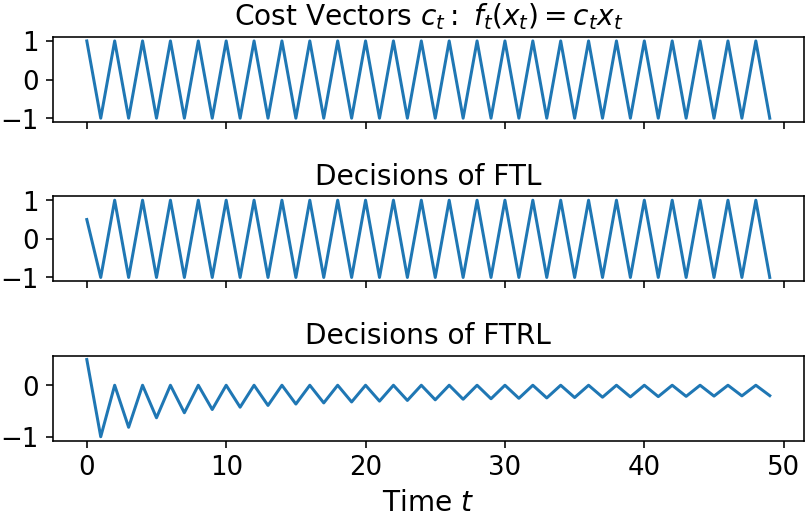}
		\caption{}
	\end{subfigure}%
	\vfill 
	\vspace{2mm}
	\begin{subfigure}[t]{0.48\textwidth}
		\centering
		\includegraphics[width=0.68\textwidth, page=1]{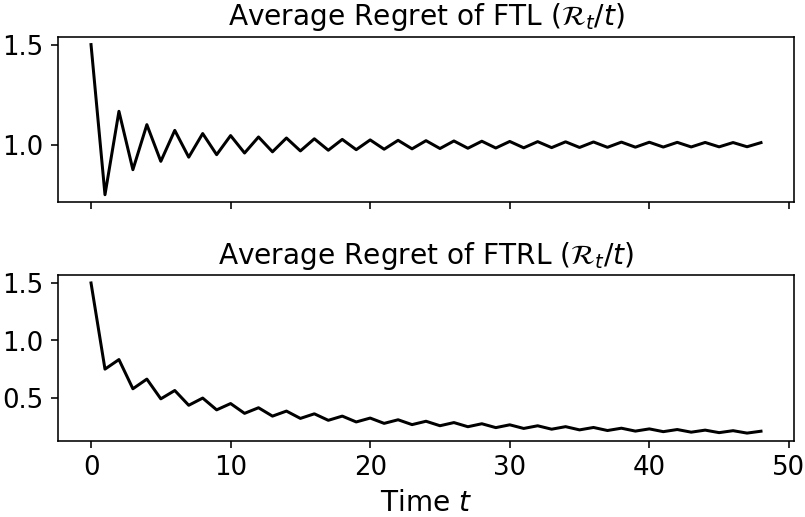}
		\caption{}
	\end{subfigure}
	\caption{\small{\textbf{(a):} The decision space is $x\in [-1,1]\subset \mathbb R$, the objective (cost) function is linear, $f_t=c_tx$, and the cost parameters $c_t$ change in each slot, alternating -1 and 1. FTL is unstable (\emph{ping-pong}) as it is heavily influenced by the sign of the aggregate cost at each slot; while FTRL is converging to a stable decision.  \textbf{(b)}: The evolution of average regret for different time windows $t\!=\!1,\ldots,T$ (with $T\!=\!50$) for the two algorithms demonstrates that FTL does not learn.} }
	\label{fig:ftl-ftrl}
\end{figure}

\subsubsection{Follow The Regularized Leader} The third key algorithm, which plays a crucial role in the optimistic learning tools presented next, is FTRL. Here, the idea stems from the observation that a natural candidate for the next decision is one that optimizes the aggregate observed costs until that slot: 
\[
\bm x_{t+1}=\arg\min_{\bm x\in \c X} \sum_{\tau=1}^tf_\tau(\bm x).
\]
However, this intuitive algorithm, known as Follow-The-Leader (FTL), might induce highly-fluctuating (i.e., unstable) decisions across time, thereby failing to learn effectively and leading to super-linear regret, as it is demonstrated with a toy example in Fig. \ref{fig:ftl-ftrl}. FTRL mitigates this issue through regularization that injects the necessary inertia: 
\begin{align}
	\bm{x}_{t+1}\! =\! \arg\min_{\bm{x} \in \mathcal{X}} \Big\{\! \dtp{\bm{g}_{1:t}}{\bm x}  +r_{1:t}(\bm{x})\!\Big\}.  \label{eq:ftrl-update}
\end{align}
Note that, unlike OMD and OGD, this decision rule relies on the entire history of gradients $\bm g_{1:t}=\sum_{\tau=1}^t \nabla f_\tau(\bm x_\tau)$.   

The function $r_{1:t}(\bm x)=\sum_{\tau=1}^tr_{\tau}(\bm x)$ represents the aggregate regularization\footnote{We follow the convention in \cite{mcmahan-survey} and define the regularization as a sum of per-slot regularizers, as this facilitates the discussion about learning rates.}, where $r_\tau(\bm x)$ is the additional regularization impoased at time slot $\tau$. Unlike OMD and OGD, the purpose of regularization in FTRL extends beyond distance measurement; it influences the attained regret bounds as well as the computational and memory requirements of the decision rule. One of the widely used regularization functions is $r_t(\bm{x})\!={\sigma_t}\|\bm{x}\|_2^2$, where parameter $\sigma_t$ decides the regularization weight. One can understand the role of this parameter by linking it to the learning rate (as those used in OGD and OMD), via the formula \cite{mcmahan-survey}: 
\[
\sigma_t=\frac{1}{\eta_t} - \frac{1}{\eta_{t-1}}, \ \forall t
\]
which implies $\sigma_{1:t}=1/\eta_t$. A different regularization choice is the proximal function\footnote{A regularizer $r_t(\bm x)$ is called proximal if $\bm x_t=\arg\min_{\bm x\in\c X}r_t(\bm x)$; and non-proximal otherwise \cite{mcmahan-survey}. This concept is related to proximal operators that are used in various optimization techniques, cf. \cite[Ch. 6]{beck-book}.} $r_t(\bm{x})\!=({\sigma_t}/2)\|\bm{x}-\bm x_t\|_2^2$, which links the next decision with the current one $\bm x_t$ (similarly to OGD), but requires more memory and more involved calculations.

It is noteworthy that with non-proximal regularizers, the FTRL rule does not depend on the previous decision, in contrast to OGD/OMD, where $\bm x_t$  directly influences $\bm x_{t+1}$. More generally, a fundamental distinction between FTRL and OGD/OMD lies in how these algorithms maintain the solution (or problem) \emph{state}. OGD and OMD track only the current optimal point $\bm x_t$ and determine the next decision $\bm x_{t+1}$ based on the most recent gradient $\bm g_t$. FTRL, on the other hand, utilizes the entire history of gradients $\bm g_{1:t}$ to define its state. To put it differently, OGD and OMD only inform us that the algorithm has reached a certain point, while FTRL conveys information about how well that point represents optimality w.r.t. the entire history of gradients. We refer the reader to \cite{mcmahan-survey} for a detailed discussion on this subtle point that, nevertheless, has significant technical and performance implications.

\subsubsection{Function Linearization} Finally, it is worth discussing the concept of linearization that affects the properties of the above algorithms; see \cite[Sec. 2.1]{mcmahan-survey} and \cite[Sec. 2.4]{shai-book}. This is more clearly demonstrated in the case of FTRL. Its original decision rule can be described as:
\begin{align}
	\bm x_{t+1}=\arg\min_{\bm x\in \c X} \Big\{ f_{1:t}(\bm x) + r_{1:t}(\bm x) \Big\}
\end{align}
where we select the minimizer of the hitherto revealed cost functions, while regularizing for stability and for improving the learning performance. This is quite intuitive, given that our ultimate goal is to learn to perform as well as the benchmark $\bm x^\star$ which is defined exactly as the minimizer of those functions. However, more often than not, one uses the formula \eqref{eq:ftrl-update} where instead of the actual functions the decision is optimized over the linearized surrogate functions $\dtp{\bm g_{1:t}}{ \bm x}$. 

This idea exploits the property of convexity:
\[
f_t(\bm x_t) - f_t(\bm x^\star)\leq \dtp{\bm g_t}{\bm x_t - \bm x^\star},
\]
which gives rise to the observation (first made in \cite{zinkevich-icml03}) that the regret of $\{f_t\}_t$ is bounded by that of $\{\bar f_t\}_t$, where $\bar f_t(\bm x)\!=\dtp{\bm g_t}{\bm x}$, i.e.,
\[
\c R_T(\bm x^\star, f_t)\leq \c R_T (\bm x^\star, \bar f_t).
\]
Hence, one can use these linear functions instead of the initial convex functions, to perform online (linear) learning. This comes with several advantages pertaining to memory requirements of the algorithm (remains constant, as $\bm g_{1:t}$ is a single vector) and to its computational complexity (often admits closed-form solutions). On the other hand, the linearization reduces the information available to the learner (tracking the gradients and not the entire function) which in turn affects the learning performance for some types of problems.

\subsection{Regret Bounds \& Adaptivity} \label{subsec:regret-adaptivity}

\begin{table}[!t]
	{\small{
			\centering
			\caption{Basic OGD/OMD Regret Bounds}
			\begin{tabular}{p{3.25cm}p{1.87cm}p{2.55cm}}				
				\toprule
				\textbf{Algorithm} & $\c R_T\leq$ & \textbf{Remarks} \\
				\midrule
				OGD,  $\eta_t\!=\!\frac{D}{L\sqrt{t}}$ & $\frac{3}{2}LD\sqrt T$  & \qquad --- \\  [1.1ex] 
				OGD, $\eta_t\!=\!\frac{1}{\alpha t}$ &   $\frac{L^2}{\alpha}(1\!+\!\log T)$ & $f_t$ is $\alpha$-strongly cvx\\	[1.1ex] 	
				OGD, $\eta_t\!=\!\frac{D}{\sqrt{2\!\sum_{\tau\!=\!1}^t\!\! \|\bm g_\tau\|_2^2}}$ &   $D\sqrt{\!\sum_{t=1}^T\!\|\bm g_t\|_2^2 }$ & \qquad --- \\	[1.9ex] 					
				OMD, entropic regul. &   ${L}\sqrt{2T\log N}$  & $\c X=\c S_N$ (simplex)\\	[0.3ex] 			
				\bottomrule
			\end{tabular}
			\label{tab:regret-bounds}}}
\end{table}

Having introduced the main algorithms, the next question concerns their performance in terms of convergence rate. The starting point is the lower bound $\c R_T=\Omega(\sqrt T)$  \cite{abernethy-lower-bound}, which establishes that, in the general case, no algorithm can achieve a faster convergence rate. In other words, an adversary can construct a sequence of convex functions that forces any algorithm to make a sufficient number of mistakes, preventing convergence at a rate faster than this bound. Still, despite this lower bound, there are important differences among the various learning algorithms. First, in terms of  performance, their regret bounds are influenced differently by problem parameters such as the decision space diameter $D_{\c X}$ and the upper bound on gradients $L$. Second, if the decision space or the objective functions possess additional properties, the regret bounds can be improved. Notably, \emph{strong convexity} leads to highly-desirable regret bounds with logarithmic dependence on the time horizon $T$ which, further, do not depend on the dimension of the decision space, see \cite[Sec. 3.3.1]{hazan-book}. 

Below, we provide an overview of methods that enhance the performance of OCO algorithms through the lens of \emph{adaptivity} and its various forms. Table \ref{tab:regret-bounds} summarizes some representative bounds for a quick reference.

\subsubsection{Adaptivity to Time Horizon} The adaptivity of the algorithms is essentially determined by the learning rates ($\eta_t$ or $\sigma_t$). The simplest way to make an OCO algorithm adaptive is by using information about the horizon $T$ during which we intend to use the algorithm, when this information is available. When the learner does not have any other information except from the problem's main parameters $D_{\c X}$ and $L$ (i.e., $T$ is unknown) it can minimize the regret upper bound of, e.g., OGD, by setting $\eta_t=D/L\sqrt{2t}$ to ensure $\c R_T\leq (3/2)DL\sqrt{T}$ \cite[Th. 3.1]{shai-book}. These bounds are often called \emph{anytime bounds}, as they are valid for any value of $T$. When there is prior information about the time horizon, we can adapt the learning rate accordingly by setting  $\eta=D/L\sqrt{2T}$, and this way reduce the regret to $\c R_T\leq DL\sqrt{T}$ (fixed horizon bound). In general, there is a streamlined method, termed the \emph{doubling trick}, for allowing algorithms to work without access to $T$, see discussion in \cite[Sec. 5.3]{mertikopoulos-tutorial}. This technique produces regret bounds which are, at most, a factor of $2/(\sqrt{2}\!-\!1)$ worse than the respective fixed horizon bound (achieved with knowledge of $T$).

\begin{table*}[t]
	\centering
	\begin{tabular}{c || c c c } 
		\textbf{Reference} & \textbf{Decision Variables} & \textbf{Goal} & \textbf{Unknown Parameters at each slot} \\ [0.5ex] 
		\toprule
		\cite{AWERBUCH200897} & Selection of routing path & Minimum-cost routing & Path delays    \\ [0.5ex] 
		\cite{paschos-infocom, tareq-omd-caching} & Cached files (continuous) & Maximum-utility caching (hit ratio) & File requests    \\ [0.5ex] 
		\cite{abishek-sigmetrics-caching, abishek-caching} & Cached files (discrete) & Maximum-utility caching (hit ratio) & File requests    \\ [0.5ex] 		
		\cite{salem-similarity-caching} & Cached files (continuous) & Maximum-utility-similarly caching & File requests    \\ [0.5ex] 		
		\cite{srikant-power-control, mertikopoulos-iot, neely-iot, mertikopoulos-allerton-powercontrol, mertikopoulos-jsac-powercontrol} & Transmission Power & Maximum throughput in wireless networks & Channel quality \\ [0.5ex] 				
		\cite{paschos-streaming} & Bitrate selection& Maximize video streaming quality & Channel quality \\ [0.5ex] 						
		\cite{oco-task-offloading} & Amount of offloaded data & Energy minimization in URLLC & Energy and delay parameters \\ [0.5ex] 	
		\cite{giannakis-TSP17} & Scheduling of workloads & Energy minimization in datacenters & Energy and bandwidth cost \\ [0.5ex] 		
		\cite{oco-carlee-cloud, oco-cloud-paschos, oco-tnsm-jb, edge-server-hosting}	& Leasing of resources & Maximize-Performance / Minimize-cost  & Prices and user demand \\ [0.5ex] 					
		\bottomrule
	\end{tabular}
	\caption{\small{Representative Applications of OCO in Network Management Problems}}
	\label{table:oco-applications}
\end{table*}

\subsubsection{Adaptivity to Decision Space} We typically select a learning rate that is proportional to the diameter $D$ of $\c X$ (the maximum distance of any two points) so as to adapt the constant factors of the regret bound to the properties of the problem. Whereas linear dependency on the diameter is satisfactory from a theoretical point of view, it may be problematic for many resource allocation problems in communication networks. This motivates finding ways to reduce further the dependency.

Both OMD and FTRL encompass this idea and select their distance-measuring method (for OMD, cf.  \cite{mertikopoulos-tutorial}) or regularization (for FTRL) based on the geometry of $\c X$. For instance, when the decision space is the unit simplex:
\begin{align}
	\c S_N=\left\{ \bm x  \ \bigg \vert \  \sum_{i=1}^N x_i=1, \ x_i\geq 0	\right\},
\end{align}
we can use the entropic regularizer:
\begin{align}
	r_{1:t}(\bm x)=\frac{1}{\eta} \sum_{i=1}^N x_i \log x_i, \quad \text{with} \ \ \eta=\frac{\sqrt{\log N}}{L\sqrt{2T}} \label{eq:entropic-reg}
\end{align}
and improve the regret to:
\begin{align}
	\c R_T \leq L\sqrt{2T\log N }.
\end{align}
This bound has a logarithmic dependency on the problem size $N$, whereas with the quadratic regularizer (or when using OGD) this dependency would have been linear.  

Going a step further, in some settings, e.g., when the benchmark lies close to the origin of $\c X$, regreet bounds that depend on $D_{\c X}$ might be considered vacuous. In such cases, a stronger and more refined notion of adaptivity on the space can be used, namely adaptivity to the norm of $\bm x^\star$. This reflects how well the algorithm performs relative to the actual solution's scale, rather than the geometry of the space, see \cite{orabona-betting}.

\subsubsection{Adaptivity to Strong Convexity} When the cost functions or the decision space have additional properties beyond being convex, we can further refine the bounds by changing accordingly the algorithm. Perhaps the most prominent such case is the property of strong convexity. A simple way to exploit this property is to decrease the learning rate faster with time, using $\eta_t=1/\alpha t$ instead of $1/\sqrt t$, which leads to highly-desirable logarithmic regret bounds that are, also, independent of the decision space diameter $D$:
\[
\c R_T\leq \frac{L^2}{\alpha}(1+\log T).
\]
Similarly, if the decision space is strongly convex, such as when defined based on the $\ell_2$ norm,  there are improved regret bounds for various OCO algorithms. For example, the regret bound of the Online Frank-Wolfe algorithm is reduced to $\c O(\sqrt T)$ \cite{strong-convex-zan-aaai21} by fine-tuning the update rules in this case, and one can leverage strong convexity to design better optimistic algorithms \cite{dekel-hint-neurips17}, as we will see in the sequel.

\subsubsection{Adaptivity to Gradients} The bounds presented so far depend on the problem properties as they are expressed in terms of the maximum possible cost gradient norms $L$ and the diameter $D$ of the decision space $\c X$. Nevertheless, not all problem instances will exhibit behavior that will involve these limits. For instance, one may encounter an \emph{easy} instance where the cost functions that the adversary selects, do not have large gradients. We would like to be able to benefit from these cases and reduce the regret.

The seminal AdaGrad algorithm \cite{duchi-jmlr-2011}, and its numerous variations, that adapt the regularization based on the observed data is a paradigm-shift that aims exactly to realize this improved performance. This approach leads to regret bounds that depend on the observed gradients of the functions, instead of their maximum possible value (upper bounds). For instance, OGD with a horizon-adaptive step $\eta_t=D/L\sqrt{2T}$ achieves $\c R_T\leq DL\sqrt T$, while if we use the following step that adapts to the observed gradients:
\begin{equation}
\eta_t=\frac{D}{\sqrt{2\sum_{i=1}^t\|\bm g_{i}\|_2^2}}, \label{eq:ada-grad-rate}
\end{equation}
OGD achieves a regret bound:
\begin{equation}
\c R_T\leq \sqrt2 D\sqrt{\sum_{t=1}^T\|\bm g_t\|_2^2 } \label{eq:ada-grad-regret}
\end{equation}
which is upper-bounded by $DL\sqrt{T}$, but in practice can be significantly improved if the actual gradients are smaller than their maximum value. Refinements of this idea extend in various interesting directions, including, for instance, the usage of different learning rates per coordinate, so as to learn faster/slower along the directions of $\c X$ where the cost functions exhibit larger/smaller differences.

\subsubsection{Adaptivity to Prediction Errors} Finally, it is reasonable to conclude, even at this early point, that optimism can be regarded as another form of adaptivity -- arguably the most comprehensive one. The concept of leveraging the problem's temporal variability or our ability to predict future cost functions to achieve lower regret is closely related to the previously discussed forms of adaptivity. In essence, optimistic learning involves adjusting learning rates based on observed prediction errors, analogous to how methods such as AdaGrad \cite{duchi-jmlr-2011} adjust learning rates based on observed cost gradients, see eq. \eqref{eq:ada-grad-rate}.

\subsection{OCO in Communication Systems}

Due to its relevance to previous NUM frameworks (stemming from convexity), its modeling versatility (requiring no strict assumptions), and its performance robustness (extending beyond static and stationary conditions), OCO-based algorithms have already been applied to a variety of resource allocation problems in communication systems. Table \ref{table:oco-applications} summarizes representative examples. One could argue that the entire spectrum of problems previously addressed using first-order convex optimization algorithms (such as the gradient descent algorithm) can be revisited and analyzed in their dynamic and more general forms using OCO.\footnote{In fact any OCO algorithm with sublinear regret can be used in stochastic optimization problems; see Online-to-Batch Conversion in \cite[Ch. 3]{orabona-book}.}

For instance, as highlighted in the motivating example in the introductory section \ref{sec:introduction}, OCO has been employed in the design of transmission control policies for (mobile and IoT) wireless networks. These networks often face unpredictable and highly volatile channel conditions caused by factors such as fast fading, intermittent interference, high-speed nodes, and other unpredictable exogenous influences \cite{srikant-power-control, mertikopoulos-iot, neely-iot, mertikopoulos-allerton-powercontrol, mertikopoulos-jsac-powercontrol}. Additionally, OCO policies have been applied to content caching problems, beginning with the work in \cite{paschos-infocom} and followed by \cite{abishek-sigmetrics-caching, abishek-caching, salem-similarity-caching, tareq-omd-caching, tareq-sigmetrics}. They have also been extensively used in edge computing to determine how, where, and when to offload computation tasks in IoT networks \cite{giannakis-oco-iot}. Other notable application domains include video streaming \cite{paschos-streaming}, URLLC schemes \cite{oco-task-offloading}, and demand-response algorithms in smart grids \cite{giannakis-smart-grid}. Furthermore, problems such as routing were among the first to benefit from OCO, as demonstrated in the seminal work \cite{AWERBUCH200897}.

%
%
\section{Optimism in Online Learning} \label{sec:optimistic-basics}

In this section we present the main mechanisms and results of optimistic learning using the FTRL framework as a basis. Accordingly, we take a step back and provide a historical perspective on the introduction of optimism in online learning and discuss the key results associated with this theory.

\begin{figure}[t]
	\centering
	\includegraphics[width=\columnwidth]{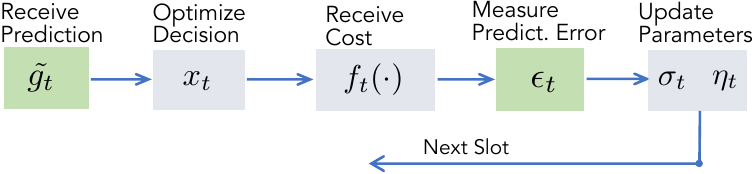}
	\caption{\small{Optimistic OCO template for the interaction between a predictions-assisted learner and an adversary. 
	}} \label{fig:oco-optimistic-sequence}
\end{figure}

\subsection{Optimistic FTRL}

Assume that we have a mechanism in place that can provide predictions for the next-slot gradient $\bm{\p g}_{t+1}$, $\forall t$. These predictions may be accurate, meaning they are close to the yet-to-be-observed gradient $\bm g_{t+1}$, or they may significantly deviate from it. In either case, the  \OL~framework does not make assumptions on the prediction errors $\{\epsilon_t\}_t$, and only requires that they can be observed at the end of each time slot. 

Using this prediction information, we can revise the FTRL decision rule to its optimistic variant (OFTRL), as follows:
\begin{align}
\bm{x}_{t+1} = \arg\min_{\bm{x} \in \mathcal{X}} \Big\{ \dtp{\bm{g}_{1:t}+\p{\bm{g}}_{t+1}}{\bm x} +r_{1:t}(\epsilon_{1:t}; \bm{x})\Big\}.\label{eq:optimistic-ftrl-update}
\end{align}

This rule differs from the standard (non-optimistic) FTRL decision rule in two ways. First, it incorporates the predicted gradient $\bm{\p g}_{t+1}$, which corresponds to the same time index as the optimized decision $\bm{x}_{t+1}$. This intuitive modification captures our hope, or optimism if you like, that if the prediction is accurate we will update the decision in a prescient fashion, i.e., as if we knew the missing next-slot function. Second, the regularization depends on the prediction errors, allowing the framework to control the level of trust in $\bm{\p g}_{t+1}$. When the predictions are inaccurate the regularization increases so as to prevent abruptly fast shifts towards the direction of these erroneous gradients. Conversely, if the predictions are found to be reliable, the learner trusts them gradually and guides its decisions towards the direction indicated by the predictions.

The learning process of this modified OCO routine is presented in Figure \ref{fig:oco-optimistic-sequence}, and its algorithmic template is summarized in Algorithm \ref{alg:optimistic-template}. Compared to the standard OCO sequence of events (Fig. \ref{fig:oco-template}), here we see the addition of the prediction before optimizing for the next decision, and the evaluation of the predictor's accuracy (measuring its error), after the cost function is revealed. The last step, as before, concerns the update of the regularization parameters using the prediction error, and is central to the performance of the algorithm. 

The regularization in \texttt{OpL} follows the regularization techniques in OCO, with the necessary modifications so as to account for the prediction errors. A typical example for the additional regularization at each slot $t$ is:
\begin{align}
&r_t(\epsilon_{1:t}; \bm x)=\sigma_t\|\bm x\|_2^2 \label{eq:oftrl-reg-l2}\\   
&\text{with} \ \sigma_t=\sigma\left(\sqrt{\epsilon_{1:t}}-\sqrt{\epsilon_{1:t-1}}\right), \ \ 
 \  \epsilon_t\!=\!\|\bm g_t \!- \bm{\p g}_{t}\|_2^2. \notag
\end{align}
which results in aggregate regularization:
\begin{align}
	r_{1:t}(\bm x)=\sigma \sqrt{\epsilon_{1:t}}\|\bm x\|_2^2. \notag
\end{align}
This regularizer simply extends the data-adaptive FTRL regularizer by replacing the gradients with the prediction errors $\{\epsilon_t\}_t$, where the latter are measured using the $\ell_2$ norm. More generally, errors should be measured using the norm relative to which the regularizer is strongly convex (see Appendix for details). For instance, if an entropic regularizer is employed, which is strongly convex with respect to the $\ell_{1}$ norm, the prediction errors should be assessed using the respective dual norm, i.e., $\ell_{\infty}$. This is a technical requirement stemming from the analysis of FTRL and serves the purpose of adapting to the geometry of the decision space $\c X$.

With these modifications--inclusion of predicted gradient and the regularization based on prediction errors--the regret is transformed into a prediction-adaptive quantity:
\begin{align}
	\c R_T=\c O\left( \sqrt{\epsilon_{1:T}}	\right). \notag
\end{align}
When all predictions are accurate, i.e, $\|\bm g_t-\bm{\p g}_t\|_2^2=0, \forall t$, the regret becomes $\c R_T\!=\!0$,  indicating that the learner matches the performance of the benchmark extremely fast. Even if few predictions are erroneous, the achieved regret is constant, $\c R_T\!=\!\c O(1)$. Conversely, even when all predictions are maximally off point, i.e., $\|\bm g_t-\bm{\p g}_t\|_2^2=2L, \forall t$, the regret remains order-optimal, meaning that it matches the best achievable $\c R_T$ bound in terms of $T$, see Table \ref{tab:ftrl-bounds}. There is a caveat however: optimistic bounds typically feature larger constant factors than their non-optimistic counterparts. This is considered (usually) an acceptable compromise, and we will see examples next.

\begin{algorithm}[t]
	\begin{small}
		\nl \textbf{Input}: $\bm x_1\!\in\!\mathcal X$; $\sigma_1=\sigma$.\\ 
		\nl \textbf{Output}: $\{\bm{x}_t\}_t$.\\%
		\nl \For{ $t=1,2,\ldots, T$  }{
		\nl Receive prediction $\tilde {\bm{g}}_{t}$;\\[0.6ex]
		\nl Calculate the next action using \eqref{eq:optimistic-ftrl-update} -- \eqref{eq:oftrl-reg-l2};\\[0.6ex]		
		\nl Apply $\bm{x}_t$ and incur cost $f_t(\bm x_t)$; \\[0.6ex]
		\nl Observe the new cost gradient $\bm g_t$;\\[0.6ex]
		\nl Calculate the $t$-slot prediction error $\epsilon_{t}=\|\bm{g}_t- \bm{\p g}_t\|_2^2$;\\[0.6ex]
		\nl Update the aggregate regularization  $\sigma_{1:t} \!=\! \sigma\sqrt{\epsilon_{1:t}} $.\\
		}
		\caption{\small{Optimistic FTRL (OFTRL)}}
		\setlength{\intextsep}{0pt} 
		\label{alg:optimistic-template}
	\end{small}	
\end{algorithm}

\subsection{Who Provides the Predictions?}

An aspect that is rarely discussed in depth in the literature, with only a few exceptions \cite{peyman-oco-tac, llp}, is the practical implications of obtaining the predictions used in optimistic learning. The vast majority of \OL~studies assume that an oracle -- some type of an exogenous mechanism -- feeds the learner with predictions for the next gradient. Upon closer examination, it becomes apparent that this assumption implies the mechanism needs to predict both the next cost function and the next decision, since it is:
\begin{align}
	\bm{\p g}_{t+1}\triangleq \nabla \p f_{t+1}(\bm{\p x}_{t+1}),
\end{align}
where $\bm x_{t+1}$ is the very point the learner needs to decide at this slot. This reveals a cyclic dependency: the optimistic algorithm relies on a prediction that inherently involves (a prediction of) its own future decision. Despite this apparent circularity, there are several practical scenarios where such predictive mechanisms are feasible and can be effectively implemented.

First, when the functions are linear, $f_t(\bm x_t)=\dtp{\bm g_t}{\bm x}$, the gradient is independent of the decision point, and hence for procuring a prediction it suffices to make a guess for $\bm g_{t+1}$ (e.g., estimate the next function parameters). In several problems in communication systems one can use even simple tools such as time-series analysis to, e.g., predict the congestion of a link, the user requests, etc. In cases where the functions are non-linear, the predictor needs to guess both $\p f_{t+1}(\cdot)$ and $\bm{\p x}_{t+1}$. In some sense, it needs to predict both the decision of the adversary and the decision of the learner. Given the plethora of training data and Deep Learning-based predictors, one can envision systems where an ML model predicts both these quantities, e.g., based on historic data, previous interactions of the learner with the adversary, and so on. The fact that \OL~does not require to know in advance the accuracy of the predictor, makes this assumption less consequential: if we happen to be able to guess correctly both terms, we improve the performance; otherwise we get the typical OCO bounds.

\begin{table}[!t]
	{\small{
			\centering
			\caption{FTRL Adaptive \& Optimistc Regret Bounds}
			\begin{tabular}{p{1.9cm}p{2.56cm}p{3.2cm}}				
				\toprule
				\textbf{Adaptivity} & \textbf{Regularizer} $\sigma_{1:t}\sim$ & $\c R_T\leq$  \\[0.5ex] 
				\midrule
				Fixed Horizon  		 &  $\frac{L\sqrt T}{D}$  & $DL\sqrt{2T}$ \\ [0.8ex] 
				Anytime (Time Adapt.)   	  &  $\frac{L\sqrt{t}}{\sqrt{2}D}$ &  $2\sqrt{2}DL\sqrt{T}$\\[0.8ex] 						
				Gradient Adaptive   	  &  $\frac{\sqrt{\sum_{\tau=1}^t\|\bm g_\tau\|_2^2}}{\sqrt 2 D}$ &  $2\sqrt 2 D\sqrt{\sum_{t=1}^T\|\bm g_t\|_2^2}$\\[0.8ex] 		
				Optimistic   &  $\frac{\sqrt{\!\sum_{\tau=1}^t\|\bm g_\tau\!-\!\bm{\p g}_{\tau}\|_2^2}}{\sqrt 2 D}$ &  $2\sqrt 2 D\sqrt{\sum_{t=1}^T\!\|\bm g_t\!-\!\bm{\p g}_t\|_2^2}$\\	[0.8ex] 	
				\bottomrule
			\end{tabular}
			\label{tab:ftrl-bounds}}}
\end{table}

\subsection{History Bits}
It is useful at this point to discuss how the concept of optimistic learning emerged, how it is used in other OCO algorithms, and overall its evolution over the course of time.

\subsubsection{Slow-varying Environments} One of the first pertinent works in this context is \cite{hazan-ml2010} which showed that FTRL attains improved bounds if the cost gradients stay close to their long-term average $\bm{\bar{g}}_T=\bm{g}_{1:T}/T$. Specifically, under such favorable conditions, the regret is bounded in terms of the function variance as:
\[
\mathcal R_T= \mathcal{O}\left(\sqrt{ \sum_{t=1}^T \left\|\bm{g}_t-\bm{\bar{g}}_T \right\|_2^2}\right).
\]
A related variance-based bound was proved in a follow-up paper by the same authors, about the portfolio management problem \cite{hazan-kale-nips09}. Using a different approach, \cite{pmlr-v23-chiang12} considered the case where successive functions have small variation and proposed an OMD-like algorithm that executes an additional update during each slot. Since the functions change slowly, this extra update resembles the calculation of the prescient action, leading to improved regret bounds:
\[
\mathcal R_T\!=\!\mathcal{O}\!\left(\!\sqrt{Q_T}\right),  \ \text{with} \   Q_T\!=\!\sum_{t=1}^T \!\max_{\bm x\in\mathcal X}\left\|\nabla f_t(\bm{x})\! -\! \nabla f_{t-1}(\bm x)\right\|_2^2.
\]
Similar bounds were studied in \cite{rakhlin-NIPS2011} from a game-theoretic perspective. Although these works do not use predictions directly, they do assume some predictability due to properties of the problem, i.e., the environment being slow changing or following a pattern (tracking an average function), and building on this condition, they achieve faster convergence compared to the legacy OCO algorithms. Despite its simplicity, this approach can have practical benefits, and has been used in network flow control and scheduling \cite{mobihoc24-optimistic}.

\subsubsection{Introduction of Optimism} A milestone for \OL~is the seminal paper \cite{pmlr-colt-Rakhlin13} which coined the term, and proposed a family of optimistic algorithms for different scenarios. First, it extended FTRL to encompass the gradient prediction:
\begin{align}
&\bm{\hat x}_{t+1} = \arg\min_{\bm{x} \in \mathcal{X}} \Big\{ \eta\dtp{\bm{g}_{1:t}+\tilde{\bm{g}}_{t+1}}{\bm x} +\Phi_{\vartheta}(\bm x)\Big\} \notag
\end{align}
where $\eta$ is a fixed learning rate, and $\Phi_{\vartheta}(\bm x)$ is a $\vartheta$-self-concordant barrier function (see Appendix for definition). This proposal departed from the study of slow-varying environments of that time, and expressed the regret in terms of prediction errors:
\[
	\mathcal R_T\leq \frac{1}{\eta} \Phi_{\vartheta}(\bm x^\star) + 2\eta\sum_{t=1}^T {\left\|\bm g_t-\tilde{\bm g}_t \right\|_{(t),\star}^2}.
\]   
where the first term is the value of the barrier function at the benchmark and, under some mild assumptions, can be upper-bounded with $\vartheta \log T$; while the second term offers the desirable dependence on the prediction errors with respect to an appropriately-defined norm (based on the problem's geometry). Second, this paper introduced the optimistic version of OMD, which consists of the following two formulas: 
\begin{align}
	&\bm{y}_{t+1} = \arg\min_{\bm y \in \c X}\left\{ \dtp{\bm g_t}{\bm y} + \frac{1}{\eta}B_\psi(\bm y, \bm x_t)	 \right\}, \notag \\
	&\bm{x}_{t+1} = \arg\min_{\bm x \in \c X}\left\{ \dtp{\bm{\p g}_{t+1}}{\bm x} + \frac{1}{\eta}B_\psi(\bm x, \bm y_{t+1})	 \right\}. \label{eq:optimistic-OMD}
\end{align}
where recall that $B_\psi$ is the Bregman divergence and quantifies the distance of its arguments. Finally, \cite{pmlr-colt-Rakhlin13} studied also the case where there are multiple predictors and the learner needs to discern which of them to trust in real-time. We discuss scenarios and tools for this important multi-predictor version of \OL~in Section \ref{sec:experts}. The authors extended their work with a game-theoretic view to the problem in \cite{rakhlin-nips13}, where they also study its bandit version.

\subsubsection{Optimism \& Adaptivity}  Another important stream of works, directly related to our treatment of \OL~here, are  \cite{pmlr-v51-mohri16} and \cite{pmlr-v76-joulani17a} which combined data-adaptivity (adaptivity to gradients) and optimism. The work \cite{pmlr-v51-mohri16} in particular, tackled this topic in quite general terms and introduced algorithms with regret bounds which shrink with the prediction errors and, additionally, adapt to the observed gradients. This idea fills a gap that was not addressed in \cite{pmlr-colt-Rakhlin13}. Their engine is FTRL and their optimistic decision rules are in the spirit of \eqref{eq:optimistic-ftrl-update}, leading to bounds:
\begin{align}
	\c R_T \leq r_{1:T}(\bm x) + \sum_{t=1}^T \frac{1}{\sigma_t}\left\| \bm g_t - \bm{\p g}_t \right\|^2
\end{align}
where $\sigma_t$ is the regularization parameter of $r_t(\cdot)$, and, as explained in the previous subsection, it can adapt to the prediction errors (thus, also to the observed gradients), leading to the desirable order-optimal prediction-improved bounds.

\subsubsection{Optimism and Be-The-Leader} Concluding this background discussion, it might be helpful for some readers to view optimistic learning as an interpolation between FTRL (or any other typical OCO algorithm) and a fictional algorithm that relies fully on predictions. This latter algorithm can be termed Be-The-Leader (BTL), and has been actually used in the analysis of FTRL, cf. \cite{mcmahan-survey}, where its decision rule is:
\begin{align}
\bm x_{t+1}=\arg\min_{\bm x\in\c X} \dtp{\bm g_{1:t}+\bm{\p g}_{t+1}}{\bm x},
\end{align}
i.e., it does not use any regularization, and essentially extends FTL with the predicted gradient. Clearly, when $\bm{\p g}_{t+1}\!=\!\bm g_{t+1}, \forall t$, the regret of BTL is $\c O(1)$. On the other hand, the regret of FTRL that uses fully-fledged regularization and no predictions is $\c O(\sqrt T)$. OFTRL in turn, uses prediction-adaptive regularizers and its performance ranges from $\c O(\sqrt T)$ (when predictions fail), to $\c O(1)$ (as BTL) for perfect predictions, while achieving $\c O(\sqrt{\epsilon_{1:T}})$ for the in-between cases.

%
%
%
%
%
%
%
%
%
%
%
%
%
%
\section{Different Approaches to Optimism}\label{sec:different-optim}

Given the impressive performance gains that predictions can potentially bring, it is not surprising that there are several, and diverse, approaches to introducing and using them in learning and optimization tools. We present below some of the main facets and discuss how they are related to the optimism framework presented in the previous section.

\subsection{Directional Predictions}
In our discussion so far, we assumed the predictions come in the form of approximate gradients $\bm{\p g}_{t+1}$  for the next gradient $\bm g_{t+1}$, and the learner measures, a posteriori, its error using some norm that depends on the problem's geometry,  $\epsilon_t\!=\!\| \bm{g}_t \!- \! \bm{\p g}_t\|$. While this is the most prevalent method for introducing optimism in learning, there are other notable approaches to optimism.  In fact, one of the first optimistic learning studies \cite{hazan-2007} assumes the learner has instead access to one coordinate, e.g., ${g}_{t+1,1}$ of the next gradient $\bm{g}_{t+1}$ for all decisions slots. In this setting, the authors showed that when $\c X$ is the $N$-dimensional Euclidean ball, this additional information allows OGD-like algorithms to achieve $\c R_T\!\leq \! \c O\big((N^2/\gamma) \log T\big)$, where $\gamma=\min\{|g_{t,1}|, \forall t\}$. They further discussed extensions on general decision sets and for the case where more than one elements of the next gradient can be predicted.

\begin{figure}[t]
	\centering
	\includegraphics[width=0.4\columnwidth]{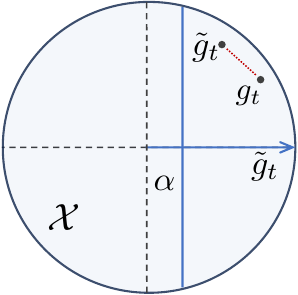}
	\caption{\small{Comparison of predictor concepts in \OL. The decision space is an $\ell_2$ ball in $\mathbb R^2$. The actual (to be predicted) gradient is $\bm g_t$, and a predictor that attempts to directly approximate it yields the point $\bm{\p g}_t$. A directional prediction provides a vector that points in the direction that $\bm g_t$ is expected to lie, with a margin of error $\alpha$, \cite{dekel-hint-neurips17}.}} \label{fig:alternative-predictors}
\end{figure}

A different line or work defines predictions as \emph{directional} (instead of approximate) hints \cite{dekel-hint-neurips17}. In particular, instead of assuming that the learner knows an approximate gradient vector that has bounded distance from the actual cost gradient, they assume the predictions satisfy the inequality:
\[
	\dtp{\bm{\tilde{g}}_t}{\bm{g}_t}\geq \alpha \| \bm{g}_t\|
\]
This means that, with a margin of error captured with the parameter $\alpha\geq 0$, the prediction points to the same direction as the gradient; or, in other words, that the learner has access to a halfspace that classifies the next gradient with some margin of error, see\footnote{To visualize this relation, recall that $\dtp{\bm{\tilde{g}}_t}{\bm{g}_t}=\|\bm{\tilde{g}}_t\|\|\bm{g}_t\|\cos \alpha$; and the bound is equivalent to the angle between the two vectors being smaller than $\cos^{-1}(\alpha)$ (assuming $\bm{\tilde g}_t\geq 1$).} Fig. \ref{fig:alternative-predictors}. It is not difficult to imagine scenarios in communication problems where instead of having a prediction within $\epsilon_t$-bounded distance from the gradient, one knows the direction of the gradient with some error. Under these conditions, and when the decision space $\c X$ satisfies some additional convexity assumptions, the authors proved regret:
\begin{align*}
\mathcal{R}_T= \mathcal{O}\big((1/\alpha)\log T\big).
\end{align*}

Finally, a follow up work \cite{pmlr-v119-bhaskara20a} dropped the assumption that the predictions are accurate at each slot, and presented a regret bound that improves with the number of within-margin-accurate predictions.

\subsection{Online Optimization with Predictions \& Competitive Ratio}

There is a parallel thrust of work that designs algorithms with predictions under different assumptions and using different metrics. Specifically, techniques of online optimization with predictions (or, with {ML ''advice''}) study sequential decisions as in OCO, but assume the cost function at each slot is revealed \emph{before} the optimizer commits its decision. This simple change in the sequence of steps turns the online learning problems to online \emph{optimization} problems. This literature uses a different performance metric than the regret, namely the competitive ratio which is defined as the total cost of the learner $\sumT f_t(\bm x_t)$ over the cost of a dynamic benchmark that is allowed to change at each slot, $\sumT f_t(\bm x_t^\star)$. The predictions in this setting refer to either an oracle that provides a prediction for the best decision in each slot $\bm{\p x}_t$, which the learner needs to trust or not; or to an oracle that provides prediction for the entire sequence of cost functions that are still to appear (i.e., from slot $t+1$ and on). The accuracy of these predictions is typically measured in terms of norm-distance with respect to the oracle $\{\bm x_t^\star\}_{t=1}^T$, \cite{competitive-smoothed-wierman}.

This prediction-assisted optimization framework has salient differences from the optimistic OCO  framework. On the one hand, it uses a more demanding benchmark, thus, algorithms that achieve satisfactory competitive ratios are expected to work well in practice (resilience to strong adversaries). On the other hand, it requires the cost functions to be known at the start of each slot, which can be an impractical assumptions for communication systems, and one that we would like to be able to drop. In general, the two frameworks, online optimization with predictions and online learning with predictions, have been used from different communities but often for similar problems. For instance, \cite{competitive-lykouris18a} studied caching using the former, while \cite{naram-ifip} studied it through the latter. Optimization with predictions has found numerous applications in dynamic decision problems, such as online bidding in auctions \cite{competitive-adaptive-bidding}, resource leasing problems \cite{competitive-ML-advised-ski-rental}, portfolio selection \cite{competitive-tsai2023datadependent}, and so on. 

An interesting discussion about online optimization and how it differs from online learning can be found in \cite{competitive-tale-of-metrics}; these remarks and differences apply naturally to the optimistic versions of these two frameworks, as well. The metrics of regret and competitive ratio can be even incompatible or conflicting, but at the same time the design of algorithms that work well in both settings is an emerging and important research area.

\subsection{SNUM with Predictions}

Finally, it is worth discussing prior efforts to incorporate predictions into the stochastic NUM framework. Indeed, a series of papers have investigated the benefits of predictions about user requests and/or the system state (e.g., channel conditions) on the system operation; see \cite{longbo-predictions-ton18, longbo-predictions-ton18b, atilla-caching-predictions, longbo-mobihoc14} and references therein, and the more recent works in \cite{mobihoc24-optimistic, modiano-optimistic-drift}. For instance, \cite{longbo-predictions-ton18b} considers the general problem of average performance optimization subject to network queues stability and the predictions refer to probability distributions of the system state over a lookahead-window of slots. The model assumes the prediction errors depend on the distance of the predicted slot from the current slot, and that the controller has a full and accurate characterization of the prediction errors.

With this information at hand, the network controller can improve the performance of the typical Backpressure policy and achieve both better performance (distance of the dynamic policy from the ideal static policy) and shorter queues, thus offering smaller delay to the served flows, compared to the typical Backpressure non-predictive algorithm. Furthermore, predictions can assist extending these SNUM tools to non-stationary (slow-changing) systems, see \cite{longbo-predictions-ton18b}. However, the presented solutions in this context, apart from the need to know the prediction errors, assume the perturbations are benign and optimize for fixed cost and constraint functions. The optimistic tools presented in this tutorial drop these assumptions.

\section{Benefiting from Multiple Predictors}\label{sec:experts}

From a practical perspective, a critical question in communication systems is whether -- and how -- the controller can effectively leverage multiple predictors, when more than one is available, particularly when these predictors exhibit potentially varying and unknown levels of accuracy. Given the widespread application of deep learning in building forecasting models,and the abundance of training data in networks, such scenarios are expected to become increasingly prevalent. In these cases, the NC must determine which predictors to trust, if any, or devise methods to combine their output to enhance their overall prediction quality.

\begin{algorithm}[t]
	\small{	
		\SetAlgoLined
		\KwIn{Function $\psi(\cdot)$, horizon $T$, learning rate $\eta$}
		\KwOut{ $\{\bm x_t\}_t$  }
		\textbf{Initialize}: $\bm x_1=\bm y_1=\arg\min \psi(\bm x)$; $w_1^p=1/P, \forall p\in\c P$\\
		\For{$t=1, \dots$, T}{
			\nl Apply $\bm x_t$ and observe the gradient $\bm g_t$;\\[1.1ex]
			\nl	Update the prediction weights: $w_{t+1}^p=w_{t}^p\exp{(-\| {\bm{\p g}_t^p} - \bm{g}_t \|_2^2)}$, $\forall p\in \c P$;\\[1.2ex]
			\nl Receive predictions $\{\bm{\p g}_{t+1}^p\}_{p=1}^P$; \\	[1.1ex]
			\nl	Calculate the next prediction: $\bm{\p g}_{t+1}=\sum_{p\in \c P} w_{t+1}^p \bm{\p g}_{t+1}^p$;\\[1.1ex]
			\nl	Perform optimistic OMD: 
			\begin{align*}
				&\bm{y}_{t+1} = \arg\min_{\bm y \in \c X}\left\{ \dtp{\bm g_t}{\bm y} +B_\psi(\bm y, \bm x_t)	/\eta \right\};\\
				&\bm{x}_{t+1} = \arg\min_{\bm x \in \c X}\left\{ \dtp{\bm{\p g}_{t+1}}{\bm x} + B_\psi(\bm x, \bm y_{t+1})/\eta	 \right\}
			\end{align*}
		}
	}
	\caption{\OL~with Multiple Predictors \cite{pmlr-colt-Rakhlin13}}\label{alg:multiple-predictors-rakhlin-colt13}
\end{algorithm}

The problem of using multiple predictors in \OL~was first studied in the seminal paper \cite[Sec. 4]{pmlr-colt-Rakhlin13}. The authors assume that the learner has access to a set $\c P$ of $P$ predictors each producing a sequence of approximate gradients $\{\bm{\p g}_t^p\}, {p \in \c P}$ in each slot $t$. Ideally, in this setting we would like to benefit from the \emph{best} predictor and upper-bound the regret as: 
\[
	\c R_T=\c O\bigg(\inf_{p \in \c P}\sum_{t=1}^T\|\bm g_t - \bm{\p g}_t^p\| \bigg).
\]	
Unfortunately, this bound is not easy to achieve, since the predictors' quality is unknown. Therefore, the learner, apart from devising the decisions $\{\bm x_t\}_t$ (as usual), needs to also discern at runtime the best predictor or, in general, decide dynamically how to utilize the set of predictors. 

To that end, \cite{pmlr-colt-Rakhlin13} proposed a 2-layer learning approach, where first a fused prediction is created by combining the $P$ predicted gradients:
\[
\bm{\p g}_t=\sum_{p=1}^P w_t^p\bm{\p g}_t^p,
\] 
which is then fed to an optimistic OMD algorithm which produces the decision $\bm x_{t+1}$. The first task amounts to learning the weights $\{w_t^p\}_p$ and is tackled as an experts problem\footnote{Experts problem have a long history in, and are very important for, online learning. Algorithms that tackle these problems can be seen as variants of FTRL (or OMD) over the unit simplex, and with entropic regularizers.}; while the OMD update proceeds as if there was a single predictor in place (providing this fused gradient). 

This tiered approach can be also seen and interpreted through the lens of \emph{meta-learning} in the sense that our goal is to learn which of the forecasters (which in turn can be learners) is best-performing. The steps of this double-learning routine are summarized in Algorithm \ref{alg:multiple-predictors-rakhlin-colt13} which leads to regret:
\begin{align}
	\c R_T\leq \frac{\psi_{max}}{\eta} + 3.2\eta \bigg(\! \log P + \inf_{p \in \c P}\sum_{t=1}^T\|\bm g_t - \bm{\p g}_t^p\|	\bigg) \notag
\end{align}
where $\psi_{max}\!=\!\max_{\bm x\in\c X} \psi(\bm x) \!- \min_{\bm x\in\c X}\psi(\bm x)$ is a constant depending on the Bregman divergence function; and $\eta$ is the learning rate that can be optimally set to $\eta=\c O(\sqrt T)$ when $T$ is known. We see that indeed this approach achieves the desired bound which, interestingly, increases only logarithmically with the number of predictors -- a reasonable price to pay for being able to benefit from them.

\begin{figure}[t]
	\centering
	\begin{subfigure}[H]{0.95\columnwidth}
		\centering
		\includegraphics[width=\columnwidth]{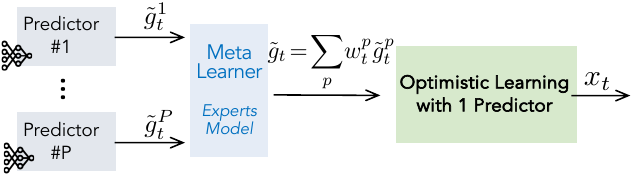}
		\caption{Meta-learning on Predictions.}
		\label{fig:meta-learning-rakhlin}
	\end{subfigure}
	\begin{subfigure}[H]{0.95\columnwidth}
		\vspace{2mm}
		\centering
		\includegraphics[width=\columnwidth]{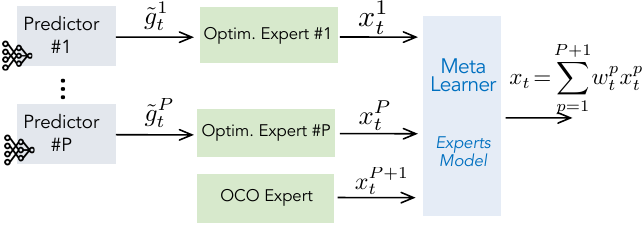}
		\caption{Meta-learning on suggested actions.}
			\label{fig:meta-learning-naram}
	\end{subfigure}
	\caption{\small{Different Approaches to handling multiple predictors. \textbf{(a)}: Synthesizing, with an experts algorithm, the weights for combining directly the $P$ available predictions, followed by a single \OL~algorithm \cite{pmlr-colt-Rakhlin13}; \textbf{(b)}: Synthesizing the suggested actions by $P$ \OL~algorithms, and one OCO algorithm, with an experts algorithm, to obtain the final action \cite{naram-tmc}.} } 	\label{fig:meta-learning}
\end{figure}

\begin{figure*}                                    
	\centering
	\includegraphics[scale=0.88]{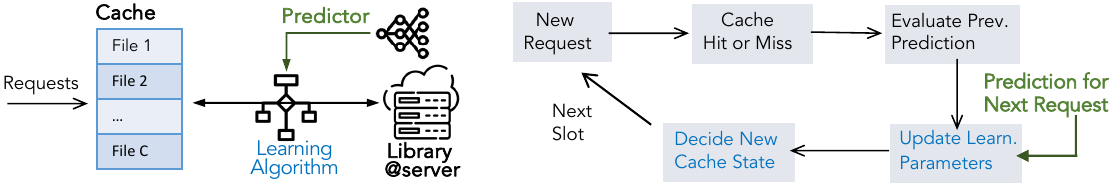} 
	\caption{Optimistic online caching with predictions: system schematic (\emph{left}) \& algorithm template (\emph{right}).}
	\label{fig:caching-fig}
\end{figure*}

The same problem was studied in \cite{pmlr-v99-cutkosky19b} using a different approach based on the observation that the combination (namely addition) of the decisions of different auxiliary algorithms can lead to regret which is the minimum of these algorithms. In particular, the following one-line proof result holds for any two learning algorithms $\c A$ and $\c B$:
\begin{align}
	\c R_T&\!\leq \!\sumT \dtp{\bm g_t}{\bm x_t\!-\bm x^\star}\!=\!\sumT \dtp{\bm g_t}{\bm x_t^{\c A}\!-\bm u^\star} \!+\! \sumT \dtp{\bm g_t}{\bm x_t^{\c B}\!-\bm v^\star}\notag \\
	&\leq \c R_T^{\c A}(\bm u^\star) + \c R_T^{\c B}(\bm v^\star) \!\leq \epsilon + \min\left\{\c R_T^{\c A}, \c R_T^{\c B}  \right\}. \notag
\end{align}
where $\bm x_t=\bm x_t^A +\bm x_t^B$ is the learner's decision which simply adds the outputs of the auxiliary algorithms. Vectors $\bm u^{\star}$ and $\bm v^\star$ are any two partial benchmark which satisfy $\bm u^\star\!+\bm v^\star\!=\bm x^\star$; and $\c R_T^{\c A}(\bm u^\star)$ (respectively  $\c R_T^{\c B}(\bm v^\star)$) denotes the regret of algorithm $\c A$ (resp. $\c B$) w.r.t. benchmark $\bm u^\star$ (resp. $\bm v^\star$). The last step holds only under the technical assumption that $\c R_T^{\c A}(\bm 0)\!\leq \!\epsilon$ and $\c R_T^{\c B}(\bm 0)\!\leq\! \epsilon$, for some constant $\epsilon$.

Transferring this result to \OL~with many predictors, the learner's decision is constructed by combining a typical learning algorithm $\c A$, like FTRL or OMD, with a fused prediction $\bm x_t - \sum_{p} w_{t}^p \bm g_{t}^p$ that synthesizes the $\c P$ predictors. Each weight is learned in parallel using a 1-dimensional learning algorithm as explained in \cite{pmlr-v75-cutkosky18a}, leading to an overall regret of:
\begin{align}
\c R_T\leq \epsilon P + \c R_T^{\c A} \label{eq:regret-kosky}
\end{align}
where the latter term is the regret of algorithm $\c A$ that produces the shifted vectors $\{\bm x_t\}_t$. This approach follows essentially the meta-learning approach, as the above tiered-learning model with the experts approach, and has less computation overhead but imposes some additional technical assumptions, cf. \cite{pmlr-v99-cutkosky19b}.

Finally, as we will see in the next section, \cite{naram-tmc} follows yet another approach where the experts formulation is used to combine predictors and learners. In this case, the predictor provides directly a decision (e.g., it can be the optimistic OMD), and the experts algorithm combines linearly the different suggestions, including suggestions from non-optimistic algorithms (e.g., a typical OGD or FTRL without predictions) to create a fused action. This allows to leverage predictors (of unknown quality) in a different way than \cite{pmlr-colt-Rakhlin13}. The difference in applying the meta-learning among these two approaches is explained in Figure \ref{fig:meta-learning}. Concluding, we note that the question of how to leverage multiple predictors in online \emph{optimization} algorithms was studied in \cite{pmlr-v139-flaspohler21a} and \cite{irvine-expert-ml-sigmetrics22} using, as usual, a competitive ratio approach.

%
%
\section{Optimistic Caching} \label{sec:opt-caching}

The previous sections presented the main elements and building blocks of the theory of optimistic learning. We proceed now to apply these ideas to important resource management problems in communication systems. These problems, we believe, are both timely and representative of larger families of problems that are commonly encountered in such systems. We start with the problem of dynamic caching.

\subsection{Background}
Caching is of paramount importance in communication and computing systems, and has appeared in different forms during the last 70 years or so. From memory systems of mainframe computing machines, to web traffic management and content delivery networks, and more recently, to edge computing and wireless communications, the design of dynamic caching policies has attracted great interest \cite{paschos-jsac-caching}. These algorithms decide \emph{reactively} which files to store at a cache so as to maximize the cache hits without knowing the future requests. 

While there are widely used algorithms for this problem, such as the Least Frequently Used (LFU) and Least Recently Used (LRU) policies, the design of algorithms that perform satisfactory in terms of cache hits (or, cache utility, in general) under many different request patterns, remains an open problem \cite{paschos-book}. The sequence of events in this problem (first cache, then observe the request) places it naturally in the context of online learning. Indeed, the dynamic caching problem was recently studied through the lens of online convex optimization \cite{paschos-infocom}, which created fresh insights and optimization opportunities for this fundamental problem. In this section we explain how one can leverage optimistic learning and how much the caching algorithms can benefit from optimism.

\subsection{Caching as an Online Learning Problem}

\begin{figure*}
	\centering
	\begin{subfigure}[t]{0.65\columnwidth}
		\centering
		\includegraphics[width=0.859\textwidth]{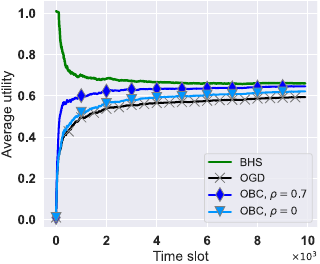}
		\caption{}
	\end{subfigure}%
	\hfill 
	\begin{subfigure}[t]{0.65\columnwidth}
		\centering
		\includegraphics[width=0.859\textwidth]{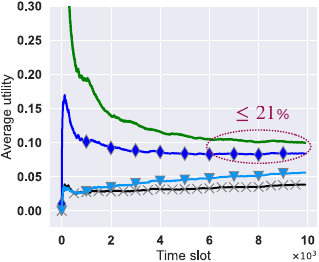}
		\caption{}
	\end{subfigure}
	\hfill 
	\begin{subfigure}[t]{0.65\columnwidth}
		\centering
		\includegraphics[width=0.859\textwidth]{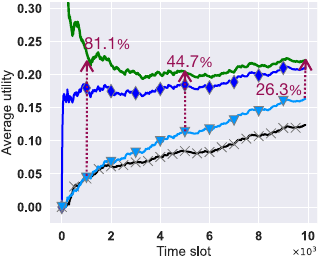}
		\caption{}
	\end{subfigure}
	\caption{\small{Evolution of utility in the single cache model for OGD-based policy \cite{paschos-infocom}, Best-in-Hindsight (BHS) policy $\bm x^\star$, and with \OL-based caching \cite{naram-ifip} with predictors of different quality $\rho$ (probability of predicting correctly the next request) for a system with: \textbf{(a)} Zipf requests with $\zeta = 1.1$; \textbf{(b)} YouTube request traces \cite{ZINK2009501}; and \textbf{(c)} Movie-Lens request traces \cite{movielens}. } }
	\label{fig:caching-simulations}
\end{figure*}

\subsubsection{Single Cache} We study a caching system serving a library $\c N$ of $N$ files of unit size\footnote{The unit file assumption is without loss of generality.}, and a cache that can store up to $C<<N$ of these files. The cache serves the file requests emanating from a set of users $\c I$. The system operation is time-slotted, and at the beginning of each slot we assume the system receives a single request that is modeled with the one-hot request vector $\bm q_t=\left( q_{t,ni}\in\{0,1\}, n\in\c N, i\in\c I \right)$. Our model does not impose any restrictions on the request pattern $\{\bm q_t\}_t$ which, in the general case, can be determined by an adversary aiming to reduce the cache hits. After observing the request, the cache updates its contents by deciding the cache vector $\bm x_t\!=\!(x_{t,n}, n \!\in\!\c N)$ that is drawn from the set of eligible caching decisions:
\[
	\c X=\left\{	\bm x \in [0,1]^{N} \ \bigg \vert \sum_{n=1}^{N}x_{n}\leq C	\right\}.
\]
In other words, we assume that the caching decisions are continuous, i.e., any arbitrary part of the file can be cached. This is technically possible with coded caching techniques which are excessively used in practice (e.g., in Amazon S3), while it is a meaningful approximation for chunked large-size files even if they are uncoded; see discussion in \cite{paschos-jsac-caching}.

The caching decision shapes the performance of the next slot, which in its most basic form, can be a linear function: 
\[
	f_t(\bm x_t)=\sum_{n\in\c N}\sum_{i\in \c I} w_{t,n} q_{t,ni}x_{t,n}
\]
where the vector $\bm w_t\!=\!(w_{t,n}\!\leq w_{max}, n\in\c N)$ determines the relative caching importance of each file. These weights might even be unknown at the start of each slot, or changing with time; as long as $\{f_t\}_t$ are convex, we can cast this problem as OCO. And one can introduce any other caching performance function as long as the convexity condition is satisfied.

For this general caching setting, \cite{paschos-infocom} proposed the use of OGD for deciding at the beginning of each slot (or, equivalently, at the end of the previous slot), the portion of each file the cache should hold. The algorithm iteration is:
\begin{align}
	\bm x_{t+1}=\Pi_{\c X}\big(\bm x_t + \eta_t\bm g_t	\big), \label{eq:ogd-caching}
\end{align}
where $\bm g_t\!=\!\nabla f_t(\bm x_t)$. Using an optimized learning rate $\eta_t\!=\!{D}/{w_{max}\sqrt{T}}$, the obtained regret bound is: 
\begin{align}
\c R_T &\triangleq  \dtp{\bm g_{1:T}}{\bm x^\star} - \sum_{t=1}^T \dtp{\bm g_t}{\bm x_t} \notag \\
&\leq w_{max}D\sqrt T=w_{max}\sqrt{2C}\sqrt T, \label{eq:regret-oco-single-cache}
\end{align}
where the last step follows since $D_{\c X}\!=\!\sqrt{2C}$ \cite{paschos-infocom}. The regret is defined w.r.t. the benchmark cache configuration $\bm x^\star\!=\!\arg\max_{\bm x\in \c X} \sum_{t=1}^T f_t(\bm x)$ (Best-in-Hindsight). Interestingly, the regret bound is not only sublinear on $T$, but also independent of the library size $N$. This is particularly important, since $N$ can take very large values, while the cache capacity $C$ that appears in the bound is typically much smaller.

\begin{figure*}[t]
	\centering
	\includegraphics[width=0.9\textwidth]{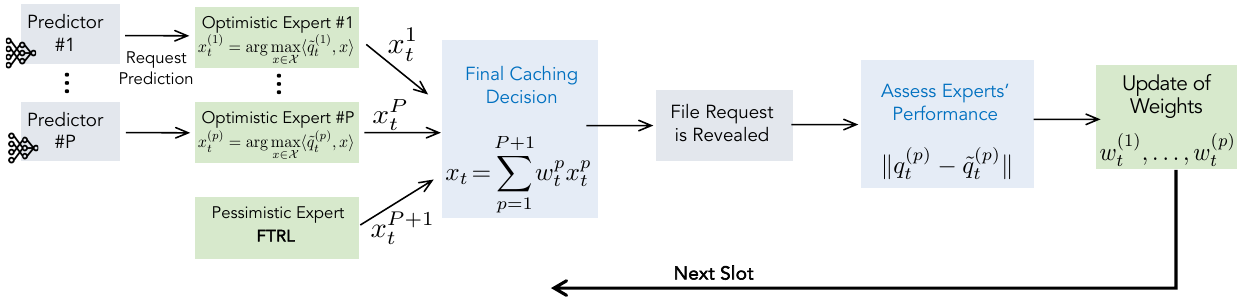}
	\caption{\small{Optimistic Caching with Multiple Predictors. The meta-learner receives suggested caching vectors, $\{\bm x_t^{(p)}\}$, from the $P$ optimistic experts, each of which uses a different forecasting model (predictor), and from one non-optimistic caching expert. The meta-learner combines the suggestions to create its final caching decision $\bm x_t$, and adjusts the combination weights accordingly to the prediction errors.}}
	\label{fig:experts-caching-naram}
\end{figure*}   

\subsubsection{Caching Networks} This approach can be extended to the problem of caching in bipartite networks where a set of caches $\c J$ serve a set of users $\c I$ through a graph $G=\big( \c J \cup \c I, (\gamma_{ij})_{i,j} \big)$, where $\gamma_{ij}\!=\!1$ if user $i$ can receive content from cache $j$, and $\gamma_{ij}\!=\!0$ otherwise. The caching decision vector is extended to account for the cache id, $\bm x_t=(x_{t,nj}, n\in \c N, j\in\c J)$; and we need to decide also the routing vector, $\bm y_t=(y_{t,nij}\in[0,1], n\in\c N, i\in\c I, j\in \c J)$, where $y_{t,nij}$ denotes the portion of file $n$ that is requested by user $i$ and served by cache $j$.

The static benchmark, $(\bm x^\star, \bm y^\star)$, for this learning problem is defined as the solution to this horizon-long caching problem:
\begin{align*}
	&\max_{\bm x, \bm y } \ \sum_{t=1}^T f_t(\bm x) \\ 
	&\text{s.t.} \quad y_{nij}\leq x_{nj}\gamma_{ij}, \ \ \ \forall i\in \c I, j\in \c J, n\in \c N \\
	& \qquad \ \ \bm{x}\in \c X, \ \bm{y} \in \c Y.
\end{align*}
The authors in \cite{paschos-infocom} observed that, for any fixed caching decision $\bm x_t$, one can find the optimal routing vector $\bm y_t$ for any given request $\bm q_t$ by solving, w.r.t. routing, the respective utility maximization problem . Therefore, in order to perform learning, it suffices to apply OGD (or some other algorithm) for the caching decisions. This leads to regret:
\begin{align}
\c R_T\leq w_{max}D\sqrt T=w_{max}\sqrt{2JC}\sqrt T  \label{bound:infocom-caching}
\end{align}
where $C\!=\!\max_{j\in\c J} C_j$ is the maximum capacity of any cache.

There are several works that extended the toolbox of OCO-based caching algorithms by penalizing the fetching of new files \cite{oco-caching-ton, 10446753}, providing lower regret bounds \cite{abishek-sigmetrics-caching}, designing tailored OMD algorithms \cite{tareq-omd-caching} and algorithms that consider the similarity of requested files  \cite{salem-similarity-caching, faizal-peva2023-online-caching}, among others.

\subsection{Adding Optimism to (Continuous) Caching}

Building on the above, \cite{naram-ifip} studied caching as an optimistic learning problem, using the same problem setting as \cite{paschos-infocom, oco-caching-ton}. The authors assumed the existence of an approximate gradient $\bm{\p g}_t$ of $f_{t}$ and designed an optimistic FTRL caching algorithm. In practice, this prediction means that the system has some prior information (of unknown accuracy) about the request pattern. There are different ways one can obtain such predictions; e.g., using a statistical model fitted on historical data; a deep learning model trained offline with representative datasets; or even leveraging the recommender systems (RecSys) that often accompany content delivery platforms. For example, Netflix and YouTube provide content recommendations to their users, which can be interpreted as predictions of what will be requested. The accuracy of this type of predictions depends on whether the users follow the recommendations or not which is not known in advance and may differ across users and platforms. Nevertheless, since \OL~ does not require knowing the predictions accuracy, such RecSys can serve as a new, potentially insightful, forecasting tool in caching.

In this setting, \cite{naram-ifip} takes a black-box approach and makes no assumptions on the accuracy of predictions. The latter are assessed only after being used, i.e., after observing whether the requested file coincides with the predicted one. The optimistic caching decision rule is:
\begin{align}
	\bm x_{t+1}=\arg\min_{\bm x\in \c X}\Big\{ r_{1:t}(\bm x) - \dtp{\bm g_{1:t}+\bm{\p g}_{t+1}}{\bm x}	\Big\}
\end{align}
where note the minus sign stemming from this being a maximization problem\footnote{Alternatively, we can express this update using the projection operation, as in the OGD caching iteration \eqref{eq:ogd-caching}.}; and, similarly to \cite{paschos-infocom, oco-caching-ton}, the routing is decided based on the caching. As Figure \ref{fig:caching-fig} demonstrates, the difference of this approach with the non-optimistic OCO caching algorithm is that before we optimize the cache state (i.e., decide which files to store for the next slot), we receive exogenously a prediction for the next request that is used in the decision making. Then, after fixing the caching, the actual request arrives, the system experiences a cache hit or miss (accruing utility, or not), we evaluate the accuracy of the predictor based on whether the request was predicted or not, and we update the regularization parameters accordingly.

In particular, \cite{naram-ifip}  proposed using the following FTRL proximal optimistic regularization scheme:
\begin{align}
	&r_t(\bm x)=\frac{\sigma_t}{2}\|\bm x-\bm x_t\|_2^2, \ \ \sigma_t=\sigma\left(\sqrt{\epsilon_{1:t}} - \sqrt{\epsilon_{1:t-1}}	\right), \notag
\end{align}	
where the $t$-slot prediction error is $\epsilon_t=\|\bm g_t - \bm{\p g}_t\|_2^2$. This optimistic caching policy achieves:
\begin{align}
	\c R_T\leq 2\sqrt{1 + JC}\sqrt{ \epsilon_{1:T}	} \label{eq:regret-oco-optimistic-cache}
\end{align}
which has the desirable property of shrinking with the accuracy of predictions. Compared to the data-adaptive non-optimistic regret bound \eqref{bound:infocom-caching} of the online caching algorithm  \cite{paschos-infocom}, this optimistic bound is worse by a factor of $\sqrt 2$, but maintains the same worst-case bound of $\c O(\sqrt T)$. Therefore we obtain the desirable best-of-both-worlds performance, without significant additional computing or memory overheads compared to vanilla OCO-based caching.

One might naturally ask at this point whether this theoretical advantage of \OL~can bring practical caching gains. Figure \ref{fig:caching-simulations} provides a quick answer to this question. Comparing, OGD-based caching from \cite{paschos-infocom} and the optimistic caching from \cite{naram-ifip} with predictors of different accuracy, we see that the gains in terms of attained utility, due to optimism, are typically in the order of $50\%$. We also observe that these gains are more pronounced during the early stages of learning, while both algorithms (with and without predictions) converge asymptotically to the same solution, as expected based on their regret bounds for $T\!\rightarrow\! \infty$. The idea of using predictions in caching via OCO, was also used in \cite{neglia-computer-networks24}, while the earlier work \cite{competitive-lykouris18a} has used predictions for caching in the context of online optimization (the requests arw observed before caching) and assessed through a competitive ratio metric.

\subsection{Optimistic Caching with Multiple Predictors}

Since there are several potential sources for providing the request predictions in caching, it is only natural to ask the question how can we design an online caching policy that utilizes, and benefits from, multiple predictors at the same, without knowing in advance their accuracy. As discussed in Sec. \ref{sec:experts}, \OL~can encompass multiple predictors and here we explain how this idea can be applied to caching.

\begin{figure*}
	\centering
	\begin{subfigure}[t]{0.5\columnwidth}
		\centering
		\includegraphics[width=0.94\textwidth]{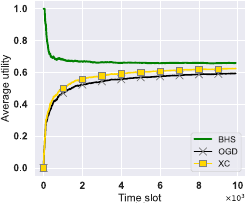}
		\caption{}
	\end{subfigure}%
	\hfill 
	\begin{subfigure}[t]{0.5\columnwidth}
		\centering
		\includegraphics[width=0.94\textwidth]{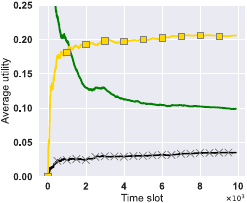}
		\caption{}
	\end{subfigure}
	\hfill 
	\begin{subfigure}[t]{0.5\columnwidth}
		\centering
		\includegraphics[width=0.94\textwidth]{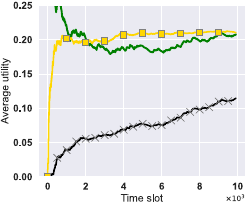}
		\caption{}
	\end{subfigure}
	\begin{subfigure}[t]{0.5\columnwidth}
		\centering
		\includegraphics[width=0.94\textwidth]{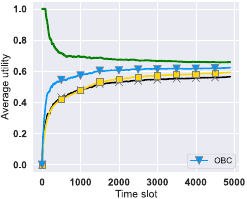}
		\caption{}
	\end{subfigure}	
	\caption{\small{Utility in a single cache with two predictors of recommendation qualities $\rho = 2\%$ and $\rho = 20\%$, each modeled as an expert within XC, in \textbf{(a)} Zipf requests with $\zeta = 1.1$; \textbf{(b)} YouTube request traces \cite{ZINK2009501}; \textbf{(c)} MovieLens request traces \cite{ZINK2009501}. \textbf{(d)}  Comparison between Optimistic FTRL (OBC) and XC using one predictor with highly-fluctuating quality.} }
	\label{fig:caching-experts-simulations}
\end{figure*}

The problem of dynamic learning-based caching with a set $\c P$ of $P$ predictors was studied in \cite{naram-tmc}. The main idea relies on the experts learning paradigm where each expert proposes a caching vector by assuming the respective predictor is fully reliable (\emph{optimistic experts}), and there is one additional expert, termed \emph{pessimistic}, that does not utilize predictions and follows a standard OCO-based algorithm for finding caching decisions. The meta-learner then combines, through a convex combination, all the $P\!+\!1$ proposed caching vectors and produces the vector that is actually applied to the system. This decision process is summarized in Figure \ref{fig:experts-caching-naram} which tailors the multiple predictors model of Fig. \ref{fig:meta-learning-naram} to the caching problem.

This algorithm essentially applies meta-learning on the provided caching suggestions from the experts, and achieves meta-regret that is upper-bounded as follows:
 \begin{align*}
	\mathcal{R}_T \ \leq \underbrace{\sqrt{\log P}\sqrt{T}}_\text{\color{blue}{find best predictor}}+ \underbrace{\min_{p\in \c P} \mathcal{R}_T^{p}}_\text{\color{red}{max hit ratio}},
\end{align*} 
where the second term indicates that the bound is shaped by the regret of the best predictor, and the first term quantifies exactly the cost (in terms of convergence delay) of finding this best expert. In other words, the more are the experts (e.g., the more predictors or learners we employ), the more we increase the chances to find a well-performing predictor, but also increase the exploration delay; yet, this cost depends only logarithmically on the number of experts $P$. This bound can be strictly negative, depending on the optimistic expert's regret. For example, in case of perfect predictions and non-fixed cost functions, the $\text{min}$ term evaluates to $-\epsilon T$ for some $\epsilon>0$, making the meta-regret negative for large enough $T$.

There are some subtle --  albeit important --  differences between the algorithm in \cite{naram-tmc} and \cite{pmlr-colt-Rakhlin13}. In \cite{naram-tmc} each expert proposes a caching vector and the meta-learner fuses them to create an eligible (implementable) caching decision for the next slot, while in \cite{pmlr-colt-Rakhlin13} the experts propose predictions and the meta-learner fuses them to create a vector that is then fed to the learning algorithm. Moreover, \cite{naram-tmc} includes a pessimistic (standard OCO) algorithm as one of the experts, which serves as the fall-back option for the case \emph{all} predictors fail. Similarly, it is interesting to compare the above bound, with the one in \eqref{eq:regret-kosky} which has a different dependence on $P$ (but independent of $T$), and holds under different assumptions, see \cite{pmlr-v75-cutkosky18a}.

The trace-based evaluation of this meta-learning caching framework in Figure \ref{fig:caching-experts-simulations} verifies its efficacy but also reveals its shortcomings. The plots in Figure \ref{fig:caching-experts-simulations}(a)-(c) compare the utility accrued by: \emph{(i)} the benchmark caching $\bm x^\star$ (BHS), \emph{(ii)} the OGD-based caching policy of \cite{paschos-infocom} and the meta-learning caching policy (denoted XC) from \cite{naram-tmc}. This latter policy utilizes an FTRL caching expert and two optimistic-caching experts with predictors of different accuracy: $\rho\!=\! 2\%$ and $\rho\! =\! 20\%$, where $\rho$ is the percentage of predictions that are accurate, on average. We can see that the meta-learning caching policy always outperforms the OGD policy, and even the benchmark policy for the dynamic real-world traces. Finally, Figure \ref{fig:caching-experts-simulations}(d) compares an OFTRL-based caching policy (in the spirit of Alg. \ref{alg:optimistic-template}) with the meta-learning caching framework  that uses two experts, namely an expert implementing FTRL and an optimistic expert with highly volatile prediction accuracy (alternates between $100\%$ and $0\%$). We see here that the meta-learning framework does not perform as well as the OFTRL algorithm, although both of them use the same predictor. This highlights the importance of the way one selects to incorporate the predictions in caching, i.e., through regularization as in OFTRL or through an expert model that uses the optimistic caching decisions as in Fig. \ref{fig:experts-caching-naram}.

%
\section{\OL~for Discrete Placement Problems} \label{sec:opt-discrete-caching}

In this section we drop the assumption of continuous caching decisions and study how to perform optimistic learning for a discrete (whole-file) caching problem. We introduce two distinct approaches. The first method uses OFTRL to learn a no-regret probability distribution over the caching decisions for the files of the library, and then applies unbiased sampling to determine exactly which files should be cached. The second method follows a different path and introduces an optimistic version of the seminal Follow-The-Perturbed-Leader (FTPL) algorithm.  Both approaches extend well-beyond caching and can address other discrete placement problems. For instance, they can be used to decide which services to deploy at an (edge) server or to select among various system configurations in softwarized (wireless) networks.   

\begin{algorithm}[t]
	\begin{small}
		\caption{{\small{OFTRL \& Sampling for Discrete Caching}}}
		\label{alg:oftrl}
		\nl \textbf{Input}:  $\sigma_1\!=\!\sigma\!=\! \sqrt{\epsilon_1/C}$, $\bm x_1\!\in \c X_d$ \\%
		\nl \textbf{Output}: $\{\bm x_t\}_t$ \\%
		\nl \For{ $t= 2, 3\ldots$ }{
			\nl Receive prediction $\bm{\p q}_{t}$ \\ [0.6ex]		 
			\nl $\bm{\hat x}_t = \arg\min_{\bm x \in \c X} \left\{ r_{1:t-1} (\bm x) - \dtp{\bm x}{\bm q_{1:t-1}+\bm{\p q}_{t}}\right\}$\label{algstep:update1}  \\[0.6ex]		 
			\nl $\bm x_t\leftarrow \texttt{MadowSample}(\bm{\hat x}_t)$ \\ [0.6ex]	 
			\nl Apply caching $\bm x_t$;\\[0.6ex]		
			\nl Receive request $\bm q_t$ and calculate utility $\dtp{\bm q_t}{\bm x_t}$;\\[0.6ex]		
			\nl Measure prediction error $\epsilon_t$;\\[0.6ex]		
			\nl Update parameter $\sigma_{1:t} = \sigma \sqrt{\epsilon_{1:t}}$  
		}
	\end{small}	
\end{algorithm}

\subsection{OFTRL and Sampling}

The discrete version of caching is naturally a non-convex problem. In the simplest case where all files have equal size, the caching decisions need to be drawn from the set of eligible discrete caching decisions:
\begin{align}
	\mathcal{X}_d =  \left\{ \bm{x} \in\left\{0, 1 \right\}^N \, \Bigg| \, \sum_{n=1}^N x_{n} \leq C \right\},
\end{align}
where we assumed, without loss of generality, that the file sizes are unitary, and recall that $C$ is the capacity of the cache (number of cached files, in this case). Furthermore, we simplify the problem by assuming all files have equal importance, i.e., $w_n^t=1, \forall n\in \c N, t$, and therefore the utility functions simply measure the cache hits, $f_t(\bm x)= \dtp{\bm q_t}{\bm x_t}$.  Since $\c X_d$ is non-convex we cannot apply off-the-shelf algorithms such as OGD or FTRL for deciding which files to cache. The first method overcomes this obstacle through sampling, which is facilitated by the fact that the caching utility function is linear.

The steps of the method are summarized in Algorithm \ref{alg:oftrl} which was introduced in \cite{naram-sigmetrics}. The main idea is to apply OFTRL on the convex relaxation of this problem, in specific to optimize the continuous caching vectors $\{\bm{\hat x}_t\}_t$ over the convex hull of the discrete caching set, $\c X\!=\! Co(\c X_d)$. Then, using these decisions, the algorithm employs unbiased sampling (in this case, with Madow's sampling \cite{madow}) to obtain the discrete caching decisions that belong to $\c X_d$. The first step essentially learns the optimal caching distribution across the possible cache states, using the regularization from \eqref{eq:oftrl-reg-l2}, and the second step makes sure that we create discrete caching vectors that, on expectation, yield the same regret as their continuous (relaxed) counterparts. This equivalence stems from the following result: 
\begin{align}
			\mathbb E\left[\c R_T\right] =  \dtp{\bm q_{1:T}}{\bm x^\star} - \mathbb E\left[{\sum_{t=1}^T \dtp{\bm q_t}{\bm x_t}}\right]= \widehat{\c R}_T.  
\end{align}
where the expectation is defined w.r.t. the randomized decisions of the algorithm; and $\widehat{\c R}_T$ is the regret of the continuous caching decisions which, as we use OFTRL, can be upper-bounded using \eqref{eq:regret-oco-optimistic-cache} with $2\sqrt{1+C}\sqrt{\epsilon_{1:T}}$, where $\epsilon_t$ is the squared $\ell_2$-based prediction error at slot $t$. 

We observe that this method achieves regret that shrinks with the prediction errors and which is only $\sqrt 2$ worse than the non-optimistic bound, see \eqref{eq:regret-oco-single-cache} for $w_{max}\!=\!1$, in the worst case where all predictions are maximally inaccurate. The drawback of this optimistic caching approach is the computational complexity for calculating $\bm{\hat x}_t$ which involves a projection operation.  The next subsection presents a less involved learning technique.

\subsection{Optimistic Perturbations}

\begin{algorithm}[t]
	\caption{\small{OFTPL for Discrete Caching}}
	\label{alg:oftpl}
	\begin{small}
		\nl \textbf{Input}: $\eta_1=0$, \ $\bm x_1\!\in \c X_d$ \\%
		\nl \textbf{Output}: $\{\bm x_t\}_t$ \\  
		\nl  $\bm \gamma \stackrel{iid}\sim \mathcal{N}(0,\boldsymbol{1}_{N\times 1})$ \hfill \emph{Sample a perturbation vector}\\
		\nl \For{ $t=2,3,\ldots$  }{
			\nl Receive prediction $\bm{\p q}_{t}$ \\[0.2ex] 
			\nl Update $\eta_t = \frac{1.3}{\sqrt{C}}\left(\frac{1}{\ln (Ne/C)}\right)^{\frac{1}{4}}\sqrt{\sum_{\tau=1}^{t-1}\|\bm q_\tau - \bm{\p q}_\tau\|^2_1}$ \\[0.6ex]	 
			\nl  $\bm x_t = \arg\max_{\bm x \in \mathcal{X}_d}\dtp{\bm x}{\bm q_{1:t-1} + \bm{\p q}_t + \eta_{t}\bm \gamma}$ \\[0.6ex]	 
			\nl Apply caching $\bm x_t$;\\[0.6ex]		
			\nl Receive request $\bm q_t$ and calculate utility $\dtp{\bm q_t}{\bm x_t}$.
		}
	\end{small}
\end{algorithm}

FTPL was introduced in \cite{kalai-2005} while \cite{abernethy-ftpl-colt14} re-framed it in the context of online convex optimization and enabled its unified view with FTRL. It can be seen as an extension of the intuitive FTL algorithm that was discussed in Sec. \ref{sec:oco-algorithms-rules}
\begin{equation}
	\bm x_t = \arg\max_{\bm x\in\mathcal{X}} \dtp{\bm g_{1:t-1}}{\bm x},  \label{eq:FTL}
\end{equation}
which is optimal  when the utility functions are sampled from a stationary statistical distribution, but has linear (non-convergent) regret in adversarial problems where successive gradients can be arbitrary far from each other. Differently from FTRL, FTPL remedies this stability issue by \emph{smoothing}, i.e., adding noise to the gradient:
\begin{align}
   \bm x_t = \arg\max_{\bm x\in\mathcal{X}} \dtp{\bm x}{\bm g_{1:t}+\eta_t\bm \gamma},
\end{align}
where $\bm \gamma  \sim\!\! \mathcal{N}(0,\mathbf{1})$ is the random noise vector, and $\eta_t$ is a scaling factor that controls the smoothing effect, i.e., it has a role similar to learning rate in OGD, OMD and FTRL.

FTPL was shown to provide optimal regret guarantees for the discrete caching problem in \cite{abishek-sigmetrics-caching}. At the same time, it has attractive  computing efficiency, as its update rule involves only an ordering operation (instead of projection, as in FTRL). This means that the FTPL decisions are derived by solving at each slot $t$ a linear program with the parameterized perturbed cumulative utility vector.

In order to obtain the optimistic FTPL (OFTPL) variant, \cite{naram-sigmetrics} introduced two twists: \emph{(i)} the prediction for the next-slot utility $\bm{\p g}_t$ is added to the cumulative utility; and \emph{(ii)} the perturbation parameter $\eta_t$ is scaled according to the accumulated prediction error. Interestingly, due to the structure of the decision set $\mathcal{X}_d$, the LP solution reduces to identifying the $C$ files with the highest coefficients, thus producing a feasible caching vector. And this step can be efficiently implemented in deterministic linear, $\c O(N)$, time.

The steps of OFTPL are summarized Algorithm \ref{alg:oftpl} where note that the prediction errors are measured in this case using the $\ell_1$ norm. Under some mild technical assumptions ($N\geq2C$, $C\geq11$), this algorithm ensures the regret bound:
\begin{align}\label{thm:oftpl}
	\mathbb{E}_\gamma[\c R_T] \leq 3.68 \sqrt{C}\ \bigg(\ln \frac{Ne}{C}\bigg)^{1/4} \sqrt{\sum_{t=1}^T ||\bm q_t - \bm{ \p q}_t||_1^2},
\end{align}
where recall that for cache-hit utility models, it is $\bm g_t\!=\!\bm q_t$. This bound has the desirable property of scaling with the prediction errors, i.e., there are no constant terms that are not modulated with the prediction errors. Compared to non-optimistic FTPL, here we have the same worst-case bound in terms of convergence rate, and only sacrifice a constant factor of $\sim2.5$. In terms of implementation, it is substantially faster than the OFTRL-based caching, but has a worst constant factor which depends (loosely) on the library size $N$. Other efforts for using FTPL in optimistic learning include \cite{suggala-oftplneurips20} and \cite{suggala20a-oftpl-colt} which prove regret bounds that have high-order dependencies on the problem dimension and therefore are ineffective for large problems such as those one typically encounters in caching.

\subsection{Handling Files of Different Size}

Finally, these methods can be extended to the setting where each file $n\in \c N$ has (different) size of $s_n$ bytes. The set of feasible discrete caching vectors in this case is redefined as:
\begin{align}
	\mathcal X_s=  \left\{ {x} \in\left\{0, 1 \right\}^N \ \bigg| \ \sum_{n=1}^N s_nx_{n} \leq C \right\},
\end{align}
and the benchmark policy is: 
\[
\bm x^\star=\arg\max_{\bm x\in \mathcal X_s}\dtp{\bm x}{\bm g_{1:T}}.
\]
The work \cite{naram-sigmetrics} introduced two solutions for this problem using OFTRL and OFTPL. We  focus  below on the latter and refer the reader to the original paper for the other technique.

The OFTPL algorithm in this case determines the next cache configuration $\bm x_{t}$ by solving the following integer programming problem at each slot $t$:
\begin{align} 
	\mathbb{P}_1:\qquad	\max_{\bm x \in \mathcal{X}_s }\,\, \dtp{\bm q_{1:t-1}+\tilde{\bm q_t} + \eta_t \bm \gamma}{\bm x}, \label{eq:oftpl-uneq-obj}
\end{align}
which is a Knapsack instance with profit vector $\bm p\! = \bm q_{1:t-1}\!+ \bm{\p q}_t+\!\eta_t\bm \gamma$; size vector $s \!=\!\left(s_n, \ n \in \c N\right)$; and capacity $C$. Since the Knapsack problem is NP-Hard, we cannot solve $\mathbb{P}_1$ efficiently (fast and accurately) at each slot, and therefore it is not practical  to use the previous approach. Instead, we rely here to an approximation scheme for solving $\mathbb P_1$, and do so in a way that these approximately-solved instances do not accumulate an unbounded regret w.r.t. $\bm x^\star$. This requires a tailored approximation analysis and a new regret metric.


Starting with the latter, due to the difficulty of the problem, we adopt an \emph{easier} benchmark which is discounted by some factor. In particular, we use the $\alpha$-approximate regret \cite{kalai-2005}:
\begin{eqnarray}
	\c R_T^{(\alpha)} =\sup_{\{\bm q_t\}_{t=1}^T} \left[ \alpha \dtp{\bm q_{1:T}}{\bm x^\star} - \sum_{t=1}^T \dtp{\bm q_t}{\bm x_t}\right], \label{eq:def-alpha-regret}
\end{eqnarray}
for some positive constant $\alpha$ that is decided by the learner and modulates the discounting effect. This generalized regret metric allows to use a parameterized benchmark, in line with prior works, see \cite{leadcache} and references therein.

 Now, it is important to see that while the Knapsack problem admits an FPTAS, due to the online nature of our caching problem, not all $\alpha$-approximation schemes for the offline OFTPL problem provide an $\alpha$-approximate regret guarantee. In light of this, we employ the stronger notion of \emph{point-wise} $\alpha$-approximation scheme \cite{kalai-2005}: a randomized $\alpha$-point-wise approximation algorithm $\mathcal{A}$ for a fractional solution $\bm{\hat x}\!=\!(\hat x_n, n\in \c N)$ of a maximizing LP with non-negative coefficients, is one that returns an integral solution $\bm x\!=\!(x_n, n\in \c N)$ such that $\mathbf{E}[x_n]\! \geq \!\alpha \hat{x}_n, \forall n \in \c N$ and some $\alpha\!>\!0$; where the expectation is taken over possible random choices made by  $\mathcal{A}$.

For the caching problem, we propose an $(\nicefrac{1}{2})$-point-wise approximation algorithm for $\mathbb{P}_1$ using the celebrated, and remarkably simple, approach of Dantzig \cite{dantzig}. This technique can solve approximately the Knapsack-style problem $\mathbb P_1$ by sorting the different \emph{items} in terms of profit-to-size ratios, and selecting the highest ranked items that fit entirely in the Knapsack, and a possibly fractional part of the last item that is required to fill the remaining capacity. The solution may inevitably include a non-binary element (the portion of the last item), and hence the algorithm requires to perform a randomized sampling so as to decide whether to include it. 

\begin{algorithm}[t]
	\caption{{\small{OFTPL for General Discrete Caching}}}
	\label{alg:oftpl-unequal}
	\begin{small}
		\nl \textbf{Input}: $\eta_1=0$, $\bm s=(s_n, n\in \c N)$, $\bm x_1\!\in \c X_d$ \\ 
		\nl \textbf{Output}: $\{\bm x_t\}_t$ \\ 
		\nl $\bm \gamma \stackrel{iid}\sim \mathcal{N}(0,\boldsymbol{1}_{N\times 1})$ \\ 
		\nl \For{ $t=2,3,\ldots$  }{
			\nl Receive prediction $\bm{\p q}_{t}$ \\[0.4ex] 
			\nl $ \eta_t = \frac{1.3}{\sqrt{C}}\left(\frac{1}{\ln(Ne/C)}\right)^{1/4}\sqrt{\sum_{\tau=1}^{t-1}\|\bm q_\tau - \bm{\p q}_\tau\|^2_1}$ \\[0.5ex]	 
			\nl $\bm p\leftarrow \bm q_{1:t-1}+\bm{\p q}_t+\eta_t\bm \gamma$ \\ [0.5ex]		 
			\nl $(\bm{\hat x}_t, k)\leftarrow$ \texttt{Dantz}$(C,\bm p,s)$ \ \ \ \ \emph{{{"Almost integral" caching}}}\\[0.4ex]		
			\nl  $\bm x_t \leftarrow$ \texttt{Rand}$(\bm{\hat x}_t, k)$ \ \qquad \qquad \  \emph{Randomized Rounding}\\[0.4ex]		
			\nl Apply caching $\bm x_t$;\\[0.6ex]		
			\nl Receive request $\bm q_t$ and calculate utility $\dtp{\bm q_t}{\bm x_t}$.
		}
	\end{small}
\end{algorithm}

The detailed steps of the OFTPL scheme are presented in Algorithm \ref{alg:oftpl-unequal}. At each slot we obtain a prediction for the next requested file $\bm{\p q}_t$ and update the perturbation parameter $\eta_t$. Then we calculate the new profits $p_n\!=\!\bm q_{1:t-1}\!+\bm{\p q}_t+\eta_t\bm \gamma$, $n\in \c N$, and solve the relaxed Knaspack by invoking \texttt{Dantz}$(C, \bm p, \bm s)$ (see  \cite{dantzig} for details) to obtain the almost-integral $\bm{\hat x}_t$ and parameter $k$. This vector has $k\!-\!1$ components equal to 1, one additional non-negative component, and $N\!-\!k$ components equal to 0. This solution is then rounded through a typical randomization scheme, denoted $\texttt{Rand}$, which selects with equal probability the first set of whole files or the last file. This creates the integral caching vector $\bm x_t$ which satisfies the capacity constraint. Finally, we observe the new gradient, update the aggregate gradient vector and repeat the process. 

Algorithm \ref{alg:oftpl-unequal} achieves expected regret:
\begin{align} 
	\mathbb{E}[\c R^{(\nicefrac{1}{2})}_T]\leq
	1.84\sqrt{C}\left(\ln{\frac{Ne}{C}}\right)^{1/4}
	\sqrt{\sumT \|\bm q_t - \bm{\p q}_t\|_1^2}  \notag
\end{align}
\noindent This bound  possesses the desirable property of being modulated with the prediction errors, and in fact it is improved by a factor of 2 compared to the equal-sizes bound. However, this regret metric is defined w.r.t. a weaker benchmark.  The complexity of Algorithm \ref{alg:oftpl-unequal} is comparable to that of Algorithm \ref{alg:oftpl}, which is quite surprising since we are able to handle integral caching decisions with arbitrary-sized files. Finally, we refer the reader to \cite{naram-sigmetrics} for the respective solution that is based on OFTRL and which has the additional benefit of a regret bound that is independent of the library size.

\section{Other Applications of OL} \label{sec:opt-applications}

In this section we present three additional problems, or rather application areas, in communication networks where optimistic learning can contribute: \emph{\textbf{(i)}} dynamic reservation of virtualized resources by a service provider (a \emph{vertical}) where there is uncertainty about the cost of resources, the demand of users and the performance-effect of each resource type; \emph{\textbf{(ii)}}  network management problems with time-average (or, \emph{budget}) resource constraints which arise in a plethora of settings and have been extensively studied using SNUM; and finally \emph{\textbf{(iii)}} the design of dynamic and fair workload assignment policies in O-RAN systems with shared computing infrastructure.

\subsection{\OL~for Network Slicing \& Leasing}

An interesting application of \OL~is in optimizing the reservation of network, compute, storage and other resources from infrastructure providers. The on-demand leasing of slices in 5G-and-Beyond networks is a typical such example \cite{paschos-slicing}. In these scenarios, \OL~can tackle both the uncertainty about the type and volume of demand that the slice requester faces (which determines the size and composition of the slice it should request), and the uncertainty of leasing costs that are fluctuating and are unknown at the time of biding, as it happens, e.g., in spot markets. Notably, there are several OCO-based solutions for such resource reservation problems, e.g., see \cite{oco-carlee-cloud, oco-cloud-paschos, oco-tnsm-jb, edge-server-hosting}. These problems are becoming increasingly important since the role of virtualization in next generation communication systems is expected to be more central than ever. Below we present a basic model of how \OL can advance resource reservation in this context. 

\begin{figure}[t]
	\centering
	\includegraphics[width=0.76\columnwidth]{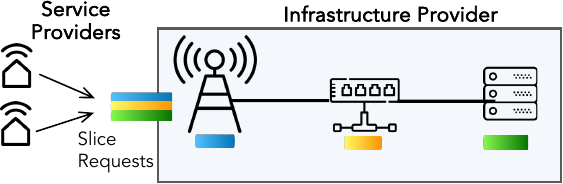}
	\caption{\small{Reservation of slices with radio, compute and storage resources in virtualized RANs; see \cite{optimistic-slicing-JB} for model details. }} \label{fig:optimistic-slicing}
\end{figure}

We consider a hybrid resource market where a vertical, namely a Service Provider (SP), leases resources from a physical network infrastructure provider (NIP) as shown in Figure \ref{fig:optimistic-slicing}. The market operates in a time-slotted fashion, where at each slot, the SP can lease a bundle of network and computing resources, and possibly complement them with additional resources leased from a spot market during the slot. We denote with $\bm p_t \!=\! (p_{t,1} , . . . , p_{t,m})$ the unit price of the $m$ resources in the advance market; and with $\bm u_t \!=\! (u_{t,1},...,u_{t,m})$ the unit price of the $m$ resources in the spot market. The SP reservation policy consists of the reservation decision $\bm x_t$ and the spot decision $\bm y_t$. At the start of each slot $t$, the SP decides its $t$-slot reservation plan $\bm z_t=(\bm x_t , \bm y_t )$, and pays the respective price at the end of the slot.  

The utility is determined by the resource configuration vector $\bm \varphi_t$, that captures the relative importance of each type of resource for the service this SP offers to its users. This vector typically depends on the requests of the users; for example, the users might need more bandwidth or more computing capacity. In the general case, the ideal mix of resources for serving the forthcoming user requests is unknown at the time of reservation, thus $\bm \varphi_t$ is unknown when $\bm z_t$ is decided. The goal of the SP is to maximize the utility from the reserved resources while minimizing the reservation costs. A general utility model that reflects this objective is:
\[
f_t(\bm z)=\alpha_t \log \big(1 + \dtp{\bm{\varphi}_t}{\bm{x}+\bm y}\big) - \dtp{\bm p_t}{\bm x} - \dtp{\bm u_t}{\bm y},
\]
where the logarithm captures the effect of diminishing returns (demand congestion), and $\alpha_t\!\geq \!0$ models the demand intensity of users which, similarly to $\bm \varphi_t$ and $\bm u_t$, is considered unknown.

The SP can employ an \OL~algorithm so as to learn how to reserve resources, both in terms of quantities and slice composition, dynamically and while trying to benefit from the availability of a forecaster. Predicting future prices of cloud and network resources is a well-explored topic \cite{patras-dl-survey}, and similarly one can assume the availability of a predictor for the user demand and ideal slice composition. Putting these together, \cite{optimistic-slicing-JB} assumes the availability of (unknown quality) predictions $\bm g_t$ for the gradient of the objective function $\bm g_t \!=\! \nabla f_t(\bm z_t)$, and proposed to optimize the reservation of resources by using the learning rule:
\begin{align*}
	&\bm z_{t+1}=\arg\max_{\bm z \in \c Z}\Big\{ r_{1:t}(\bm z)+ \dtp{\bm g_{1:t} + \bm{\p g}_{t+1}}{\bm z}	  \Big\}, \ \text{with}\\
	& r_t(\bm z)=\frac{\sigma_t}{2}\|\bm z - \bm z_t\|_2^2, \ \ \sigma_t=\sigma(\sqrt{\epsilon_{1:t}}-\sqrt{\epsilon_{1:t-1}}).
\end{align*}
where $\epsilon_t\!=\| \bm g_t\! - \bm{\p g}_t \|_2^2$ is the prediction error.

\begin{figure}[t]
	\centering
	\includegraphics[width=0.94\columnwidth]{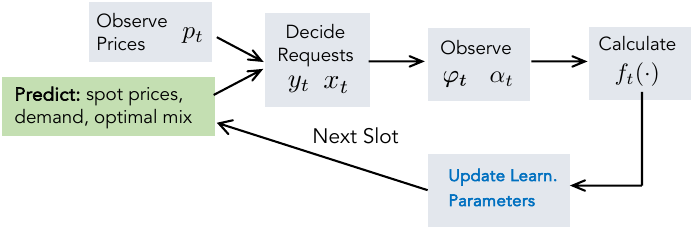}
	\caption{\small{Decision steps for optimistic learning in multi-resource slice synthesis and reservation.}} \label{fig:optimistic-slicing-events}
\end{figure}

The sequence of events is presented in Figure \ref{fig:optimistic-slicing-events}. The SP receives predictions for the future spot market prices and the intensity and type of the demand it will need to serve; decides accordingly its advance purchase and spot purchase strategy (amounts and types of resources), observes the demand intensity and type of requests, and calculates the full-information cost function. The steps are repeated after the SP updates its learning decision rule parameters.

The benefits of using \OL~in this problem is to ensure sublinear $\c O(\sqrt T)$ regret w.r.t. the ideal reservation plan $\bm z^\star$ at all cases, which ensures that, as $T$ grows, we will reach a performance that is at least as good as the best fixed reservation plan; and additionally, whenever the predictions are relevant, this convergence will be expedited expedited, all the way to achieving $\c O(1)$. For example, in many cases the demand of users, but also the pricing of resources, follows a diurnal pattern, with small deviations, and therefore one can benefit from their predictability and improve the convergence of the learning algorithm.

\subsection{Optimistic Constrained-OCO: Handling Budgets \& Queues}

Another important application domain of \OL~is in network resource allocation problems where the goal is to optimize a performance criterion while bounding a set of other criteria, or metrics, in a time-average fashion. The starting point here is the \emph{constrained} OCO extension of the OCO framework, where the decisions of the learner must satisfy $d$ long-term constraints of possibly time-varying functions:
\begin{equation}
	\bm c_t(\bm x)=\Big(c_{t,1}(\bm x), c_{t,2}(\bm x), \ldots, c_{t,d}(\bm x) \Big)\preceq 0, \notag
\end{equation}
which might as well be unknown when $\bm x_t$ is decided. Formally, additionally to sublinear regret, the learner is interested in achieving sublinear total constraint violation:
\begin{equation}
	\c V_T = \left \| \left[ \sum_{t=1}^T  \bm c_t(\bm x_t)\right]_+ \right\|. \label{const-viol}
\end{equation}

Constrained OCO (COCO) algorithms have applications in the control of capacitated communication systems, network queuing problems \cite{neely-book}, and network management with multiple constraints and performance criteria \cite{giannakis-TSP17}. In fact, much of the NUM framework addresses problems that can be cast in this format -- optimizing an objective function subject to a set of time-average budget constraints \cite{chiang-layering}. Modeling and solving this broad family of problems using \OL~algorithms is, therefore, particularly appealing.

\begin{algorithm}[t]
	\begin{small}
	\SetAlgoLined
	\KwIn{$\bm x_1 \!\in\! \c X_{T}$,  $\bm \mu_1\!=\!0$.}
	\KwOut{ $\{\bm x_t\}_t$  }
	\For{$t=1, \dots$, T}{
		\nl Decide $\bm x_t$ using \eqref{primal-update};\\	[0.5ex]
		\nl	Incur cost $f_t(\bm x_t)$ and violation $\bm c_t(\bm x_t)$;\\ [0.5ex]
		\nl Decide $\bm z_t$ using \eqref{prescient-update}; \\[0.5ex]				
		\nl	Receive predictions $\bm{\p g}_{t+1}$, $\bm{\p c}_{t+1}(\cdot)$, $\bm{\p c}_{t+1}(\bm{\p x}_{t+1})$;\\ [0.5ex]
		\nl	Decide $\bm \mu_{t+1}$ using \eqref{dual-update};
	}
	\caption{{\small{Lazy Lagrangians w. Predictions ({\texttt{LLP}})}}}
	\end{small}
	\label{alg:llp}
\end{algorithm}

Unfortunately, such multi-objective learning problems are notoriously hard to tackle. Specifically, \cite{tsitsiklis} showed that no algorithm can achieve sublinear regret and constraint violation relative to the ideal benchmark $\bm x^\star$ defined as: 
\begin{align}
	&\bm x^\star=\arg\max_{\bm x \in \c X_T^{\text{max}}} \ \sumT f_t(\bm x), \,\quad \text{where:} \\
	&\c X_T^{\text{max}}=\left\{\bm x\in \c X \, \bigg\vert \, \sum_{t=1}^T \bm c_t(x)\preceq 0 \right\}. \notag
\end{align}
Subsequent works considered more restricted benchmarks that maximize the performance but respect the constraints for shorter time windows \cite{paschos-icml}; or dynamic benchmarks $\{\bm x_t^\star\}$ that satisfy separately each $t$-round constraint $\bm c_t(\bm x_t^\star)\!\preceq 0$ \cite{johansson-TSP2020}, \cite{giannakis-TSP17}; or benchmarks \cite{kapoor-icml17}, \cite{neely-nips17} restricted in:
\begin{equation}
	\c X_T=\left\{\bm x\in \c X \, \Big\vert \, \bm c_t(\bm x)\preceq 0,\,\forall t\leq T \right\}.  \notag
\end{equation}
In all these cases, it holds $\c X_T\subseteq \c X_T^{\text{max}}$, which facilitates achieving sublinear $\c R_T$ and $\c V_T$. Other special cases of $\c X_T$ are considered in \cite{mahdavi-jmlr2012,jennaton,lamperski-nips2018, bandit-knapsacks-2019} where $\bm c_t(\bm x)=\bm c(\bm x)$, $\forall t$; and in \cite{victor} which focuses on linearly-perturbed constraints. 

The first work that studied the question of how predictions for the objective and constraint functions can benefit COCO algorithms was \cite{llp}. First, the authors presented an impossibility result, showing that even if one knows accurately the objective and constraint functions before optimizing for $\{\bm x_t\}_t$, it is impossible to compete with benchmarks in $\c X_T^{\text{max}}$. Following that, they proposed a primal-dual OFTRL algorithm, based on the following doubly-regularized Lagrangian function:
\begin{align}
	\c L_t(\bm x, \bm \mu)&=\frac{\sigma_t}{2}\|\bm x- \bm x_t\|^2+ \dtp{\bm g_t}{\bm x} \notag \\
	&+\dtp{\bm \mu}{\bm c_t(\bm x)}-\frac{\phi_t}{2}\|\bm \mu\|^2. \label{eq:lagrangian} 
\end{align} 
where $\bm \mu\succeq \bm 0$ are the dual variables for relaxing the constraints, and $\sigma_t$ and $\phi_t$ are the $t$-slot regularization parameters for the primal and dual space, respectively. The proposed primal and dual decision rules are then:
\begin{equation}
	\!\!\bm x_t\! =\! \arg \min_{\bm x\in \c X}  \left\{ \sum_{\tau=0}^{t-1} \c L_\tau(\bm x,\bm \mu_\tau) \!+\dtp{\bm{\p g}_t}{\bm x} \!+\dtp{\bm \mu_t}{\bm{\p c}_t(\bm x)} \!\right\} \! \label{primal-update}
\end{equation}
\begin{equation}
\bm \lambda_{t+1}\!=\! \arg \max_{\bm \mu\in\mathbb R_+^d} \left\{ \sum_{\tau=0}^t \c L_\tau(\bm z_\tau,\bm \mu)\! + \dtp{\bm \mu}{\bm{\p c}_{t+1}(\bm{\p x}_{t+1})}\right\}  \label{dual-update}
\end{equation}
where the dual rule uses the prediction $\bm{\p c}_{t+1}(\bm{\p x}_{t+1})$. Note also that it relies on the \emph{prescient} decision $\bm z_t$, which is calculated after the objective and constraint functions are revealed: 
\begin{equation}
	\bm z_t = \arg \min_{\bm x\in \c X} \left\{ \sum_{\tau=0}^{t} \c{L}_\tau(\bm x, \bm \mu_\tau)\right\}. \label{prescient-update}
\end{equation}
This quantity is typically used in the analysis of OFTRL algorithms (e.g., see \cite{mohri-aistats}), but in this case has a specific role in the algorithm's implementation, as well.

\begin{figure*}[t!]
	\centering
	\begin{subfigure}[t]{0.48\textwidth}
		\centering
		\includegraphics[width=0.8\textwidth]{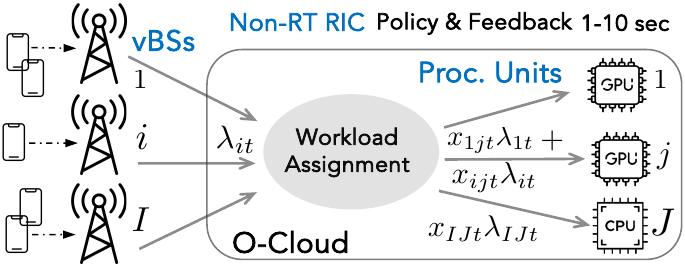}
		\caption{}
	\end{subfigure}%
	\hfill 
	\begin{subfigure}[t]{0.48\textwidth}
		\centering
		\includegraphics[width=0.92\textwidth, page=1]{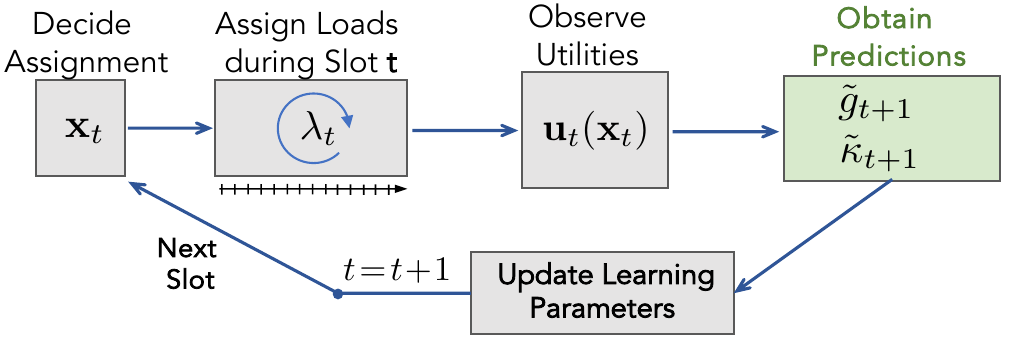}
		\caption{}
	\end{subfigure}
	\caption{\small{\textbf{(a):} A RAN controller devises the load assignment policy every $\sim \!1\!-\!10$ seconds and sends it to the vBSs which send their loads accordingly. \textbf{(b)}: Timing diagram of assignment implementation and learning policy.} }
	\label{fig:fairness-vran}
\end{figure*}

In COCO, the regularization depends on a new quantity, the prediction errors for the cost and the constraint functions modulated by the dual variables: 
\begin{align}
\xi_t=\big\|(\bm g_t-\bm{\p g}_t) + \dtp{\bm \mu_t}{\nabla \bm c_t(\bm x_t)- \nabla\bm{ \p c}_t(\bm x_t)}\big\|, \ \forall t,
\end{align}
and the primal regularization parameters are defined as:
\begin{align}
	\sigma_t\!=\sigma \left( \sqrt{ \xi_{1:t} }\! -\! \sqrt{\xi_{1:t-1} }\right),  \label{primal-reg}
\end{align}
while for the dual variables we use regularization parameters:
\begin{align}
	&\phi_t\!=\frac{1}{a_t}\!-\!\frac{1}{a_{t-1}} \quad \text{with rates} \notag \\
	& a_t=\frac{a}{ \max\left\{ \sqrt{ \sum_{\tau=1}^t \|\bm c_\tau(\bm z_\tau)\!-\bm{\p c}_\tau(\bm{\p x}_\tau)\|_2^2},\ t^\beta\right\}}, \notag  
\end{align}
where $\beta\!\in[0,1)$ is a tunable hyperparameter. 

The performance of this algorithm is summarized in \cite[Theorem 2]{llp}, where we see a trade-off between the learning convergence w.r.t. the objective and the constraints of the problem. The network controller can set the tuning parameter $\beta$ so as to prioritize the regret or the constraint violation, based on its priorities, namely:
\begin{align*}
	&\c R_T=\c O\left(T^{\frac{5}{8}}\right),\qquad\,\,\,\,\, \c V_T= \c O\left(T^{\frac{3}{4}}\right) \quad\,\,\,\,\, \text{when}\quad \beta\!<\!1/2, \\
	&\c R_T=\c O\left(T^{ \frac{3-\beta}{4}}\right),\qquad \c V_T= \c O\left(T^{ \frac{1+\beta}{2} }\right) \,\,\,\,\, \text{when}\quad \beta\!\geq\!1/2.
\end{align*}
Furthermore, when the predictions are accurate, the regret shrinks to $\c R_T=\c O(1)$ and the constraint violation becomes $\c V_T=\c O(\sqrt T)$. Interestingly, the subsequent work \cite{jordan-coco} has improved this result by a constraint violation bound that is fully adaptive to prediction errors. 

Equipped with the optimistic COCO framework, one can revisit the entire gamut of problems studied in, e.g., \cite{neely-book}, and develop learning algorithms that leverage predictions to achieve accelerated sublinear regret and constraint violation. This enhances the network control framework by introducing robustness (via OCO) and predictive capabilities (via optimism) to a wide range of resource management problems.

\subsection{\OL~for Load Assignment in O-RAN} 
The next application of \OL~pertains to achieving dynamic fairness in OCO problems, with applications to workload assignment in O-RAN platforms \cite{optimistic-fairness-fatih}. Consider a typical O-RAN comprising a set $\mathcal I$ of virtualized Base Stations (vBSs), and a set $\mathcal{J}$ of Processing Units that are co-located at some central facility (termed, O-Cloud) and process the workloads (signals) from the vBSs. We wish to design non-Real-Time policies where the operation of the system is time-slotted with slot duration of $\sim(1-10)$ secs, Fig. \ref{fig:fairness-vran}(a). During each slot $t$, every vBS $i\in \mathcal I$ sends to O-Cloud the input streams from its users, and we denote with $\lambda_{t,i}\!\geq\! 0$ the computing cycles that are required for extracting the respective user payloads (after FEC decoding, etc.). These quantities depend on the data volume but also on radio parameters such as the SNR~\cite{bib:vbs-experiments}. Hence, the value of $\bm \lambda_{t}$ is practically revealed at the end of slot $t$. Each server $j\! \in\! \mathcal J$ has computing capacity of $C_{t,j}$ cycles, and we define $\bm C_t\!=\!(C_{t,j}, j\in \c J)$. In such virtualized computing platforms, the effective capacity of each PU might change over time unpredictably.

A RAN controller (or, RIC) decides the workload \emph{assignment policy}, $\bm x_t\!=\!(x_{t,ij}\in [0,1], \forall i \in \c I, j\in \c J)$, where $x_{t,ij}$ is the portion of load of vBS $i$ that is sent to server $j$ during slot $t$. By definition, these decisions belong to a multi-simplex:
\begin{align}
	\c S_{JI}=\Bigg\{\bm{x}\in [0,1]^{J\cdot I}\ \bigg \vert \ \sum_{j\in \c J}x_{ij}=1, \forall i\in \c I \Bigg\}. \label{eq:X-simplices}
\end{align}
The assignment policy is updated at the beginning of each slot and shapes the O-RAN performance during that slot. If the RIC assigns more load to a server than its capacity, then part of this data will not be processed before its deadline and the vBSs will suffer reduced throughput \cite{bib:fluidran}. We model this effect through a utility vector function $\bm u_t(\bm x)\!=\!\big(u_{it}(\bm x), i\in \mathcal I\big)$, where $\bm u_t:\!\mathbb R^{I\times J} \mapsto \mathbb R^{I}_{+}$ is assumed non-negative and concave and each element $u_{it}(\bm x)$ denotes the performance for vBS $i$.

The goal of the controller is to devise a sequence of assignment policies $\{\bm x_t\}_t$ so as to achieve long-term fairness w.r.t. the average utilities over the entire horizon $T$ of the system operation. For the fairness criteria, we employ the \emph{generalized} $\alpha$-fairness function \cite{bib:walrand-fairness}:
\begin{align}\notag
	F_{\alpha}(\bm{u})\!=\! \sum\limits_{i\in\mathcal{I}}f_\alpha(u_i), \ \  f_\alpha(u_i)\!=\!
	\begin{cases} 
		\frac{u_i^{1-\alpha} - 1}{1-\alpha}, & \alpha \! \in\! \mathbb{R}_{\geq0} \backslash \{1\}, \\
		\log(u_i), & \alpha\! =\! 1.
	\end{cases}
\end{align}
where parameter $\alpha$ determines the fairness type, e.g., $\alpha\!=\!1$ leads to proportional fairness.  We evaluate the efficacy of the assignment policies using the metric of \emph{fairness regret}:
\begin{equation}\notag
	\begin{split}
		\mathcal{R}_T^{fr}\!=\! \sup\limits_{\!\{\bm{u}_t\}_{t}}\Bigg\{\! F_\alpha\left(\!\frac{1}{T}\sum\limits_{t=1}^T\bm{u}_t(\bm{x^\star})\right) 
		\!-\!F_\alpha\left(\frac{1}{T}\sum\limits_{t=1}^T\bm{u}_t(\bm{x}_t)\right)\!\Bigg\}. \label{eq:fairness_regret}
	\end{split} 
\end{equation}
This metric evaluates the policy that decides $\{\bm x_t\}$ by using a hypothetical benchmark $\bm {x}^\star$ that, as before, maximizes the aggregate performance (i.e., the first term in $\c R_T$). 

It is important to observe the difference of this regret definition compared to the standard regret introduced in \eqref{def:static-regret}. Here, we have a vector of utility functions instead of a single function, and there is a static encapsulating function $F_a(\cdot)$ that measures the effect of the aggregate (over time) vector of utilities. This difference has conceptual and technical ramifications. The former pertain to the multi-criteria nature of the problem where we need to find a solution that satisfies all vBSs, while the latter means that off-the-shelf OCO algorithms cannot be applied since the time-averaging inside the argument of $F_\alpha(\cdot)$ does not allow the necessary decomposition over time; see also \cite{bib:tareq_fairness, bib:agrawal2014EC} for a detailed discussion on the difficulties stemming from the presence of $F_{\alpha}$.

To overcome this obstacle, we use the Fenchel conjugate of  $F_{\alpha}\big(\bm{u}_t(\bm x)\big)$ \cite[Ch. 4]{bib:beck-book} and define the \emph{proxy function}:
\begin{align}\label{eq:psi-analytical}
	\Psi_t(\bm \theta,\bm x)=\sum_{i=1}^I \frac{\alpha(-\theta_i)^{1-1/\alpha}-1}{1-\alpha}-\dtp{\bm \theta}{ \bm{u}_t(\bm x)},
\end{align}
where $\bm \theta$ are the dual conjugate variables and belong to a compact and convex set $\Theta$ with bounded diameter $D_{\Theta}$. This function is linear on the utility values and thus, with this transformation, we return to the standard OCO setting where we optimize a (separable) sum of functions instead of a (non-separable) concave function of them. Namely, using these proxy functions we can express our problem with the help of the following per-slot program \cite[Th. 4.8]{bib:beck-book}:
\begin{align}\label{saddle-point}
	\max_{x\in \c S_{JI}}\Big\{ F_\alpha\big( \bm{u}_t(\bm{x}_t) \big) \Big\}=\max_{x\in \c S_{JI}} \Big\{\min_{\bm \theta \in \Theta} \Psi_t(\bm \theta, \bm x)   \Big\}.
\end{align}
We tackle \eqref{saddle-point} with a saddle-point algorithm that updates the primal and dual variables successively, performing independent (but coordinated) learning in the primal and dual space. In particular, we will be running an OCO algorithm on $\bm x$ to bound the primal-space regret:
\begin{align} \label{primal-space-regret-xx}
	{\mathcal{R}}_{T}^{\bm x} \doteq \sumT \Big( \Psi_{t}(\bm \theta_t, \bm x) - \Psi_{t}(\bm \theta_t, \bm x_t)\Big),\ \ \forall \bm x\in\c X,
\end{align}
and similarly, we will learn using the proxy function in the dual spaces, to bound:
\begin{align}\label{dual-space-regret-xx}
	{\mathcal{R}}_{T}^{\bm \theta} \doteq \sumT\Big( \Psi_{t}(\bm \theta_t, \bm x_t) - \Psi_{t}(\bm \theta, \bm x_t)\Big),\forall \bm \theta \in \Theta
\end{align}
As it was shown in \cite{bib:tareq_fairness}, it holds $\c R_T^{fr}\leq \c R_T^\theta + \c R_T^x + B_T$, where $B_T$ is a problem-dependent perturbation bound that is beyond the control of the learner, and therefore the primal-dual updates ensure the convergence of the targeted regret metric.

In this scenario, the predictions can expedite the learning both in the primal space (when optimizing for $\bm x$) and in the dual conjugate space (when optimizing for $\bm \theta$). For instance, we can use OFTRL for both spaces and perform:
\begin{align}
\bm \theta_{t+1}=\arg\min_{\bm  \theta \in \Theta}\Big\{r_{1:t}(\bm\theta)+\dtp{\bm \theta}{\bm  \kappa_{1:t}+{\bm{\p  \kappa}}_{t+1}} \Big\}, 
\end{align}
where vector $\bm \kappa_{1:t}=\sum_{\tau=1}^t \nabla_{\bm \theta} \Psi_{\tau}(\bm \theta_{\tau}, \bm x_{\tau})$ is the aggregate dual gradient of the proxy function w.r.t. $\bm \theta$, and $\bm{\p \kappa}_{t+1}$ the respective gradient prediction for $t+1$. Since $\Theta$ is a hyperplane, we can use a typical, general or proximal, $\ell_2$ regularizer here. It is interesting to notice that obtaining this prediction requires making a guess for the next dual decision itself, and for the next utility function value as well. 

Similarly, the primal update is performed using OFTRL: 
\begin{align}
	\bm x_{t+1}=\arg\min_{\bm x\in \c X}\Big\{  b_{1:t}(\bm x) - \dtp{\bm x}{ \bm g_{1:t}+ \bm{\p g}_{t+1}}\Big\},  \notag
\end{align}
where $b_{1:t}(\bm x)$ is a tailored entropic regularizer that extends \eqref{eq:entropic-reg} -- that was designed for the unit-simplex --  to account for the multi-simplex structure of $\c S_{JI}$;  and the gradients are defined w.r.t. the primal variables, i.e., $\bm g_{1:t}=\sum_{\tau=1}^t \nabla_{\bm x} \Psi_{\tau}(\bm \theta_{\tau}, \bm x_{\tau})$. The vector $\bm{\p g}_{t+1}$ is the gradient prediction that involves only the next-slot gradient of the utility functions and is independent of $\bm \theta$.

The detailed steps of this primal-dual optimistic algorithm can be found in \cite{optimistic-fairness-fatih}. The attained regret can be upper-bounded as follows:
\begin{align}
\c R_T^{fr} &\leq  \frac{ 2^{\nicefrac{3}{4}} I\sqrt{\log J}}{T} \sqrt{\sum_{t=1}^{T}\|\bm{g}_{t}-\Tilde{\bm{g}}_{t} \|_{\infty}^2} \quad \textit{\color{blue}\small{(primal-space regret)}} \notag \\
	&+ \frac{4\sqrt{2}{D}_{\Theta}}{T}\! \sqrt{\!\sum_{t=1}^{T}\|\bm{\kappa}_{t} - \Tilde{\bm{\kappa}}_{t}\|_{2}^2}  \qquad \ \ \ \textit{\color{blue}\small{(dual-space regret)}} \notag \\
	&+ \frac{1}{T}\sumT\left(\bm \theta_t \!- \bar{\bm \theta}_T\right)^\top \bm u_t(\bm x^\star). \qquad \ \ \textit{\color{red}\small{(perturbation budget)}} \notag
\end{align}
The first RHS term is essentially the bound for the regret $\c R_T^{\bm x}$ due to the primal OFTRL entropic learning. It depends on the diameter of the multi-simplex (written explicitly here) and diminishes to $0$ for perfect predictions. The second RHS term is the bound for the regret $\c R_T^{\bm \theta}$ due to the dual OFTRL with quadratic regularizer. This term depends on the diameter of the conjugate dual space (variables $\bm \theta$, which is bounded). The last term is a residual that depends on the time-average value of the dual variables, $\bar{\bm \theta}_T=\sumT \bm \theta_t/T$, and the utility at the optimal point. This term is beyond the learner's control and reflects the severity of the perturbations. Naturally, for certain extreme adversaries, long-term fairness learning may fail to converge, cf. \cite{bib:tareq_fairness}. Despite this inescapable inefficiency, we observe that \OL~can accelerate the learning process, despite the complexities introduced by the encapsulating fairness function.

Clearly, this technique can be used to tackle a variety of (many-to-many) load assignment problems under a satisfactory range of perturbation models that go beyond stationary i.i.d. conditions of the previous NUM tools. Furthermore, the utility model can be extended to handle also cost functions and multiple fairness criteria, e.g., an interesting twist is to learn how to achieve fair performance for the base stations and fair cost allocation for the servers over the horizon $T$. Finally, it is important to stress that the long-term fairness problem differs fundamentally from the per-slot fairness problem, which simply ensures a fair outcome in each round, and comes with higher price of fairness, see proof in \cite{bib:tareq_fairness} and discussion in \cite{optimistic-fairness-fatih}. Nevertheless, \OL~can be also used in this latter case.

\section{OL for Systems with Memory} \label{sec:opt-memory}

In this final section of applications, we return first to the theory of \OL~and discuss its extension to problems where each slot's decision affects also the future functions. Notable examples include network control problems with reconfiguration delays / costs, and problems where past actions have an accumulated effect on the objective, e.g., networks with moving nodes (e.g., drones) where past decisions impact their current position, or with storage (e.g., battery) where past decisions determine the current resource availability. These problems are technically challenging due to inter-slot decision dependencies but have numerous applications. 

\subsection{OCO with Memory \& Switching Costs}
We first discuss the class of OCO problems where the objective function at each slot $t$, denoted as $f_t(\bm x_{t-m}, \ldots, \bm x_t)$, depends directly on past decisions $\bm x_{t-m}, \bm x_{t-m+1}, \ldots, \bm x_{t-1}$ as well as on $\bm x_t$. As explained in Sec. \ref{sec:oco-metrics}, this memory property can be interpreted as the adversary being non-oblivious, meaning it selects functions (i.e., the environment) after observing past learner decisions. We refer to these problems as \emph{OCO with memory} (OCO-m).

\begin{figure}[t]
	\centering
	\includegraphics[width=\columnwidth]{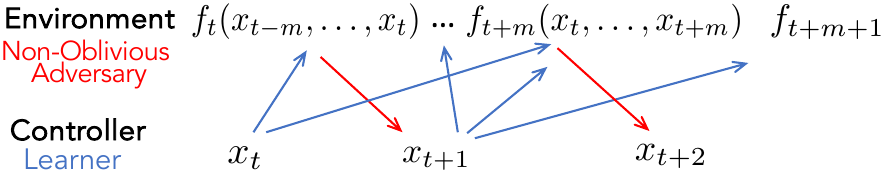}
	\caption{OCO with memory model, where each action affects current and future cost functions.} 	\label{fig:oco-m-sequence}
\end{figure}

The first study on OCO-m appeared as early as in 2002 \cite{oco-m-TIT02}, with important follow-up works few years later, such as \cite{dekel-oco-m-icml12}. A seminal contribution in this area is \cite{oco-m-hazan15} which introduced OCO-m and demonstrated that the FTRL algorithm can be applied to a modified memoryless cost function in a way that ensures sublinear regret for the original function with memory. Formally, the regret for OCO-m is defined as:
\begin{align}
	R_{T}^m=\sum_{t=m}^T f_t(\bm x_{t-m}, \ldots, \bm x_t) - \min_{\bm x\in\c X}\sum_{t=m}^T f_t(\bm x,\ldots,\bm x) \label{regret-oco-m}
\end{align}
where the benchmark is defined based on $m$-dimensional functions $f_t:\c X^m\mapsto \mathbb R$, yet uses the same argument $\bm x$ for each of the $m$ dimensions. The rationale here is that one can bound this regret in two steps. First, using Lipschitz continuity, we bound the difference: 
\begin{align*}
&|f_t(\bm x_t, \ldots, \bm x_t) - f_t(\bm x_{t-m}, \ldots, \bm x_t)|\leq \\
&\leq  L\| (\bm x_t, \ldots, \bm x_t) - (\bm x_{t-m}, \ldots, \bm x_t)\|,
\end{align*}
through the pairwise differences of their arguments\footnote{The distance of successive FTRL decisions $\bm x_t$ and $\bm x_{t-1}$ is a well-known bound often used in OCO analysis \cite{mcmahan-survey}.}; and secondly, since the right-side is an $m$-argument function with no memory, we apply the standard FTRL regret bound w.r.t. the benchmark in \eqref{regret-oco-m}. Combining these results, we obtain the desirable augmented bound for the memory function. 

A widely studied special case of OCO-m considers switching costs, where we measure the decision changes using the $\ell_1$ (or some other) norm. In this case, the system has memory of $m=1$ slot, and specifically we are interested in the quantity:
\begin{align}
f_t(\bm x_t)+\|\bm x_t - \bm x_{t-1}\|_1, \ \ \forall t.
\end{align}
The term OCO \emph{with switching costs} is coined to describe this case that captures important practical problems. Examples include datacenters where decisions relate to assignment or scaling of VMs and the switching cost captures the reconfiguration delay \cite{8626768}; optical networks where wavelength reassignments induce transmission delays \cite{7828146, 1230204};  caching systems where fetching new content incurs transmission delays \cite{switching-caching-abhishek}, \cite{8999784}; and wireless networks where handovers between base stations introduce delays \cite{6689353}.

Switching costs were first studied in \cite{oco-m-hazan15}, which proposed a modification to FTRL so as to bound the decision changes while maintaining sublinear static regret for the function (that might depend on the past in other ways, too). Later studies refined this approach, i.e., modifying standard OCO algorithms to account for the switching cost. For example, \cite{switching-caching-abhishek} enriched FTPL, and \cite{zhang-smoothed-ol} applied a meta-learning approach using OGD. However, these works did not incorporate predictions, even for this restricted class of OCO-m problems.

\subsection{Optimistic OCO-M}

Applying optimism to OCO-m requires an idea that was introduced in \cite{naram-cdc}. Instead of viewing the learning problem as one where the function at each slot depends on past decisions, we can remodel it as a problem where the cost functions have delayed gradients. To put it simply, when the learner decides $\bm x_t$, it does not get to see the induced gradient $\nabla f_t(\bm x_t)$ at the end of this slot, but instead has to wait for the next $m$ cost functions, i.e., all those that are (partially) affected by $\bm x_t$. This conceptual equivalence of memory in costs with delay in observations, is pivotal from a technical point of view.

\begin{figure}[t]
	\centering
	\includegraphics[width=0.9\columnwidth]{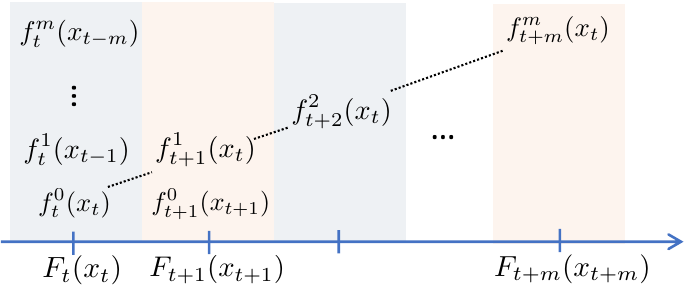}
	\caption{\small{\textbf{Diagonal}: The effect of the $t$-slot decision $x_t$ extends to slot $t+m$; thus $\nabla F_t(\bm x_t)$ is fully revealed at $t+m$. \textbf{Vertical:} The cost at each slot $t$ depends on the $t$-slot decision through $f_t^0(\bm x_t)$, on the $(t\!-\!1)$-slot decision through $f_t^1(\bm x_{t-1})$, and so on, until the past $(t\!-\!m)$ slot via $f_t^m(\bm x_{t-m})$.}} \label{fig:oco-memory}
\end{figure}

To formalize this idea, let us first assume the memory-based functions can be decomposed into components, each depending on a decision from a different slot:
\begin{align}
	f_t(\bm x_{t-m},\ldots, \bm x_t)=&f_t^m(\bm x_{t-m})+f_t^{m-1}(\bm x_{t-m+1})\notag \\
	&+\ldots+f_t^{0}(\bm x_t).\label{eq:memory-function-transf}
\end{align}
The superscript here intends to mark how far back in time the learner had decided the argument that affects each component. We stress that if the function is not separable, it can always be linearized (assuming joint convexity) to achieve this structure. Now, let us define the \emph{forward function} that quantifies the current and future impact of a decision made as slot $t$:
\begin{align}
	F_t(\bm x_t)=\sum_{i=0}^m f_{t+i}^{i}(\bm x_{t}).
\end{align}
In other words, $F_t$ collects the components from the different cost functions in the interval of slots from $t$ to $t+m$, which depend on the decision in slot $t$. The key observation here is the following result from \cite[Lemma 2]{naram-cdc} that connects the memory functions and forward functions:
\begin{align}
&\sum_{t=m}^T f_t(\bm x_{t-m}, \ldots, \bm x_t) \stackrel{\eqref{eq:memory-function-transf}}=\sum_{t=m}^T\left(\sum_{i=0}^m f_t^i(\bm x_{t-i})\right) \nonumber \\
&\stackrel{(\alpha)}=\sum_{t=m}^T\sum_{i=0}^mf_{t+i}^i(\bm x_t)\triangleq\sum_{t=m}^T F_t(\bm x_t),
\end{align}
where $(\alpha)$ follows by reordering terms and adjusting indices accordingly. The implication of this observation is that the cost accumulated over time by the memory functions is equal to the cost accumulated over time by the forward functions. The proof is surprising simple as it only requires a careful index manipulation, please see Fig. \ref{fig:oco-memory} for an illustration.

This result opens the road to use in OCO-m algorithms that are designed for problems with delayed gradients, see \cite{delay-oco-giannakis, pmlr-v28-joulani13, delay-oco-joulani-aaai16, orabona-delay-opt-icml21}. Based on this, \cite{naram-cdc} extended the regularization mechanism from \cite{orabona-delay-opt-icml21}  to develop the first OCO-m optimistic learning algorithm. In detail, \cite{orabona-delay-opt-icml21}  observed that optimism is related to delay since one can use predictions, apart from the next-slot gradient, also for the not-yet-observed gradients of past costs. Extending this idea, \cite{naram-cdc} proposed using a hybrid prediction matrix $H_t$ that accumulates the gradient of the next-slot cost function (as usual) and those gradient components of past functions that are currently unavailable due to delay. In detail, defining the gradient of the forward function $F_t$ as:
\begin{align}
	G_t = \sum_{i=0}^m \bm g_{t+i}^{(i)}
\end{align}
we can write succinctly the (hybrid) prediction vector for $t$:
\begin{equation}
	H_t\doteq\underbrace{\sum _{i=0}^{m-1}\bigg( \sum_{j=0}^{m-i-1}\bm g_{t-m+i+j}^{(j)}}_{\text{{\color{blue}{available at $t$}}}} 
	\ +
	\underbrace{\sum_{j=m-i}^m \tilde{\bm g}_{t-d+i+j}^{(j)}\bigg)+ \tilde{\bm G}_t}_{\text{{\color{red}{future predictions}}}} \notag
\end{equation}
where the first term in the parenthesis indicates the information already available, and the second term the unobserved (delayed) gradients related to past actions from $\bm x_{t-1}$ all the way to $\bm x_{t-1-m}$, while the last term is the prediction for the gradient of the forward function. To shed some light on this expression, note that for each past action $\bm x_{t-k}$ we need to predict all respective gradients for function components starting at slot $t\!-k+\!k+\!1=t+\!1$ up to $t-\!k+\!m$, after which $\bm x_{t-k}$ does not affect the functions. On the other hand, for actions farther in the past, e.g., $\bm x_{t-m-1}$, we already observed at $t$  all their component gradients (the maximum delay is $m$ slots) and therefore they do not need to be predicted.

Equipped with this prediction scheme that incorporates all the missing information, we can use, e.g., the  FTRL rule:
\begin{align}
	\bm x_{t+1}=\arg\min_{\bm x\in\c X} \Big\{ \langle (G_{1:t-m} + H_t) , \bm x\rangle +r_{1:t}(\bm x)\Big\}, \notag
\end{align}
where the regularization parameters can be defined as $\sigma_{t}\!=\sigma\sqrt{\epsilon_{1:t}}$, with the prediction error being:
\begin{align}
	\epsilon_t = \|G_{t-m:t} - H_t\|_2^2. \label{eq:errors-oco-m}
\end{align}
Comparing this expression with the respective definitions in Sec. \ref{sec:optimistic-basics}, e.g. \eqref{eq:oftrl-reg-l2}, we observe that at each slot we accumulate a truncated prediction error that reflects the memory window.

The challenge with the expression in \eqref{eq:errors-oco-m} however is that at the end of slot $t$, when we would like to calculate $\bm x_{t+1}$, we are not in position to measure this prediction error, since we will not have yet observed the delayed gradients. There are different ways to handle this issue. The brute force approach is to upper bound this missing information and regularize aggressively by assuming the worst case scenario. This, in fact, is the standard non-adaptive approach for setting the learning rate in OGD, FTRL and other OCO algorithms. A more elegant approach, first proposed in \cite{orabona-delay-opt-icml21} for OCO with delayed gradients, and further expanded in \cite{naram-cdc} for OCO-m, is to enable a cautious data-dependent regularization so that to avoid unnecessary over regularization (which would lead to slower learning). The result of this technique is a regret bound that is commensurate with the prediction errors, i.e., $\c R_T=\c O(\sqrt{\epsilon_{1:T}})$, which becomes $\c R_T=\c O(1)$ when all predictions are accurate and remains sublinear even when all predictions fail, $\c R_T=\c O(m\sqrt T)$.

\subsection{Optimistic Non-Stochastic Control}

A distinct variant of OCO-m arises when the studied system is stateful. In this case, the learner's decisions influence future functions through their effect on the system state, which evolves based on a predetermined rule that takes as input the decisions over a time window (memory $m$) and some time-evolving disturbances. The work \cite{pmlr-v97-agarwal19c} introduced the first OCO model for this class of problems where the learner observes the state $\bm{s}_t \in \mathbb{R}^{d_s}$ of a time-slotted dynamical system, then decides $\bm{x}_t \in \mathbb{R}^{d_x}$, and finally receives the cost $f_t(\bm{s}_t, \bm{x}_t)$. A fundamental class of systems studied in this setting are Linear Time-Invariant (LTI) systems, where the state transition is parameterized by fixed and known matrices and a time-varying {disturbance} vector:
\begin{align}
	\bm{s}_{t+1} = A \bm{s}_{t} + B \bm{x}_{t} + \bm{w}_{t}. 	\label{eq:LTIdynamics}
\end{align}
The adversary determines both the functions $\{f_t\}_t$ and the disturbance vectors $\{\bm w\}_t$, and does so after the learner commits its decision. This framework enables the modeling of a wide range of dynamic stateful systems that were previously intractable, see \cite{pmlr-v242-mhaisen24a} for a comparative discussion.

The goal of the learner here is to devise a decision \emph{policy} that achieves sublinear \emph{policy regret} w.r.t. the benchmark that is selected from a certain policy class $\Pi$:
\begin{align}
	\label{eq:regret-first-def}
	\mathcal{R}_T \doteq \sumT f_t\left(\bm s_t, \bm x_t\right) 
	- \min_{\pi\in\mathbf \Pi} \sumT
	f_t\big(\bm s_t(\pi), \bm x_t(\pi)\big).
\end{align}
In \cite{pmlr-v97-agarwal19c}, the class $\Pi$ comprises policies that are parameterized by a stabilizing matrix $K$ and a sequence of $m$ matrices $M\doteq\big[M^{[1]}, M^{[2]}, \dots, M^{[m]}\big]$, leading to the decision:
\begin{align}
	\label{eq:dac-action}
	\bm{x}_{t} = K \bm{s}_t + \sum_{j=1}^m M^{[j]}_t \bm{w}_{t-j}.
\end{align}
Matrix $K$ is fixed, while the matrix variables $M^{[j]}_t, \forall j, t$, are selected from a bounded set $\c M$. In other words, this class contains policies where the decisions are linearly dependent on the state and the previous $m$ disturbances, i.e., the time-window of the policy, and the learner needs to decide how to adjust these weights over time. It is not difficult to observe that this becomes a learning problem over the weight matrices $M_t^{[j]}$ with a memory effect of $m$ slots, as each matrix affects (possibly) all next $m$ functions; hence, we can return to the results of the previous subsection. 

With this in mind, \cite{pmlr-v97-agarwal19c} proposed to learn these variables using a standard OGD algorithm: 
\begin{align}
	M_{t+1}=\Pi_{\c M} \Big( M_t - \eta \nabla_{M} f_t(M, \ldots,M)	\Big) \label{eq:ogd-nsc}
\end{align}
where $\nabla_{M} f_t(M, \ldots,M)$ is the gradient of the extended $m$-slot function w.r.t. $M$. The main result in \cite{pmlr-v97-agarwal19c} establishes that \eqref{eq:ogd-nsc} achieves sublinear regret, specifically $\c R_T=\c O(m\sqrt T)$, indicating that asymptotically, the learner incurs no more cost than the benchmark policy $\pi^\star$ selected from $\mathbf \Pi$. Follow up works have extended this control-learning framework by incorporating constraints, considering stronger benchmarks (e.g., dynamic regret), and by making it adaptive to gradients (see definitions in Sec. \ref{subsec:regret-adaptivity}) so as to avoid over-regularization \cite{pmlr-v242-mhaisen24a}. Lastly, adding optimism to this framework can be achieved with the hybrid prediction vector (as explained above) and using an algorithm such as OFTRL. This will enable regret bounds that shrink commensurately with the prediction accuracy, and remain sublinear in the worst case.

%
%
%
%
%
%
%
%
%
%
%
%
%
%
%
%
%
%
%
%
%
%
%
%
%
%
%

\section{Future directions}\label{sec:future-directions}
In this section, we outline promising avenues for extending \OL~in the context of communication network optimization.

\subsection{Integrating Distance-based \& Directional Optimism}
As highlighted in Sec. \ref{sec:different-optim}-A,  an alternative approach in \OL~ is to assess the effectiveness of predictions using their directional alignment with the true costs rather than their norm-based distance. This error measurement method leads to regret bounds of a different nature ranging from $\log T$ to $\sqrt{T}$, \cite{pmlr-v119-bhaskara20a, bhaskara2023competing}, which suggests that predictions correlated with actual costs can be leveraged for expediting learning, even if they are not point-wise accurate (the norm-based error is large). This, in turn, allows us to incorporate a broader set of prediction mechanisms in \OL~and therefore extend the set of network management problems that can benefit from this toolbox. For instance, returning to the motivating example in Section \ref{sec:introduction}, Fig. \ref{fig:motivating-example}, we might be able to use predictions that merely inform us about the best channel instead of predicting with accuracy its exact gain.

In this context, an important research direction is to investigate how correlation-based optimism can be seamlessly integrated with the norm-based approach presented in this paper. From a practical point of view, this extension will allow network controllers to use concurrently a broader range of predictors and benefit from the best of those for each use case. The primary challenge in doing so lies in the structural assumptions imposed by correlation-based methods; most notably, the technical requirement for strong convexity in the decision set. This restriction limits the applicability of directional optimism to a narrower class of problems, whereas norm-based optimism is generally more flexible. Future work could explore ways to relax these convexity requirements to eventually develop hybrid approaches that switch dynamically between norm-based and directional-based optimism depending on the problem characteristics. The ideal outcome here would be to obtain bounds that depend on the minimum error among these two approaches.

Besides integration, a pertinent future direction is the theoretical refinement and empirical evaluation of correlation-based optimism in real-world network applications. So far, research in this area has been limited, with only few exceptions. The work in \cite{wen2024augment} explored directional optimism for communication, applying it to opportunistic channel selection and mobile crowd sensing. Similarly, \cite{wen2024robust} investigated the same applications but in a decentralized setting, where predictions take the form of messages from potentially malicious neighbors, making them inherently unreliable. Such works show the potential of this idea and can be further explored.

\subsection{Unifying Adversarial \& Stochastic Environments}
Another future direction for \OL-based NUM is to leverage the stochastically extended adversarial (SEA) model \cite{sarah-between}, which interpolates between stochastic and adversarial environments. This model essentially assumes the environment is not entirely adversarial but instead exhibits some stochastic structure, albeit with occasional distributional shifts. This condition allows for designing algorithms that leverage stochastic regularity while remaining robust to adversarial variations. It also subsumes the fully adversarial environments (addressed in this tutorial) and the fully stochastic settings (stochastic optimization), and the in-between spectrum. Recent works such as \cite{sarah-between, chen2024optimistic, wang2024online}, have explored regret bounds that depend on both the stochastic variance and the adversarial variation of the gradients. These results indicate that in a predominantly stochastic environment with occasional adversarial perturbations, it is possible to improve the regret bounds compared to those achieved in fully adversarial settings. This is very useful for those communication networks that operate under benign conditions most of the time (stochastic setting), with occasional disruptions due to, e.g., an attack, that makes the conditions adversarial for only a certain time window. To the best of our knowledge, the SEA model has not yet been applied to NUM problems, leaving it a promising future step.

Moreover, in this context it is intriguing to investigate how predictions should be designed or, put differently, what type of predictions are beneficial. Unlike typical \OL~where predictions focus on the next gradient (or function), in SEA the learner could use forecasts about the underlying distributional structure, its parameters, the timing of distribution shifts, or the presence of stochastic elements. Investigating how these different types of predictions affect the learning performance of the \OL~algorithm remains an open and important question, especially since in practical network problems one may have different type of predictive capabilities. 

\subsection{Reductions among Metrics}
Other important learning metrics, such as the competitive ratio and adaptive regret (see discussion in Sec. \ref{sec:different-optim}) offer valuable perspectives worth exploring in future research. Some studies have explored the connection between the competitive ratio and various regret-based metrics. For instance, \cite{pmlr-v30-Andrew13} examined the relationship between competitive ratio and static regret, while \cite{zhang-smoothed-ol} investigated online learning approaches that simultaneously address competitive ratio and dynamic regret with switching costs, illustrating the potential for algorithms to balance (or hedge on) multiple learning criteria. Furthermore, the relationship between competitive ratio and policy regret was studied in \cite{goel2023best}; and several works examined the interplay among different regret notions per se. For example, \cite{zhang2020minimizing} proposed algorithms that simultaneously achieve guarantees in both adaptive and dynamic regret; and \cite{jacobsen2024equivalence} provided insights into the relationship between static and dynamic regret through a generalized ''path-length" complexity measure.

Future research could delve into the relationships among these learning metrics. Namely, investigate the suitability of these metrics in different applications and potentially devising further reductions among them, particularly through the lens of optimistic learning (i.e., considering untrusted predictions). Understanding these intricate relationships can eventually guide the design of universal \OL~NUM algorithms that guarantee effective learning with respect to multiple criteria at the same time. This is important since different learning metrics are suitable for different network problems, or even for the same problem under different conditions\footnote{E.g., recall that if the cost functions are revealed before the decision, then one can explore competitive ratio algorithms instead of regret.} and algorithms of this nature relieve the network controller from the duty to select a different algorithm for each scenario.

%
%
%
%

\subsection{Systems with Moving Comparators}
As with any learning framework, a foundational question in optimistic learning is: what exactly are we trying to learn? In OCO and by extension, in \OL, the answer to this question is embedded in the choice of the benchmark -- a comparator against which algorithmic performance is evaluated. Most existing works in \OL~focus on static regret, comparing performance to a fixed action $\bm{x}^\star$ that can be chosen only with access to all future functions (best-in-hindsight). However, this choice may be misaligned with the structure of some practical network problems, especially for networks operating under volatile and highly dynamic conditions. Consider, for example, the user-to-base-station association problem in cellular networks, where users move frequently, and maintaining a fixed association is both suboptimal and unrealistic \cite{mike-handover}. In such settings, dynamic regret (see Sec.\ref{subsec:definitions-of-regret}) -- measuring performance against a changing comparator sequence ${\bm{x}_t^\star}_t$ -- is a more meaningful objective. Yet, achieving tight dynamic regret bounds in the optimistic setting introduces new technical challenges, including managing comparator path lengths. This issue has been only recently studied, e.g., see \cite{peyman-oco-tac}, and future research is needed to establish general-purpose frameworks that can handle dynamic comparators in diverse problem classes that span NUM.

\subsection{Systems with States and Memory}
	
In Sec. \ref{sec:opt-memory}, we explored \OL~for stateful problems where the decisions influence an evolving system state and, through that, also the cost function, and we focused on linear time-invariant dynamical systems. However, this important first step captures only a subset of broader stateful decision-making problems. Therefore, it remains an open question whether it is possible, and how, to extend \OL~to a broader class of stateful problems, e.g., when the system dynamics vary with time or based on a non-linear rule, or when the system behavior is governed by an underlying Markov Decision Process. Indeed, MDPs and the associated Reinforcement Learning (RL) algorithms have an extremely wide application range in communication systems and in that regard any such extension of \OL~will be impactful. For an introduction to MDPs in the context of wireless communication networks, we refer the reader to the survey \cite{alsheikh2015markov} and for an overview of modern reinforcement learning techniques in networks to  \cite{yang2024beyond}. Interestingly, OCO algorithms such as the FTRL and FTPL have been studied in the context of MDPs \cite{dai2022follow}, which shows promise for their applicability of their optimistic variants.

\subsection{Systems with Non-convex Decisions \& Functions} 

Another promising area of improvement is the extension of optimistic learning to handle non-convex problems, particularly those involving discrete action spaces (e.g., binary or integer decisions) or non-convex objective functions. This is a long-standing challenge inherited from the OCO framework on which optimistic learning is built. While certain structured instances -- such as those addressed in Section~\ref{sec:opt-discrete-caching} -- can already be handled using relaxed convex surrogates and approximation techniques, a general framework for non-convex optimistic learning remains elusive. One important direction lies in leveraging the connection between convex and submodular optimization \cite{10.5555/2602000}, which could yield tractable formulations and sublinear regret guarantees in combinatorial decision spaces. Bridging this gap would unlock optimistic learning for a wide range of important applications, including scheduling, routing, and resource allocation in discrete decision domains.

\subsection{Improving \OL: Fully Constant-aware Bounds} It is crucial to further develop the theoretical foundations of \OL~to ensure that regret bounds do not degrade unfavorably when predictions are highly inaccurate. In particular, optimistic algorithms should retain performance guarantees comparable to those of legacy OCO methods in worst-case scenarios. While this robustness is already ensured in some settings, in others, the introduction of optimism can come at the cost of worse constants in the regret bound. Now, in theoretical OCO studies and even in ML applications the focus has been predominately on the convergence rate of the regret bounds, i.e., on the dependency of the regret on the time horizon $T$. Nevertheless, in communication problems we are often interested as much for the dependency of the regret on other system parameters. A notable example, discussed in Sec. \ref{sec:opt-caching}, is caching where a regret bound that increases with the number of files $N$ (library size) is highly undesirable.

A notable recent step in this direction is the work of \cite{orabona-delay-opt-icml21}, which replaces the typical quadratic dependence on the prediction error with a Huber-style loss. The distinction becomes especially relevant when prediction errors are large: both approaches yield sublinear regret in $T$, but the Huber loss leads to significantly milder degradation, with constants scaling only with the square root of those in the standard case. Further exploration of such techniques may help ensure that optimistic learning remains competitive even in poorly predicted network environments.

\subsection{Improving \OL:  Joint design with Predictors}
Finally, a natural next step that has not been considered by theoretical nor application studies, is the joint design of predictors and online algorithms. In most \OL~formulations, predictors are treated as black-box inputs that are generated by some exogenous ML models and fed into the online learner without feedback or coordination. While this modularity has advantages, it also limits the performance potential of the system as a whole. There is an opportunity to explore how predictor design and learning dynamics can be co-optimized for improved outcomes. Key questions in this direction include: when is it beneficial to acquire predictions at all? In which cases do predictions meaningfully improve learning performance, and under what conditions do they become redundant or even harmful? Furthermore, given that prediction models are often learned or updated via costly offline procedures, it is essential to develop principled criteria for when to train or re-train predictors, potentially based on system feedback, performance degradation, or measures of prediction utility. Ultimately, moving beyond the static predictor-learner pipeline towards a more holistic design in \OL~ has the potential to improve both its efficiency and performance.

\section{Conclusions} \label{sec:conclusions}

Optimistic learning leverages predictive models within online learning frameworks in an elegant and efficient fashion. In particular, \OL~ integrates seamlessly predictions from one or more ML models about the future system functions or their gradients into the learning rules and devises its decisions with enhanced information. This way, it expedites the learner's convergence to the benchmark performance while maintaining the robustness of traditional OCO methods in worst-case scenarios. In this sense, \OL~is a best-of-both-worlds solution as it achieves the sweat spot between offline and online learning, utilizing dynamic forecasters that are found to be accurate and dismissing them when they turn out to be ineffective. This versatility, surprisingly, comes without noticeable computation and communication overheads and often without (significant) compromise in terms of learning rates for worst-case scenarios.

This tutorial provides a systematic overview of the theory and foundations of \OL~ and presents how it can be applied to important network management problems such as caching, edge computing, network slicing, O-RAN workload assignment and others. These examples are representative of different families of problems that have been extensively studied using static and stochastic NUM approaches, yet their manifestation in future communication systems raises non-trivial challenges that require a new approach that \OL~can offer. By leveraging predictive models, \OL~enhances decision-making in these contexts and allows for unprecedented improvement of KPIs, as was demonstrated through comparisons with other optimization and learning-based techniques.

\section{Appendix} \label{sec:appendix}

This section contains a condensed collection of basic optimization elements in order to facilitate the reader accessing this tutorial. For a detailed overview of related background material we refer the reader to \cite{boyd-book} and \cite{beck-book}.

\subsection{Lipschitz Continuity and Strong Convexity}

$\bullet$ A function $f:\c X \mapsto \mathbb R$ is $L$-Lipschitz continuous if $|f(\bm x')-f(\bm x)|\leq L\|\bm x'-\bm x\|$, $\forall \bm x',\bm x\in \c X$. 

$\bullet$  A function $f:\c X\mapsto \mathbb R$ is $\sigma$-strongly convex if:
\begin{align}
	f(\bm y)\geq f(\bm x) + \dtp{\nabla f(\bm x)}{\bm y-\bm x} + \frac{\sigma}{2}\|\bm y-\bm x\|_2^2, \ \forall \bm x, \bm y \in \c X \notag
\end{align}
which, intuitively, means that it has additional curvature beyond that of convexity. In this definition we have used the $\ell_2$ norm to measure the distance of $\bm y$ and $\bm x$. In general, one can define strong convexity with respect to any norm $\|\cdot \|$. 

Based on this, if a function is $\sigma$-strongly-convex with respect to norm $\|\cdot\|$, then it is $1$-strongly-convex (i.e., $\sigma=1$) with respect to the norm $\sqrt \sigma \|\bm x\|$. Furthermore, if we have a set of functions $\{r_t(\bm x)\}_t$, where each one is $\sigma_t$-strongly-convex with respect to norm $\|\cdot\|$, then their sum, $r_{1:T}(\bm x)$, is $\sigma_{1:T}$-strongly-convex with respect to that same norm $\|\cdot\|$.

$\bullet$  A set $\c X$ is called strongly convex when for every $\bm x, \bm y \in \c X$ and $\gamma \in [0,1]$, it holds:
\begin{align}
	\gamma \bm x + (1-\gamma)\bm y + \gamma (1-\gamma)\frac{\alpha}{2}\|\bm x - \bm y\|_2^2\bm z \ \in \c X.
\end{align}
Intuitively, this definition states that not only the direct line connecting any two points of $\c X$, but also lines with some curvature, lie within $\c X$.

\subsection{Norms}
Typical norms that are used in OCO and \OL~ include:
\begin{itemize}
	\item The Euclidean $\ell_2$ norm: $\|\bm x\|_2= \sqrt{\sum_{i=1}^N |x_i|^2}$
	\item The infinity $\ell_{\infty}$  norm: $\|\bm x\|_\infty= \max\{|x_1|, \ldots, |x_N| \}$
	\item The Manhattan $\ell_1$ norm: 	$\|\bm x\|_1= |x_1|+|x_2|+ \ldots + |x_N|$
\end{itemize}

Another norm-related concept that appears frequently in the analysis of \OL~algorithms is the associated dual norm of some norm $\|\cdot\|$, which is denoted $\|\cdot\|_{\star}$ and defined as:
\begin{equation}
	\|\bm z\|_{\star}= \max_{\bm x} \dtp{\bm z}{\bm x} \,\,\,\,\,\text{s.t.}\,\,\,\,\,\|\bm x\|\leq 1. \notag
\end{equation}
Some examples of dual norms that used in this tutorial are: 
\begin{itemize}
	\item The dual $\ell_2$ norm:  $\|\bm x\|_{2,\star}=\|\bm x\|_2$;
	\item The dual $\ell_1$ norm: $\|\bm x\|_{1, \star}=\|\bm x\|_\infty$;
	\item The dual $\ell_{\infty}$ norm $\|\bm x\|_{\infty, \star}=\|\bm x\|_1$.
\end{itemize}
Finally, the norms in OCO are often weighted with (or defined based on) different time-varying parameters, and we denote them with, e.g., $\|\bm x\|_{(t)}=\sqrt \sigma_t \|\bm x\|$, which has dual norm $\|\bm x\|_{(t),\star}=\frac{1}{\sqrt \sigma_T}\|\bm x\|$. This notation is useful in OCO as typically we have regularizers with weights that adapt to time-varying cost functions, and norms of the type $\|\cdot\|_{(t)}$ can capture this in a compact way.

\subsection{Self-concordant functions}

Self-concordant barrier functions have been mainly used in interior point optimization and specifically in the analysis of Newton methods. Their properties render them a useful algorithm design and analysis tool for OCO as well. For a detailed discussion on this topic we refer the reader to \cite{nesterov-book04} and in \cite[Ch. 6]{hazan-book} for their application to OCO.

A function $\Phi\!:\!\text{int}(\c X)\!\mapsto\! \mathbb R$ defined on the interior of a convex set $\c X\!\subset\!\mathbb R^n$ is a $\nu$-self-concordant barrier for $\c X$ if:
\begin{itemize}
	\item It is three times continuously differentiable and convex, and approaches infinity for any sequence of points that are approaching the boundary of $\c X$.
	\item For every $\bm u\in \mathbb R^n$ and every $\bm x\in \text{int}(\c X)$, it satisfies:
	\begin{align}
		&|\nabla^3 \Phi(\bm x)[\bm u, \bm u, \bm u]|\leq 2\left( \nabla^2\Phi(\bm x)[\bm u, \bm u]	\right)^{3/2} \notag \\
		&||\nabla \Phi(\bm x)[\bm u]|\leq \nu^{1/2}\left( \nabla^2\Phi(\bm x)[\bm u, \bm u]	\right)^{1/2} \notag
	\end{align}
\end{itemize}
This definition in essence corresponds to Lipschitz continuity of the Hessian of $\Phi$.

Self-concordant functions can be used to define norms:
\begin{align}
	&\|\bm u\|_{f,\bm x}=\sqrt{\bm u \nabla^2f(\bm x)\bm u } \qquad \ \ \ \  \text{and} \notag\\
	&\|\bm u\|_{(f,\bm x),\star}=\sqrt{\bm u \nabla^2f(\bm x)^{-1}\bm u }  \notag
\end{align}
which is the local norm of direction $\bm u$ induced by $f$ at $\bm x$ and its dual. These in turn, have several useful properties such as:
\begin{align}
	\|\bm x - \bm x^\star\|_{f,\bm x} \leq 2\|\nabla f(\bm x)\|_{(f,\bm x),\star}. \notag
\end{align}

One example of a self-concordant barrier function is:
\begin{align}
	&\Phi(\bm x)=-\sum_{i=1}^m\log\left( \dtp{\bm a_i}{\bm x} - b_i\right) \qquad \text{on the set}:\notag \\
	&\c X=\Big\{ \bm x: \dtp{\bm a_i}{\bm x}-b_i\geq 0, i=1,\ldots, m	\Big\} \notag
\end{align}


\bibliographystyle{IEEEtran}
\bibliography{references-optimistic-tutorial} 

\end{document}